\pdfoutput=1
\documentclass[12pt,twocolumn]{article}

\usepackage{docmute}
\usepackage{amsmath}
\usepackage{amssymb}
\usepackage{graphicx}
\usepackage{bm}
\usepackage{booktabs}
\usepackage{placeins}
\usepackage[colorlinks=true,citecolor=blue,linkcolor=blue]{hyperref}
\usepackage{xcolor}
\usepackage[margin=2cm]{geometry}
\usepackage{authblk}

\usepackage{setspace}
\usepackage[numbers,sort&compress]{natbib}
\usepackage{caption}
\newcommand{\ff}{{\bf f}}
\newcommand{\rr}{{\bf r}}
\newcommand{\xx}{{\bf x}}

\newcommand{\RR}{{\bf R}}
\DeclareMathOperator{\erf}{erf}

\begin{document}

\title{LeIFF: A localized-electron and ion force field for coupled nuclear and electronic dynamics}
\author[1,2]{Anthony Mannino}
\author[3]{Isidro Losada}
\author[7]{Simon Divilov}
\author[3]{Michelle Fritz}
\author[1,2]{Younos Hashem}
\author[1]{Ashraf Mrabet}
\author[4]{Paula Mori S\'anchez}
\author[5]{Javier Junquera}
\author[1,2,*]{Mariv\'i Fern\'andez-Serra}
\author[3,6,*]{Jose M. Soler}
\affil[1]{Physics and Astronomy Department, Stony Brook University, Stony Brook, New York 11794-3800, United States}
\affil[2]{Institute for Advanced Computational Science, Stony Brook University, Stony Brook, New York 11794-3800, United States}
\affil[3]{Dep.\ de F\'isica de la Materia Condensada, Univ.\ Aut\'onoma de Madrid, 28049 Madrid, Spain}
\affil[4]{Dep.\ de Qu\'imica, Univ.\ Aut\'onoma de Madrid, 28049 Madrid, Spain}
\affil[5]{Dep.\ de Ciencias de la Tierra y F\'isica de la Materia Condensada, Fac.\ de Ciencias, Univ.\ de Cantabria, 39005 Santander, Spain}
\affil[6]{Inst.\ de F\'isica de la Materia Condensada (IFIMAC), Univ.\ Aut\'onoma de Madrid, 28049 Madrid, Spain}
\affil[7]{Thomas Lord Department of Mechanical Engineering and Materials Science, Duke University, Durham, North Carolina, United States}
\affil[*]{\href{mailto:jose.soler@uam.es}{jose.soler@uam.es}; \href{mailto:marivi.fernandez-serra@stonybrook.edu}{marivi.fernandez-serra@stonybrook.edu}}
\date{\today}

\twocolumn[
\begin{@twocolumnfalse}
\maketitle
\begin{abstract}
Classical and machine-learned interatomic potentials enable large-scale simulations by replacing electronic structure with an effective energy surface for nuclear motion.
This excludes electronic dynamics, electronic excitation and energy exchange between electrons and ions.
First-principles electron dynamics can describe these processes, but its computational cost restricts system sizes and simulation times.
Electron force fields offer an intermediate description by treating electrons as semiclassical dynamical particles interacting with ions.
We introduce the localized-electron and ion force field (LeIFF), a compact physical model for both ground-state bonding and excited electron dynamics.
Valence electrons are represented by spin-labeled spherical charge distributions with dynamical centers and widths.
The energy combines analytic electronic kinetic, Coulomb, and dispersion terms with spline-interpolated short-range interactions fitted to a large molecular data set.
%
%
The resulting energy surface describes a variety of ground state properties across diverse bonding environments and charge states.
It also provides a qualitative description of electronic behavior away from equilibrium.
It shows improved robustness, accuracy, and transferability relative to earlier electron force-fields.
Without any solid-state training, the same model describes dielectric screening and collective electronic response in silicon.
LeIFF thus opens a route to large-scale simulations of nonadiabatic processes in molecules and solids, combining chemical bonding and nonequilibrium electronic motion within a common physical framework.

\end{abstract}
\end{@twocolumnfalse}
]

\section{Introduction}

First-principles electronic-structure methods provide a systematic description of bonding and electronic response, but their computational cost restricts the length and time scales accessible to molecular dynamics~\cite{martin2004electronic,MarxHutter2009AIMD}.
Classical force fields extend these scales by replacing the electronic problem with effective interactions between atoms~\cite{allen1989computer,frenkel2002understanding}.
Polarizable models, including induced-dipole and Drude-oscillator approaches, describe electronic response through auxiliary variables, while charge-equilibration and reactive models permit changes in charge distribution or bonding~\cite{applequist1972dipole,lamoureux2003drude,rappe1991qeq,van2001reaxff}.
These approaches capture important effects of electronic redistribution without representing individual electrons.

Machine-learned interatomic potentials (MLIPs) replace fixed analytic forms with flexible representations fitted to electronic-structure data~\cite{behler2007nn,bartok2010gap,batzner2022nequip,batatia2022mace}.
They can reproduce potential-energy surfaces obtained from density-functional theory (DFT) with high accuracy within their training domain~\cite{unke2021mlff}.
The distinction relevant here is between learning an effective nuclear potential-energy surface and constructing a model in which electronic coordinates themselves can move.
The latter offers a direct route to coupling nuclear motion to an evolving, albeit approximate, electronic charge distribution.

Explicit-electron models occupy an intermediate position between atomistic force fields and electronic-structure methods.
Su and Goddard's electron force field (hereafter \mbox{SG-eFF}) represents individual electrons by floating Gaussian wave packets with dynamical centers and sizes~\cite{SuGoddard2007,SuGoddard2009}.
In its original formulation, the energy combines Gaussian-wave-packet kinetic and electrostatic terms with an effective spin-dependent Pauli interaction~\cite{SuGoddard2009}.
Applications of \mbox{SG-eFF} included the equation of state and shock compression of warm dense hydrogen, as well as assessments of ground-state systems with covalent, ionic, multicenter, and metallic bonding~\cite{SuGoddard2009}.
Later work addressed electronic effects during silicon fracture~\cite{theofanis2012fracture}.
More broadly, wave-packet molecular dynamics uses restricted electronic wave functions to approximate nonadiabatic dynamics in quantum plasmas; the treatment of exchange and the choice of wave-packet shape are central approximations~\cite{grabowski2014wpmd}.
Recent anisotropic wave-packet models illustrate the sensitivity of electronic structure and transport to the spherical-packet restriction~\cite{svensson2023wpmd}.
Recent benchmarking of \mbox{SG-eFF} for water radiation chemistry illustrates its ability to describe electronic stopping and radiation-induced dynamics, while also highlighting limitations in molecular energetics and quantitative reaction predictions~\cite{ovad2026radiation}.
These findings highlight the need to constrain and validate equilibrium chemistry and nonequilibrium electronic dynamics within a common framework.

A complementary line of work develops semiclassical electrons for molecular chemistry.
The LEWIS model uses independently mobile valence-electron pairs to describe water's polarization, flexibility, and dissociation~\cite{kale2012lewis}; subsequent models and applications are reviewed by Bai \textit{et al.}~\cite{bai2017semiclassical}.
Other approaches include eReaxFF, which incorporates explicit electrons into a reactive force-field framework~\cite{islam2016ereaxff}, and C-GeM, which represents atomic cores and electronic shells by Gaussian charge distributions~\cite{leven2019cgem}.
Cools-Ceuppens \textit{et al.} introduced the electron machine-learning potential (eMLP), whose explicit electron-pair positions are connected to localized molecular orbitals or Wannier centers~\cite{cools2022emlp}.
That model demonstrates accurate electronic response properties for molecular and crystalline benchmarks.
In its published molecular dynamics formulation, eMLP is restricted to the Born--Oppenheimer approximation: electron-pair positions are minimized at each nuclear time step and remain at their instantaneous electronic energy minimum~\cite{cools2022emlp}.
This treatment describes adiabatic electronic response but does not propagate independent nonequilibrium electronic motion.

We introduce the localized-electron and ion force field (LeIFF)~\cite{losada2022thesis,mannino2026thesis}, designed to describe molecular chemistry and nonequilibrium electronic dynamics within a common, physically grounded energy model.
%
%
Like \mbox{SG-eFF}, LeIFF propagates individual spin-labeled electron centers and widths together with the nuclei, allowing electronic motion away from its instantaneous energy minimum and energy exchange with nuclear motion.
%

The distinction from \mbox{SG-eFF} lies in more flexible electron-electron and electron-ion interactions and in their optimization for a wide range of molecular properties: electron centers and widths are fitted to localized Boys~\cite{boys1960localized} orbital calculations, while molecular geometries, energetics and response properties are fitted to a large reference data set.

LeIFF treats valence electrons explicitly and represents frozen-core ions for elements in the first three periods of the periodic table.
Gaussian charge distributions determine the electrostatic interactions, while spline functions describe the width-dependent kinetic coefficient and the short-range core--electron and spin-dependent electron--electron potentials.
Together with an explicit dispersion contribution, this representation allows a common functional form to accommodate different electron sizes, charge states, and bonding environments.

Boys-localized orbitals~\cite{boys1960localized} and maximally localized Wannier functions (MLWFs)~\cite{marzari1997mlwf,marzari2012review} provide the electronic-structure interpretation of the electron centers and widths.
In fitting, MLWF centers and spreads serve as explicit targets for electron positions and widths, anchoring the electronic configurations to first-principles reference data.
We fit these electronic targets jointly with molecular structures and energetics, polarizabilities, and vibrational and electronic normal-mode energies for 731 atoms, molecules, and ions containing elements through argon.
Together, these targets constrain the preferred electronic configurations, the relative energies of molecular states, and the local response of the coupled nuclear--electronic energy landscape.
This joint calibration seeks stable, chemically meaningful electronic configurations and consistent nuclear and electronic forces across diverse chemistries within one parameterization.
Gradients of equilibrium properties are evaluated through linear response at each relaxed geometry, capturing the effect of relaxation without differentiating through the geometry optimization or repeating it for each parameter perturbation.

We test chemical transferability on QM9~\cite{ramakrishnan2014qm9} molecules without refitting the force-field interactions and examine collective electronic response in silicon without fitting to solid-state data.
These tests probe the ability of a parameterization developed from atoms, molecules, and ions to describe both molecular chemistry and nonequilibrium electronic motion in a solid.

Section~\ref{sec:theEFF} presents the energy function and its spline representation.
Section~\ref{sec:training} describes initialization, relaxation, and parameter optimization.
Section~\ref{sec:results} assesses molecular ground-state properties, compares LeIFF with SG-ECP, discusses electronic normal modes, and tests transferability to QM9~\cite{ramakrishnan2014qm9} and crystalline silicon.
Section~\ref{sec:conclusions} summarizes the results and directions for further development.


\section{The LeIFF Model}
\label{sec:theEFF}

\subsection{Dynamical variables and LeIFF functional form}

The complete set of dynamical variables for a system of $N_I$ nuclei and $N_e$ electrons consists of the nuclear positions $\{\mathbf{R}_I\}$, the electron centers $\{\mathbf{r}_i\}$, and the electron widths $\{w_i\}$.
The radial spread of our electrons is $\langle r^2\rangle=3w_i^2$, providing a direct connection to the spread of Boys orbitals~\cite{boys1960localized} and maximally localized Wannier functions~\cite{marzari1997mlwf}.
Each electron also carries a fixed spin label $\sigma_i \in \{\uparrow,\downarrow\}$.
The total energy $E_\mathrm{tot}(\{\mathbf{R}_I\},\{\mathbf{r}_i\},\{w_i\})$ governs the coupled evolution of nuclei and electrons.
The nuclear positions and electron centers are assigned masses $M_I$ and $m_e$, respectively, whereas each width coordinate is assigned an effective mass $3m_e$.
For a spherical Gaussian, the width describes an isotropic dilation in three spatial directions, giving a width kinetic term $3m_e\dot w_i^2/2$.
This choice is consistent with Gaussian wave-packet dynamics~\cite{grabowski2014wpmd}.

In the present model, all interactions are \mbox{pairwise}.
The total energy is
\begingroup
\allowdisplaybreaks[1]
\begin{align}
E &= \sum_i \frac{c_\mathrm{K}(w_i)}{w_i^2} + \sum_{I<J} \frac{z_I z_J}{r_{IJ}} \nonumber \\
 &\quad + \sum_{i<j} \frac{1}{r_{ij}}
\,\mathrm{erf}\!\left(\frac{r_{ij}}{\sqrt{2}\,w_{ij}}\right) \nonumber \\
 &\quad -\sum_{I,j} \frac{z_I}{r_{Ij}}
\,\mathrm{erf}\!\left(\frac{r_{Ij}}{\sqrt{2}\,w_j}\right) \nonumber \\
 &\quad + \sum_{i<j} V_{\mathrm{vdW}}(r_{ij},w_i,w_j) \nonumber \\
 &\quad + \sum_{Ij} V^\mathrm{cv}_{Z_I}(r_{Ij},w_j) \nonumber \\
 &\quad + \sum_{i<j} V^\mathrm{vv}_{\sigma_i\sigma_j}(r_{ij},w_i,w_j).
 \label{eq:Etot}
\end{align}
\endgroup
where uppercase and lowercase indices label ions and valence electrons, respectively, $Z_I$ and $z_I$ are the atomic number and the valence charge of atom $I$, $r_{ij}=|\rr_i-\rr_j|$, and $w_{ij}=\sqrt{w_i^2+w_j^2}$.
Atomic units ($\hbar = m_e = e = 4\pi\varepsilon_0 = 1$) are used throughout.

The first term is the kinetic self-energy of the independent electrons, and the $w^2$ denominator captures its main width dependence.
We allow an additional width dependence of the coefficient $c_\mathrm{K}(w)$, parameterized with a cubic spline interpolation.
This dependence is discussed in Section~\ref{sec:si-ck} of the Supplementary Information (SI), and the fitted coefficient is shown in Fig.~\ref{fig:ck}.

Terms (2-4) are long-range electrostatic interactions between point-like ions, between localized electrons, and between ions and electrons, respectively.
In these terms, we use Gaussian electron distributions that yield convenient analytic formulas. However, we are {\em not} assuming that our localized electrons have a Gaussian radial dependence because the short-range potentials $V^\mathrm{cv}$ and $V^\mathrm{vv}$ absorb corrections to the Gaussian shape.

The fifth term is the van der Waals energy.
The last two terms are short-range electron-ion and electron-electron interactions parameterized with 2D and 3D cubic spline interpolations, respectively.

\subsection{Van der Waals Interaction}

Though frequently treated with atom-atom potentials, van der Waals (vdW) interactions arise from electron-electron correlations.
Indeed, they can be described as a change of zero-point energy of the electron vibrations \cite{London1930,Tkatchenko2012}.
Then, the vdW attraction between two subsystems arises from the decrease of the (sum of) vibration frequencies produced by their dipole-dipole electrostatic coupling.
LeIFF offers a straightforward implementation of this method, by constructing and diagonalizing the dynamical matrix of all the electronic degrees of freedom and subtracting the sum of (the square root of) its diagonal terms from the sum of its eigenvalues.

Although this approach accounts for many-body correlations, it requires an expensive $O(N^3)$ diagonalization to calculate a relatively minor energy term.
Therefore, we have also implemented an alternative treatment in Eq.(\ref{eq:Etot}) of the vdW energy with (valence) electron-electron pairwise potentials, neglecting the small effect of ion-core polarization.
However, the $-1/r^6$ decay of the vdW potential does not allow to include it in our short range e-e potential and we use a specific term:
\begin{equation}
    E_\mathrm{vdW}(r_{ij},w_i,w_j) = -\frac{1}{4}\sum_{i<j}\frac{D_{ij}^2}{\Omega_i\Omega_j(\Omega_i+\Omega_j)}
    \label{eq:Evdw}
\end{equation}
with $\Omega_i=\tfrac12\sqrt{w_i^{-4}+w_0^{-4}}$, $w_0=3$ Bohr, $D_{ij}^2 = f(r_{ij}/w_{ij})/w_{ij}^6$, and
\begin{eqnarray}
    f(x) &=& \frac{1}{x^6} \left( 6\erf^2\left(\frac{x}{\sqrt{2}}\right) \right. \nonumber \\
         &&  -4\sqrt{\frac{2}{\pi}}e^{-x^2/2}\erf\left(\frac{x}{\sqrt{2}}\right)(x^3+3x) \nonumber \\
         &&  \left. + \frac{2}{\pi}e^{-x^2}(x^6+4x^4+6x^2) \right).
\end{eqnarray}
For $r_{ij}\rightarrow\infty$ and $w_{ij}<<w_0$, this reduces to $-C_{6,ij}/r_{ij}^6$ with
$C_{6,ij}=12w_i^4w_j^4/(w_i^2+w_j^2)$.
A full derivation of Eq.~\eqref{eq:Evdw} is given in Section~\ref{sec:si-vdw} of the SI.

\subsection{Short-range interactions}

In a cubic spline interpolation of a function $f(x)$, we must specify $n$ knots $x_i$, $i=1,\dots,n$, the endpoint derivatives $f'=df/dx$ at $x_1$ and $x_n$, and the function value $f_i=f(x_i)$ at each knot. The interpolating function is a cubic polynomial in each interval $[x_i, x_{i+1}]$, whose coefficients enforce continuity of the function and its first two derivatives at each interior knot~\cite{Press2007}.
It can be written as
\begin{equation}
  f(x) = p_0(x) + \sum_{i=1}^n f_i ~p_i(x)
  \label{eq:spline1}
\end{equation}
where $p_i(x)$ is a piecewise cubic polynomial, with continuous value and first two derivatives, obeying $p_i(x_j)=\delta_{ij}$
and $p'_i(x_1)=p'_i(x_n)=0$.
$p_0(x)$ is also a piecewise cubic polynomial obeying $p_0(x_i)=0$ and the specified end derivatives for $f(x)$.
Alternatively, one may specify natural endpoint conditions, $f''(x)=0$, for one or both of $x_1$ and/or $x_n$, in which case $p''_i(x)=0$ at the endpoints.
In our interpolations, either $f'(x)=0$ (when this is known by symmetry) or $f''(x)=0$ at all endpoints, so that $p_0(x)=0$ in all cases.
This representation provides analytic derivatives $\partial^k f(x)/\partial x^k$ and $\partial f(x)/\partial f_i$.

A two-variable function $f(x,y)$ can also be expanded as
\begin{equation}
  f(x,y) = p_0(x)~q_0(y) + \sum_{i=1}^{n_1}  \sum_{j=1}^{n_2} f_{ij} ~p_i(x) ~q_j(y)
  \label{eq:spline2}
\end{equation}
where $q_j(y)$ are piecewise cubic polynomials in $[y_1,y_{n_2}]$.
This tensor-product construction extends directly to three or more dimensions.

The functions $c_\mathrm{K}(w)$, $V^\mathrm{cv}_{Z_I}(r_{Ij},w_j)$, and $V^\mathrm{vv}_{\sigma_i\sigma_j}(r_{ij},w_i,w_j)$ are parameterized as 1D, 2D, and 3D spline interpolations, respectively.
To improve smoothness and interpolation ranges, the independent variables are $\{\log(w)\}$, $\{r_{Ij}/w_j,\log(w_j)\}$, and $\{r_{ij}/w_{ij},\log(w_{ij}),\log(w_i/w_j)\}$ and the interpolated functions are $c_\mathrm{K}$, $\log(V^\mathrm{cv}_{Z})$, and $\log(|V^\mathrm{vv}_{\sigma\sigma'}|)$.
There are 18 functions $V^\mathrm{cv}_Z$ for $Z=1,...,18$ (first three periods of the periodic table) and two different functions $V^\mathrm{vv}_{\sigma\sigma'}$, for $\sigma=\sigma'$ (parallel spin) and $\sigma=-\sigma'$ (opposite spin).
We use 6 knots for $c_\mathrm{K}$, $5\times5$ knots for $V^\mathrm{cv}_Z$, and $5\times5\times3$ knots for $V^\mathrm{vv}_{\sigma\sigma'}$ and the function values at the knots are our LeIFF optimization parameters (606 parameters in total).

$V^\mathrm{vv}_{\sigma\sigma}$ (parallel spins) is positive due to Pauli repulsion.
$V^\mathrm{cv}_Z$ is also repulsive for $Z>2$ because of the Pauli repulsion between the ion core and the valence electron.
For $Z=1,2$ the interaction is negative for small $r/w$ and positive afterwards because of the deviation of the electron shape from the Gaussian used in the electrostatic term.
In this case, we use $\sinh(V^\mathrm{cv}_Z/E_\mathrm{c})$ with $E_\mathrm{c}=10^{-2}$ Ha, rather than $\log(V^\mathrm{cv}_Z)$, as the interpolated function.

\begin{figure}[t]
  \centering
  \includegraphics[width=0.9\columnwidth]{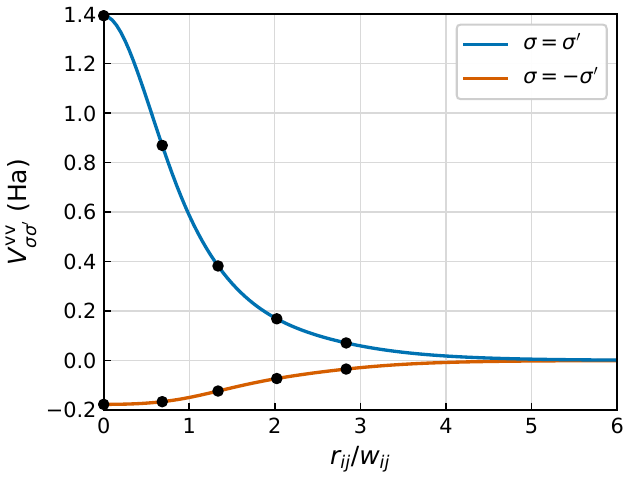}
  \caption{Short-range interaction potentials between two valence electrons with $w_i=w_j=1$ Bohr. The parallel-spin interaction is positive due to Pauli repulsion, whereas the antiparallel-spin interaction is negative due to correlation effects, favoring electron pairing.}
  \label{fig:Vee}
\end{figure}

$V^\mathrm{vv}_{\sigma\sigma'}$ is negative for $\sigma=-\sigma'$ (opposite spins) due to correlation effects.
Figure \ref{fig:Vcv} in the SI shows $V^\mathrm{cv}_Z(r/w)$ for $Z=1,\ldots,18$ and $w=0.95$ Bohr
(most $w$ values are in the range $0.5<w<1.5$ Bohr).
Figure \ref{fig:Vee} shows $V^\mathrm{vv}_{\sigma\sigma'}$ for parallel and antiparallel spins with $w_i=w_j=1$ Bohr.
Notice that $|V^\mathrm{vv}_{\sigma\sigma'}|$ is much larger for parallel than antiparallel spins. Thus, short-range correlation is much weaker than Pauli repulsion.

\section{Training Procedure}
\label{sec:training}

\subsection{Loss Function}

The LeIFF parameters are the function values $f_i$ at the spline interpolation knots $x_i$ of functions $c_\mathrm{K}$, $V^\mathrm{cv}_{Z}$, and $V^\mathrm{vv}_{\sigma\sigma'}$.
They are optimized by minimizing a multi-target loss function
\begin{equation}
\mathcal{L}(\ff) = \sum_I l\!\left(y_I(\ff),\,\sigma_I\right),
\end{equation}
where $\ff=\{f_i\}$ is the vector of all LeIFF parameters and $y_I(\ff)$ is the residual of the $I$-th target. $\sigma_I$ is the associated tolerance (not to be confused with the electron spins), with $1/\sigma_I^2$ playing the role of a weight assigned to the $I$-th term.
The single-target loss (shown in Fig.~\ref{fig:cutoff} in the SI) is
\begin{equation}
\label{eq:loss}
l(y,\sigma) = \frac{(y/\sigma)^2}{1 + |y/\sigma|/\lambda}
\end{equation}
with $\lambda=5$. This smooth robust loss is quadratic for $|y|/\sigma\ll\lambda$ and asymptotically linear for $|y|/\sigma\gg\lambda$. Its derivative approaches a finite magnitude, reducing the influence of extreme residuals. The quadratic-to-linear crossover follows the principle of robust estimation~\cite{huber1964robust}, although the rational form used here is distinct from the standard Huber loss.

The training targets and regularization terms fall into four groups:
\begin{enumerate}
\item{Geometric properties}: relaxed LeIFF atomic positions (compared with NIST CCCBDB B3LYP reference geometries~\cite{NIST_CCCBDB_2022}); electron positions and widths (compared with those of Boys-Wannier functions); deviations from reference electronic symmetries; dipole and quadrupole moments.
\item{Energies and response properties}: atomization energies; ionization energies; reaction energies (energy differences); isomer energy differences; vibrational frequencies; static molecular polarizabilities; and electronic normal-mode energies compared with QUEST excitation energies~\cite{veril2021questdb,loos2025quest}.
\item{Forces at reference geometries}: atomic LeIFF forces at the NIST CCCBDB B3LYP geometry~\cite{NIST_CCCBDB_2022} (compared with SIESTA/PBE forces calculated at that same geometry); LeIFF electronic forces at the Wannier-function positions and spreads (compared with zero).
\item{Regularization terms}
to enforce smoothness:
\begin{equation}
y_f[f] =
    \sum_{i,j=1}^{n} \int_{\Omega_f}
    \left( \frac{\partial^2 f}{\partial x_i \, \partial x_j} \right)^{\!2} d^n x
\end{equation}
for each $n$-dimensional spline-interpolated function $f$ over its interpolation domain $\Omega_f$.
\end{enumerate}
The DFT reference calculations use the SIESTA code~\cite{soler2002siesta,garcia2020siesta} with the Perdew--Burke--Ernzerhof (PBE) exchange--correlation functional~\cite{perdew1996generalized} and a double-zeta polarized basis set.

\subsection{Initialization of electronic degrees of freedom}

Like most force fields, LeIFF has multiple local minima as a function of the electron positions and widths.
Although a good initialization does not guarantee convergence to the global minimum, physically motivated initial electronic configurations are important for both training and simulation.
We construct initial configurations from Lewis--Linnett rules~\cite{lewis1916,linnet1961} and valence-shell electron-pair repulsion (VSEPR) theory~\cite{gillespie1972}.

To set the Lewis structure for given charge, spin, and atomic positions, we first identify two types of possible electron sites: 1) bonding sites between pairs of atoms separated by less than the sum of their covalent radii plus a distance tolerance; 2) lone electron sites on each atom.
The electron occupations of these sites are determined by minimizing a loss function that penalizes deviations from: 1) filled atomic valence shells (octet and spin-resolved quartet rules); 2) atomic formal charges; and 3) uniform electron distributions among equivalent sites. This is done while conserving the total number of valence electrons for each spin orientation.

The site occupations with lowest loss values define the most likely Lewis--Linnett configurations.
For each configuration, the electron positions are determined according to a generalization of VSEPR rules~\cite{gillespie1957} to open shells for spin-polarized systems.
Electron widths are assigned using empirical relations depending on bond length, bond order, and atomic electronegativities.

\begin{figure}[!]
  \centering
  \includegraphics[width=\columnwidth]{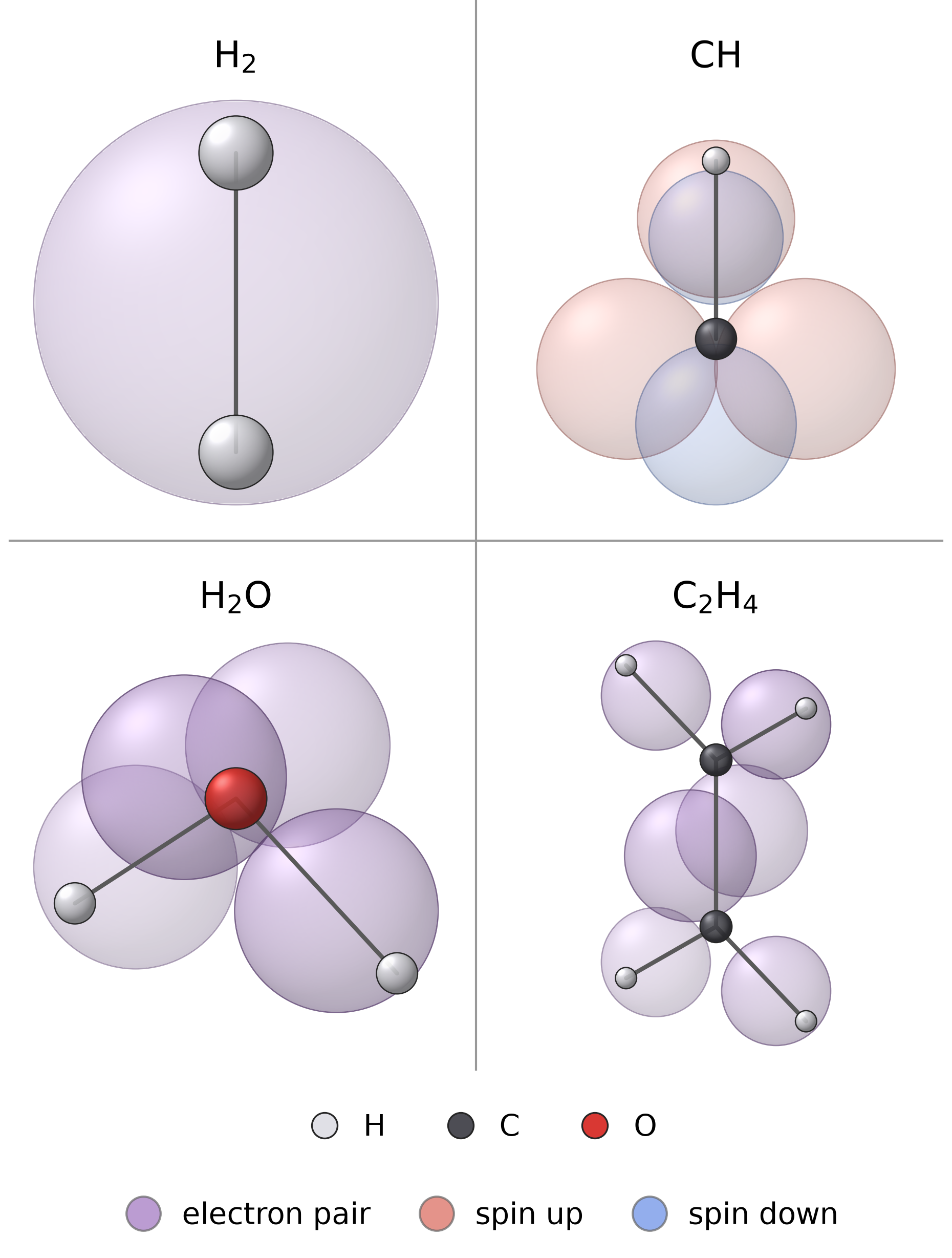}
  \caption{LeIFF localized electrons for four representative molecules at their relaxed geometries: H$_2$, CH, H$_2$O, and C$_2$H$_4$. Nuclei are shown as small spheres colored by element and connected by bonds, and localized valence electrons as translucent spheres. Red and blue denote opposite spins; coincident opposite-spin electrons are represented by a single purple sphere. Sphere centers and radii correspond to the LeIFF electron positions and widths $w_i$, respectively. Each molecule is scaled independently to fill its panel, and nuclear radii are schematic.}
  \label{fig:training-molecules}
\end{figure}

These Lewis--Linnett-VSEPR electron sites serve as initial projectors for MLWF calculations using the \textsc{Wannier90} code ~\cite{mostofi2014updatedwannier90}.
The Lewis-Linnet configuration leading to the most localized Wannier functions is finally selected and the converged Wannier centers and spreads provide the first-principles electronic training data for LeIFF.
The LeIFF localized-electron configurations for four representative small molecules are shown in Fig.~\ref{fig:training-molecules}.

\subsection{Molecular Relaxation}

Equilibrium-property targets are evaluated at relaxed geometries, which depend on the current LeIFF parameters. Force targets are instead evaluated at the specified reference geometries.
Equilibrium configurations are obtained by minimizing the total energy with respect to all molecular degrees of freedom (atomic positions $\{\mathbf{R}_I\}$, electron centers $\{\mathbf{r}_i\}$, and electron widths $\{w_i\}$) using the Newton--Raphson method~\cite{Nocedal2006},
\begin{equation}
\xx_{k+1} = \xx_k - H^{-1}(\xx_k)\,\nabla E(\xx_k),
\end{equation}
where $\xx=\{\RR_I,\rr_i,w_i\}$ and $H$ is the Hessian matrix, $H_{ij}=\partial^2E/\partial x_i/\partial x_j$.
First and second derivatives are calculated analytically from Eqs.~\eqref{eq:Etot}, \eqref{eq:Evdw}, and \eqref{eq:spline2}.

In general, the relaxed LeIFF geometry may be rotated with respect to the reference one.
Therefore, comparison requires alignment with respect to translations and rotations.
We alternate optimization of the rotation angles with matching of equivalent particles using the Hungarian assignment algorithm~\cite{kuhn1955hungarian,munkres1957assignment}.

\subsection{Linear Response Gradients}

Because the equilibrium geometry $\xx$ depends implicitly on the LeIFF parameters $f_\alpha$, computing $\partial\mathcal{L}/\partial f_\alpha$ requires the response of the relaxed geometry to parameter changes.
Differentiating the equilibrium condition $\partial E/\partial x_i = 0$ with respect to $f_\alpha$ and applying the chain rule yields
\begin{equation}
\frac{\partial x_i}{\partial f_\alpha}
= -\sum_j H^{-1}_{ij}\,\frac{\partial^2 E}{\partial x_j\,\partial f_\alpha},
\end{equation}
where $H$ is the Hessian evaluated at the relaxed geometry. In the implementation, $H^{-1}$ denotes a numerically regularized pseudoinverse, so that translational and rotational zero modes do not produce a singular response~\cite{golub1973pseudoinverse}. This local response assumes a stable stationary configuration and a fixed retained Hessian subspace; it does not describe discontinuous switches between minima.
Parameter gradients of any equilibrium property $y(\xx(\ff),\ff)$ then follow as
\begin{equation}
\frac{dy}{df_\alpha}
= \frac{\partial y}{\partial f_\alpha}
- \sum_{ij}
\frac{\partial y}{\partial x_i}
H^{-1}_{ij}
\frac{\partial^2 E}{\partial x_j\,\partial f_\alpha}.
\end{equation}
Here $\partial^2 E/(\partial x_j \partial f_\alpha)$ can be computed analytically from Eqs.~\eqref{eq:spline1} and \eqref{eq:spline2}.

Parameter gradients of polarizabilities and vibrational frequencies require derivatives up to third order,
$\partial^3 E/(\partial x_i \partial x_j \partial x_k)$ and $\partial^3 E/(\partial x_i \partial x_j \partial f_\alpha)$, also computed analytically.
For alignment-dependent targets, the coordinate response is additionally multiplied by the Jacobian of the alignment transformation as $f_\alpha$ changes.
Further details are given in the SI.

\subsection{Search over Local Minima}

As previously mentioned, LeIFF generally has multiple local minima for a given molecule.
To improve robustness, we augment the training loop with an explicit geometry search: candidate structures are generated by applying random perturbations to the current equilibrium geometry or to the LeIFF parameters, and each candidate is relaxed to a local minimum.
Geometries are then weighted by Boltzmann-like factors
\begin{equation}
P_i = \frac{\exp(-L_i/\sigma_L)}{\sum_j \exp(-L_j/\sigma_L)}, \quad \sigma_L \approx 1,
\end{equation}
where $L_i$ is the structural and response loss of the $i$th-minimum geometry. These weights favor low-loss candidates while retaining contributions from competing minima; they are not thermodynamic populations. 
An additional energy-ordering penalty discourages competing candidates whose energies lie below the energy of the lowest-loss minimum. This encourages consistency between the preferred reference-like structure and the energy landscape.

\begin{figure}[t]
  \centering
  \includegraphics[width=\columnwidth]{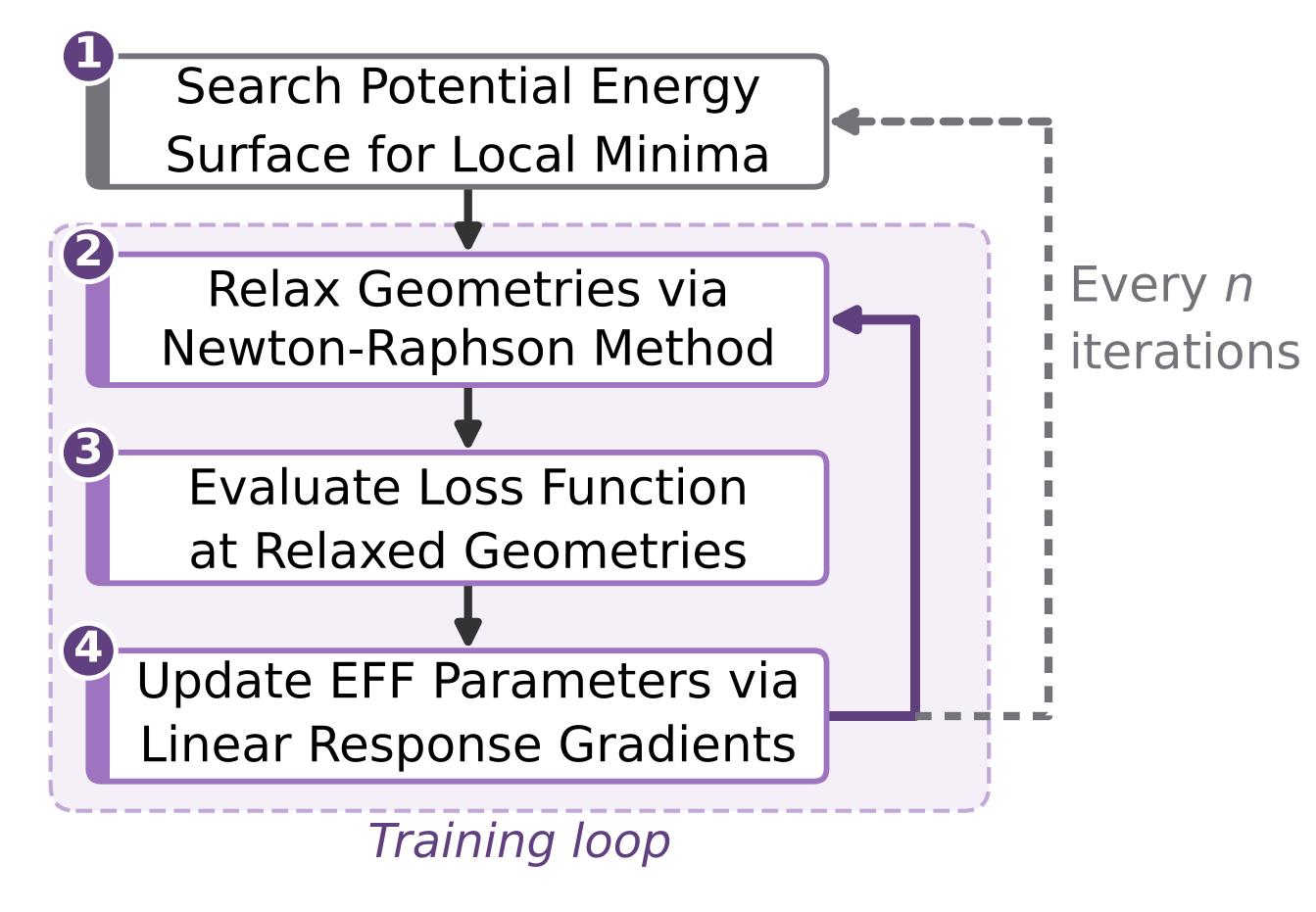}
  \caption{Schematic of the LeIFF training procedure.
  At every iteration, (i) molecular geometries are relaxed by Newton--Raphson minimization, (ii) the loss function is evaluated at the relaxed geometries, and (iii) parameters are updated via linear response gradients.
  A search of alternative geometries is performed every $n$ iterations.}
  \label{fig:training}
\end{figure}

The complete training procedure involves a search over candidate geometries, molecular relaxation of each initial geometry, loss evaluation, linear response gradient computation, and parameter update via the conjugate-gradient or Adam~\cite{Kingma2015} algorithms. It is summarized in Fig.~\ref{fig:training}.

\subsection{Training Data}

Our model was trained on 731 atoms, molecules, cations and anions from the NIST Computational Chemistry Comparison and Benchmark Database (CCCBDB)~\cite{NIST_CCCBDB_2022} with atoms from the first three periods of the periodic table (up to Ar).
Reference-data sources and the fitting protocol are fully explained in the SI.
For each system, the reference nuclear geometry was taken from the NIST B3LYP calculations with the largest available basis set~\cite{NIST_CCCBDB_2022,Becke1993,Lee1988}.
The energies and atomic forces were subsequently calculated with SIESTA/PBE at those fixed nuclear coordinates, using a double-zeta plus polarization basis.
Wannier90 calculations based on the SIESTA/PBE electronic states supplied the localized-orbital centers and spreads.
Thus, the structural targets are NIST B3LYP geometries, whereas the energy, force, and localized-orbital targets are based on SIESTA/PBE calculations at those geometries.
Note that because of this choice, the reference PBE atomic forces need not vanish.
Reference excitation energies from the QUEST database~\cite{veril2021questdb,loos2025quest} were also included in the fitting loss, a discussion about the connection between these and the electronic energy surface is discussed in section~\ref{sec:electronic-excitations}.

\section{Results}
\label{sec:results}

We begin by examining the fully relaxed states of the 731 atoms, molecules, and ions used in fitting.
Of the 731 systems, four anions did not converge because they self-ionize.
Conventional MLIPs are typically fitted to energies and nuclear forces that sample a Born--Oppenheimer potential-energy surface.
LeIFF instead constructs an energy surface in both nuclear and explicit electronic coordinates; minimization over electron centers and widths determines its effective nuclear potential-energy surface.
The ground-state DFT reference calculations used here provide electronic configurations and their properties, but do not map the energy for arbitrary electron configurations.
Fitting must therefore constrain both the properties of the reference states and their stability and energetic ordering within this larger landscape.
Because of this, including a system in the fit does not by itself ensure recovery of its reference electronic and nuclear ground state.

This requirement leads to the double optimization described in Section~\ref{sec:training}: for each parameter set, nuclear and electronic coordinates are relaxed in an inner energy minimization, while an outer optimization adjusts the interactions to reproduce properties of the resulting states.
The comparisons below assess the structures and properties supported by the fitted electronic energy surface across diverse chemistries.
They use full nuclear and electronic relaxation, with force residuals evaluated separately at the prescribed reference configurations.
We first present these molecular results and the electronic normal-mode comparisons included in fitting, then assess transferability to systems not included in the fitting.
\subsection{Ground-State Structures and Properties}
\label{sec:molecular-ground-states}

The optimized spline interaction potentials are shown in Fig.~\ref{fig:Vee} and Fig.~\ref{fig:Vcv} in the SI.
The fitted potentials are smooth over the plotted ranges, consistent with the purpose of the smoothness regularization.

Root-mean-square (RMS) errors for the reported properties are listed in Table~\ref{tab:spline-rms}.

\begin{table}[t]
\centering
\caption{Root-mean-square (RMS) errors and fitting tolerances $\sigma_I$ of the properties included in the loss function for training LeIFF on 731 atoms, molecules, and ions.}
\label{tab:spline-rms}
\small
\setlength{\tabcolsep}{3pt}
\begin{tabular}{lcc}
\toprule
Property & Error & $\sigma_I$ \\
\midrule
Atomic position (\AA)          & 0.128 & 0.053 \\
Electron position (\AA)        & 0.321 & 0.265 \\
Electron width (\AA)           & 0.065 & 0.053 \\
Dipole moment (D)               & 1.71 & 0.508 \\
Quadrupole moment (D~\AA) & 7.75  & 13.45 \\
Polarizability (\AA$^3$)        & 9.45  & 7.41 \\
Atomization energy (eV)         & 4.36  & 2.72 \\
Isomerization energy (eV)       & 1.19  & 0.54 \\
Reaction energy (eV)            & 0.876 & 0.544 \\
Ionization energy (eV)          & 2.99  & 2.72 \\
Excitation energy (eV)          & 2.49  & 2.72 \\
Vibration frequency (cm$^{-1}$) & 430 & 329 \\
Atomic force (eV/\AA)           & 17.8  & 5.1 \\
Electron force (eV/\AA)         & 53.9  & 51.4 \\
Smooth $V^\mathrm{vv}_{\sigma\sigma}$ (arb.\ units) & 15.8 & 3.0 \\
Smooth $V^\mathrm{vv}_{-\sigma\sigma}$ (au) & 17.8 & 3.0 \\
Smooth $V^\mathrm{cv}_Z$ (au)   & 7.3 & 1.0 \\
\bottomrule
\end{tabular}
\end{table}

Parity plots for bond lengths and angles, electron positions and widths, dipoles, polarizabilities, atomization, reaction and ionization energies, and vibrational frequencies are shown in Figs.~\ref{fig:bondLength}--\ref{fig:vibrations} in the SI. The electron-width distributions for bonding and lone pairs in the relaxed training-set structures are shown in Fig.~\ref{fig:widthHist}.
LeIFF gives semi-quantitative agreement for several structural and energetic properties.
These comparisons quantify the fidelity of the states obtained by coupled nuclear and electronic relaxation under a common parameterization.
Transferability beyond these fitted systems is assessed in Sections~\ref{sec:qm9} and~\ref{sec:silicon}.
Typically, about 30 systems (roughly 5\% of the set) are clear outliers with errors substantially above the average.
Such property outliers motivate more flexible interactions and further analysis of initialization and competing minima.
In particular, the present spherical, pairwise ansatz cannot represent general angular or environment-dependent electronic effects.
Extensions of LeIFF to address these deficiencies are currently being explored and will be the subject of a future publication.


%
\subsection{Comparison with other electron force fields}
\label{sec:sg-comparison}

We compare LeIFF with the effective-core-potential implementation of
SG-eFF (SG-ECP~\cite{jaramillo2011eff}) distributed with
LAMMPS~\cite{thompson2022lammps,lammpsEffManual}.
Like LeIFF, SG-ECP describes valence electrons explicitly and represents
the cores implicitly.
The supplied ECP parameters cover C, N, O, Al, and Si, with H and He
treated explicitly, restricting the shared comparison to 262 systems.
Table~\ref{tab:sg-ecp-comparison} reports LeIFF errors over its full
available set alongside comparisons on identical property-specific
subsets.

\begin{table*}[t]
\begin{center}
\captionof{table}{Force convergence and property RMSEs for LeIFF and SG-ECP. The full LeIFF column covers the 731-state benchmark inventory. For convergence, the matched LeIFF column uses the chemical domain supported by the comparator; for each property, the last two columns use identical available systems and features. Parentheses give system counts.}
\label{tab:sg-ecp-comparison}
\small
\setlength{\tabcolsep}{5pt}
\begin{tabular}{lrrr}
\toprule
Quantity & LeIFF: full & LeIFF: matched & SG-ECP \\
\midrule
Force convergence & 99.5\% (727/731) & 99.2\% (260/262) & 28.2\% (74/262) \\
Bond length (\AA) & 0.126 (668) & 0.0702 (60) & 0.266 (60) \\
Bond angle ($^\circ$) & 12.3 (460) & 8.55 (51) & 16.7 (51) \\
Atomization energy (eV)$^\dagger$ & 4.08 (310) & $4.78^{+0.07}_{-0.11}$ (88) & $64.5^{+302.6}_{-64.0}$ (88) \\
Dipole magnitude (D) & 1.59 (422) & 1.31 (39) & 5.15 (39) \\
Polarizability (\AA$^3$) & 17.5 (639) & 7.53 (58) & 36.6 (58) \\
\bottomrule
\end{tabular}
\par\smallskip
\begin{minipage}{0.97\textwidth}\footnotesize
Force-convergence and stability diagnostics are defined in the Supplementary Information.
$^\dagger$ Matched atomization estimates use the expanded 88-molecule energy assessment.
Asymmetric bars give propagated sampled-state/evaluator sensitivity ranges, not confidence intervals.
\end{minipage}
\end{center}

\end{table*}
\vspace{0.5em}

LeIFF combines broader chemical coverage with substantially more reliable
relaxation under the tested workflows.
Within the shared 262-system domain, 260 LeIFF relaxations satisfied the
force criterion (99.2\%), compared with 74 for SG-ECP (28.2\%).
On matched subsets, LeIFF also gives smaller errors in bond lengths,
bond angles, dipole magnitudes, and polarizabilities.
The bond-length RMSE is 0.0702~\AA\ for LeIFF, compared with
0.266~\AA\ for SG-ECP, while the fixed-nuclei polarizability RMSEs
are 7.53 and 36.6~\AA$^3$, respectively.
Atomization-energy estimates are additionally reported for all
88 reference-eligible neutral molecules. 
As some of the isolated atoms ground states are difficult to converge for SG-ECP,
we provide values, with the uncertainty in the atomic and molecular energies
retained explicitly.

Together, these results demonstrate the broad applicability and practical
reliability of LeIFF across this benchmark.
Full computational details and the additional  comparison with the original SG-eFF model
are provided in the SI.

\subsection{Transferability evaluation}
\label{sec:qm9}

To assess how well the model transfers beyond its training data, we evaluated it on 112,292 molecules from the QM9 database~\cite{ramakrishnan2014qm9}, 
which contains organic molecules with up to nine heavy atoms (C, N, O, and F).
Reference Wannier centers and spreads, energies, and dipole moments were obtained from SIESTA/PBE calculations.
Reference polarizabilities and vibrational frequencies were taken from the published QM9 dataset~\cite{ramakrishnan2014qm9}.
Reference sources and evaluation protocols are detailed in the SI.
The overlap between the training set of molecules and the QM9~\cite{ramakrishnan2014qm9} set is less than 0.03$\%$.

Atomization energies are computed as $E_{\mathrm{at}} = \sum_a \varepsilon_{Z_a} - E_{\mathrm{mol}}$, where the five atomic reference energies $\varepsilon_Z$ (H, C, N, O, F) 
are determined by a least-squares fit to the reference atomization energies.
These five offsets calibrate the atomic energy reference without changing the LeIFF interactions, molecular structures, or response properties; the atomization-energy error therefore measures residuals after this calibration.

Root-mean-square (RMS) errors for the reported properties are listed in Table~\ref{tab:qm9-rms}.

\begin{table}[t]
\centering
\caption{Root-mean-square (RMS) errors for 112,292 QM9~\cite{ramakrishnan2014qm9} molecules, using the property-specific reference sources described in the text. Atomization energies use the fitted atomic references described in the text.}
\label{tab:qm9-rms}
\small
\setlength{\tabcolsep}{3pt}
\begin{tabular}{lc}
\toprule
Property & Error \\
\midrule
Bond distance (\AA)             & 0.047 \\
Bond angle ($^\circ$)           & 4.56 \\
Electron width (\AA)            & 0.029 \\
Dipole moment (D)               & 2.07 \\
Polarizability (\AA$^3$)        & 1.42 \\
Atomization energy (eV)         & 2.21 \\
Vibration frequency (cm$^{-1}$) & 250 \\
\bottomrule
\end{tabular}
\end{table}

\begin{figure}[!ht]
  \centering
  \includegraphics[width=0.9\columnwidth]{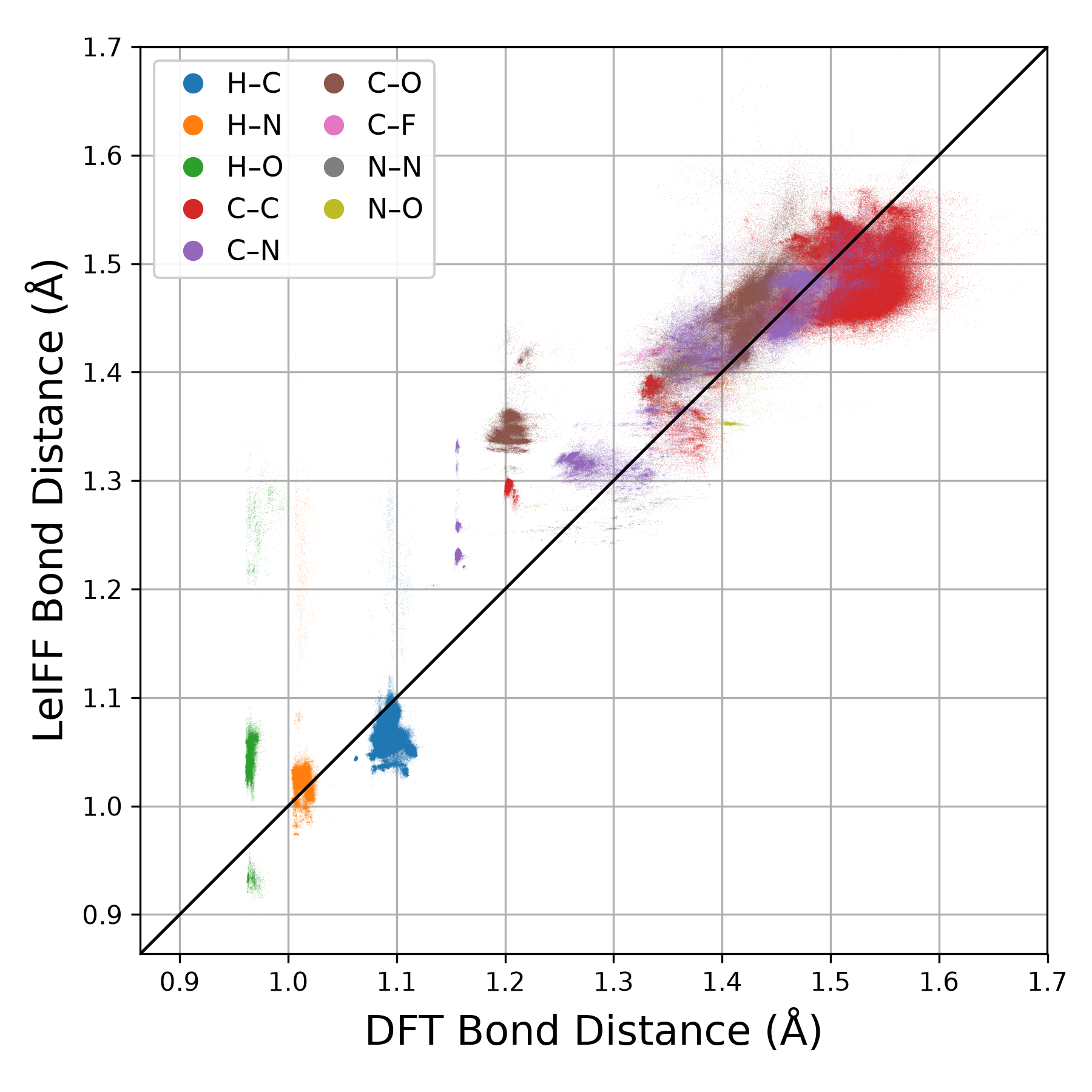}
  \caption{Bond distances calculated with LeIFF compared with published QM9~\cite{ramakrishnan2014qm9} reference structures.}
  \label{fig:bondDistanceQM9}
\end{figure}

\begin{figure}[!ht]
  \centering
  \includegraphics[width=0.9\columnwidth]{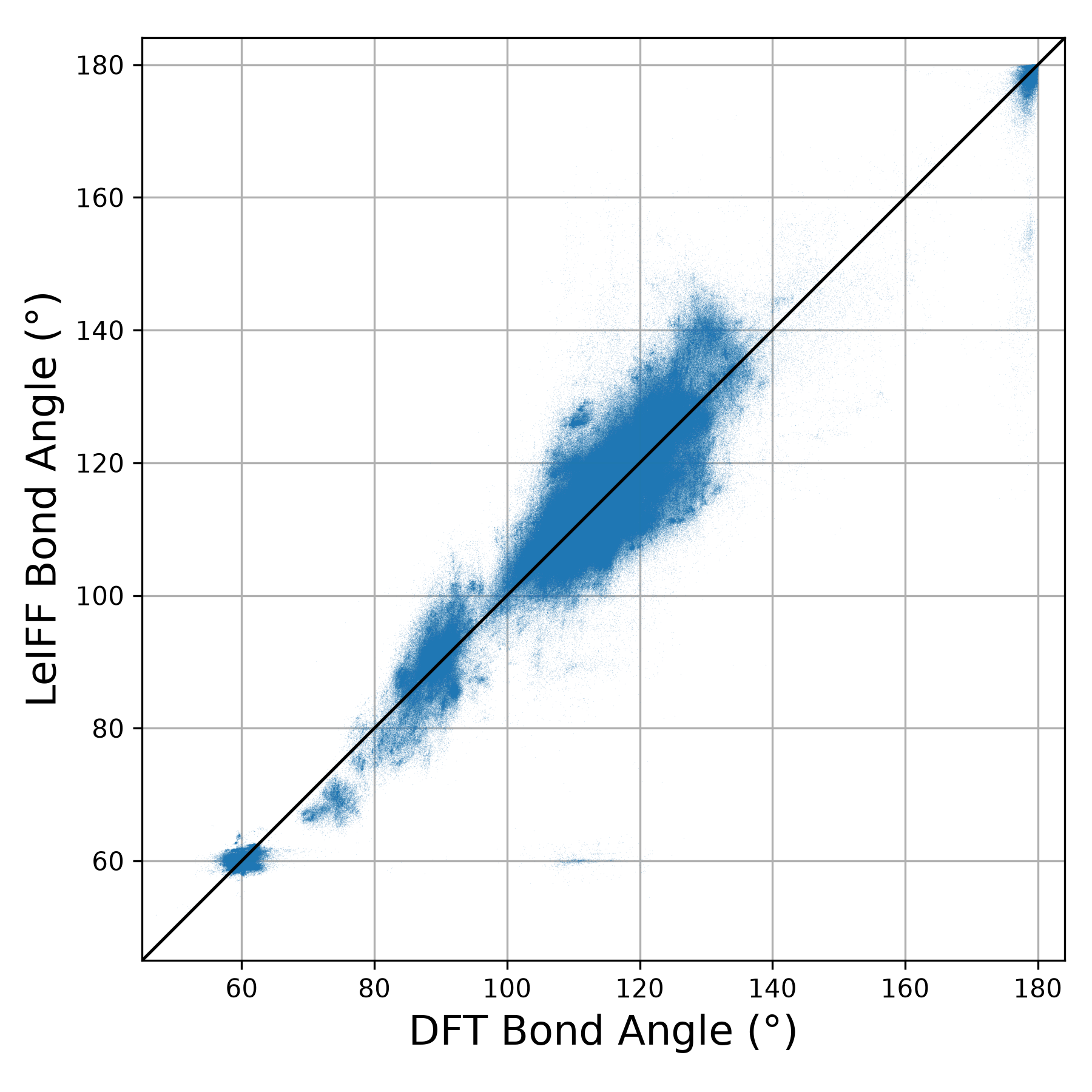}
  \caption{Bond angles calculated with LeIFF compared with published QM9~\cite{ramakrishnan2014qm9} reference structures.}
  \label{fig:bondAngleQM9}
\end{figure}

\begin{figure}[!ht]
  \centering
  \includegraphics[width=0.9\columnwidth]{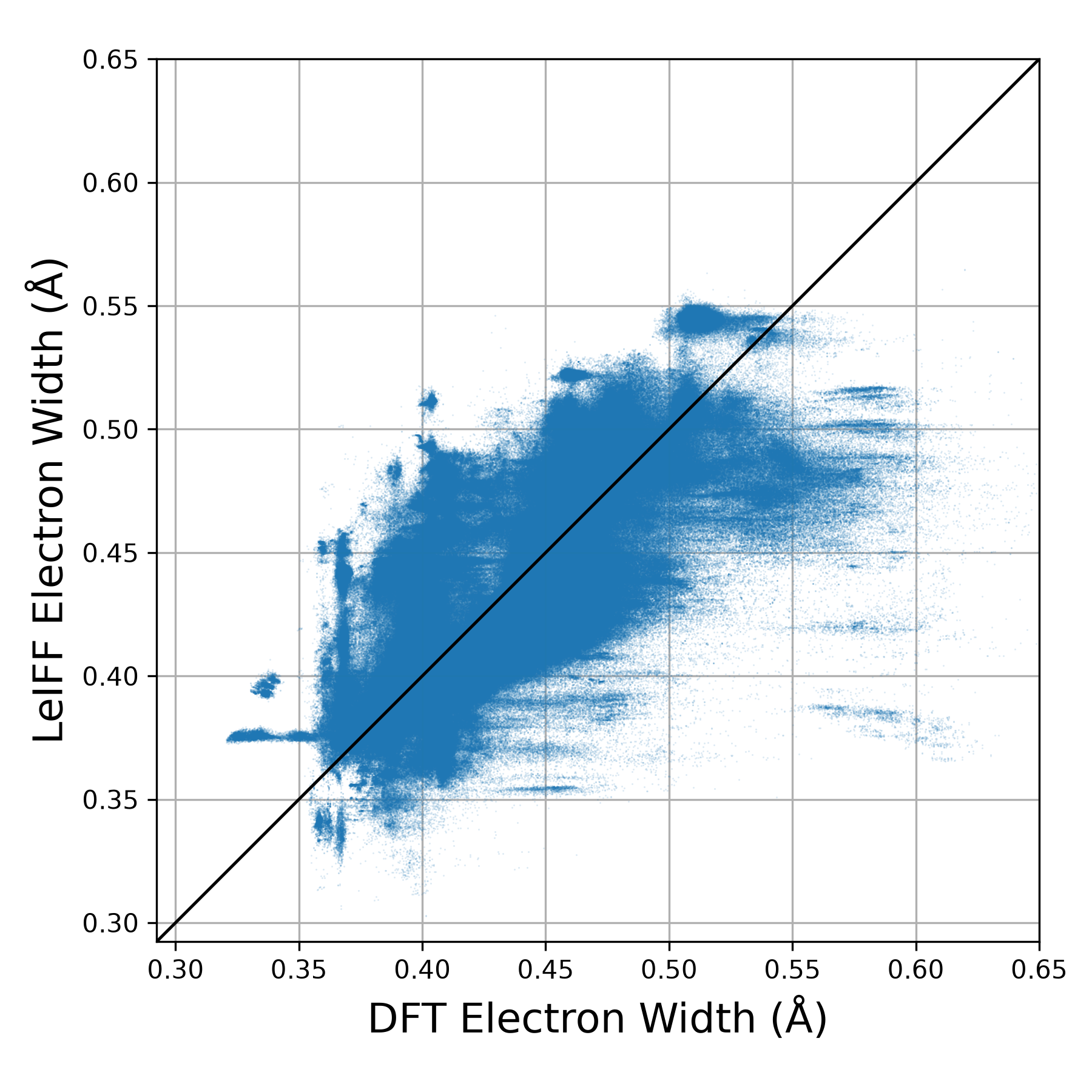}
  \caption{Electron widths calculated with LeIFF compared with DFT Wannier-function radii of the published QM9~\cite{ramakrishnan2014qm9} structures.}
  \label{fig:electronWidthQM9}
\end{figure}

\begin{figure}[!ht]
  \centering
  \includegraphics[width=0.9\columnwidth]{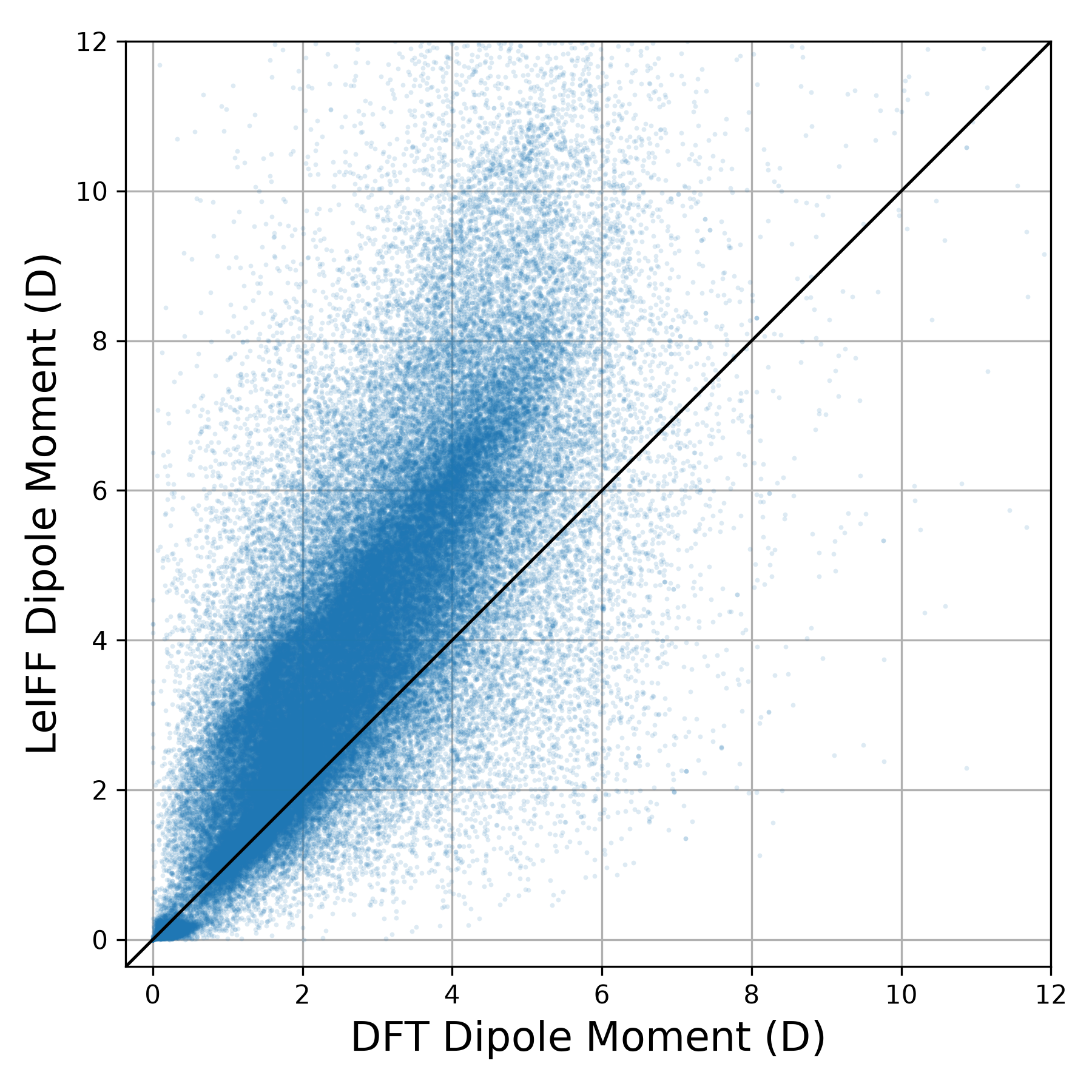}
  \caption{Dipole moments calculated with LeIFF compared with dipole moments calculated from the published QM9~\cite{ramakrishnan2014qm9} reference structures.}
  \label{fig:dipoleQM9}
\end{figure}

\begin{figure}[!ht]
  \centering
  \includegraphics[width=0.9\columnwidth]{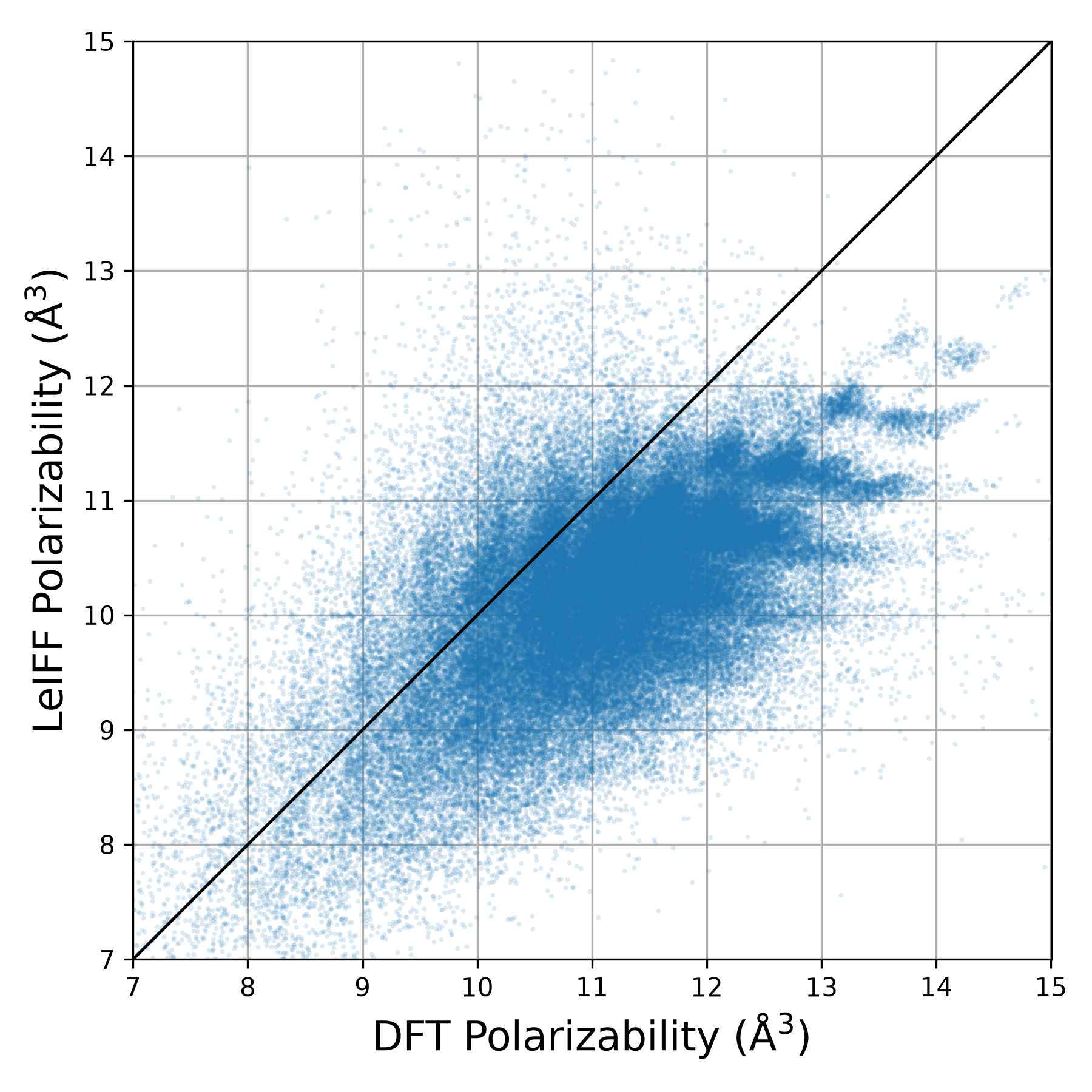}
  \caption{Polarizabilities calculated with LeIFF compared with published QM9~\cite{ramakrishnan2014qm9} reference values.}
  \label{fig:polarizabilityQM9}
\end{figure}

\begin{figure}[!ht]
  \centering
  \includegraphics[width=0.9\columnwidth]{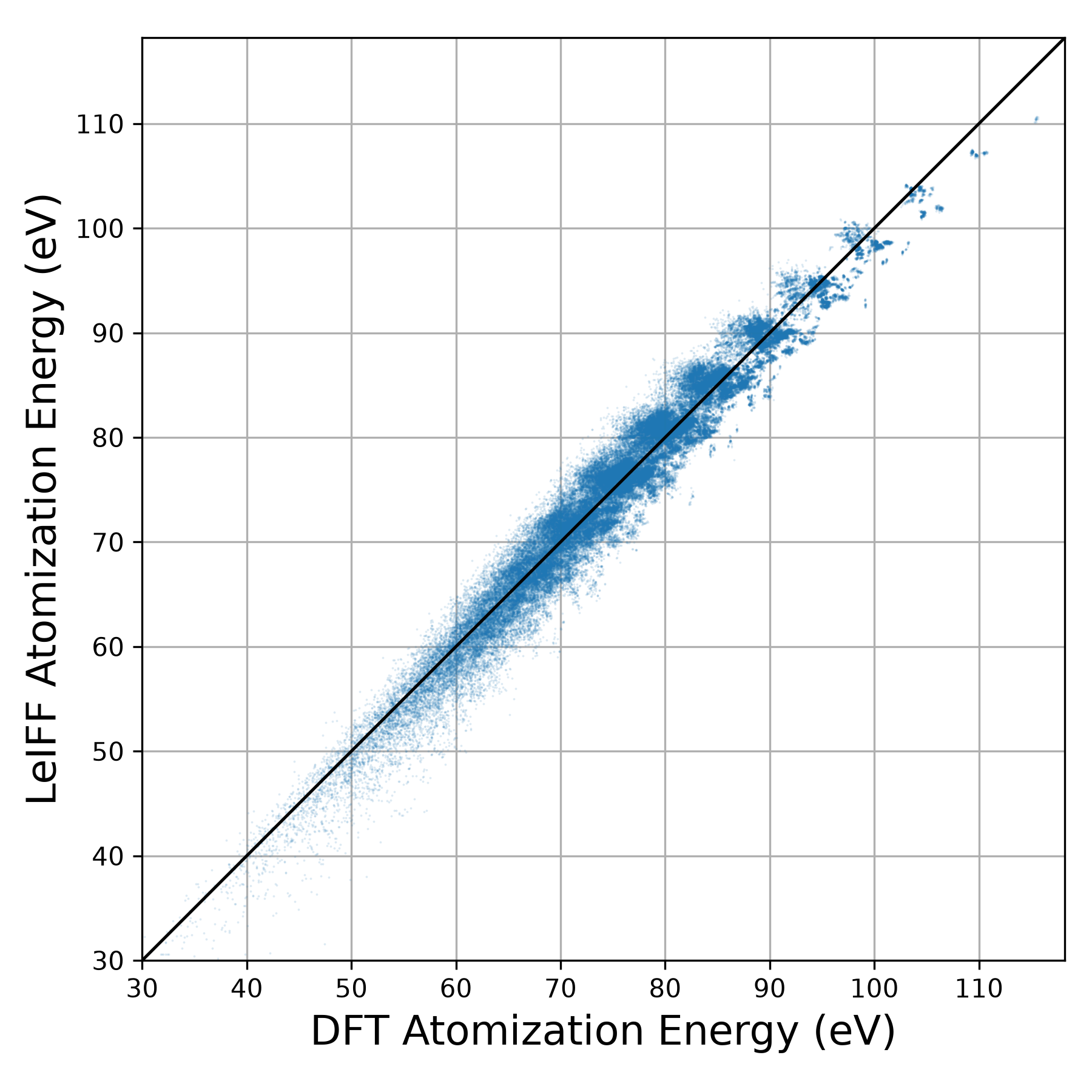}
  \caption{Atomization energies calculated with LeIFF compared with SIESTA/PBE reference values for the QM9 molecules~\cite{ramakrishnan2014qm9}. LeIFF atomic reference energies for H, C, N, O, and F are fitted by least squares to the reference atomization energies.}
  \label{fig:atomizationQM9}
\end{figure}

\begin{figure}[!ht]
  \centering
  \includegraphics[width=0.9\columnwidth]{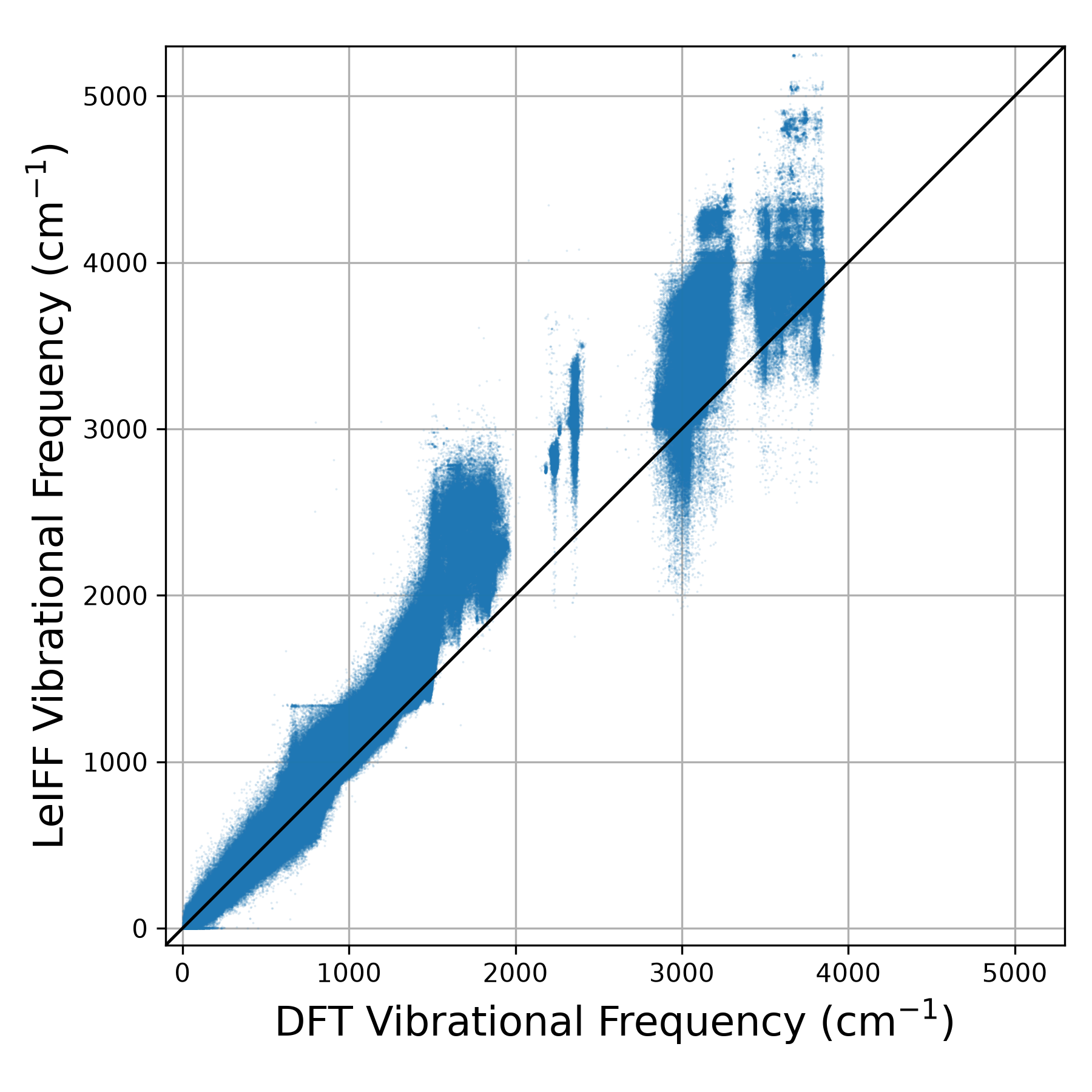}
  \caption{Vibrational frequencies calculated with LeIFF compared with published QM9~\cite{ramakrishnan2014qm9} reference frequencies. For each molecule, both sets are sorted and compared position by position.}
  \label{fig:vibrationsQM9}
\end{figure}

Parity plots for selected properties are shown in Figs.~\ref{fig:bondDistanceQM9}--\ref{fig:vibrationsQM9} and Figs.~\ref{fig:si-bondDistanceQM9}--\ref{fig:si-vibrationsQM9} in the SI.

Together, these property comparisons test transferability to molecules substantially larger than those in the fitting set. The structural benchmark shows that multiple bonds are too long and that some nominally linear geometries bend, while the vibrational comparisons show overly stiff multiple bonds. Systematic discrepancies also remain in the response properties: dipole moments are overestimated, and the polarizabilities of the most polarizable molecules are underestimated. These discrepancies point to specific targets for improving the localized-electron representation.

\subsection{Electronic Excitation Energies}
\label{sec:electronic-excitations}

Electronic excitation energies can be obtained from the resonances of the electronic response to a weak perturbation.
In time-dependent density-functional theory (TDDFT), the Casida equations determine these energies from the linearized response of the electronic system~\cite{casida1995response,casida2012progress}.
The Casida equations describe this response through orbital-transition amplitudes.

LeIFF describes changes in the electronic density through the motion of electron centers and variations of their widths.
We therefore examine the resonances of this response by linearizing the coupled electron--ion equations of motion about the fully relaxed LeIFF minimum.
This gives
\begin{equation}
H\,\mathbf{u}_{\nu}
=
\omega_{\nu}^{2}M\,\mathbf{u}_{\nu},
\label{eq:electronic-response}
\end{equation}
where $H$ is the full energy Hessian, $M$ is the diagonal mass matrix, and $\mathbf{u}_{\nu}$ contains nuclear displacements, electron-center displacements, and electron-width variations.
Electron centers and widths have masses $m_e$ and $3m_e$, respectively.
Modes associated primarily with electronic motion generally occur at higher frequencies than nuclear vibrations.

The normal-mode frequencies define possible resonances of the model's electronic response.
Their contributions to a particular observable depend on the mode character and its coupling to the applied perturbation.
This provides a physical motivation for comparing the electronic mode energies $\hbar\omega_\nu$ with molecular excitation energies.
The comparison assesses how well LeIFF's reduced electronic description reproduces these energies; it does not assume a correspondence between individual modes and quantum excited states.

In atomic units, an electronic mode of angular frequency $\omega$ is compared with an excitation energy through $\hbar\omega=\omega$.
Width oscillations resemble radial breathing, whereas center displacements resemble dipolar motion.
For hydrogen, these motions have the qualitative symmetry of $1s\rightarrow2s$ and $1s\rightarrow2p$ excitations, respectively; 
this analogy does not make them equivalent to quantum transitions.

\begin{figure}[t]
  \centering
  \includegraphics[width=\columnwidth]{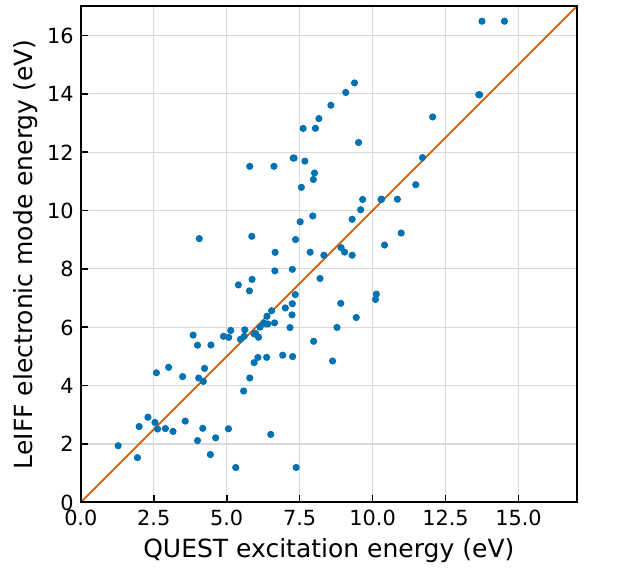}
  \caption{Comparison of LeIFF electronic normal-mode energies with QUEST reference excitation energies~\cite{veril2021questdb,loos2025quest} included in fitting. Near-zero frequencies of modes that correspond to rotations about a cylindrical symmetry axis have been excluded.}
  \label{fig:excitation}
\end{figure}

Figure~\ref{fig:excitation} compares electronic normal-mode energies with reference excitation energies from the QUEST database~\cite{veril2021questdb,loos2025quest}.
Because these excitation energies were included in fitting, the qualitative correspondence measures the ability of the representation to reproduce the supplied electronic energy scales; it is not an independent test of predictive accuracy.
%
%
A systematic exploration of how far LeIFF can represent molecular excited states will be the focus of a dedicated future study, including state assignments and coupling to external perturbations.

The silicon study in Section~\ref{sec:silicon} provides a complementary assessment by following the induced electronic density and extracting dielectric and energy-loss spectra.
These observables connect the reduced LeIFF dynamics directly to the density-response description used in TDDFT~\cite{weissker2010silicon}, extending the discussion from molecular electronic modes to collective excitations in a solid.

\subsection{Transfer to crystalline Silicon}
\label{sec:silicon}

\begin{figure}[!t]
  \centering
  \includegraphics[width=0.99\columnwidth]{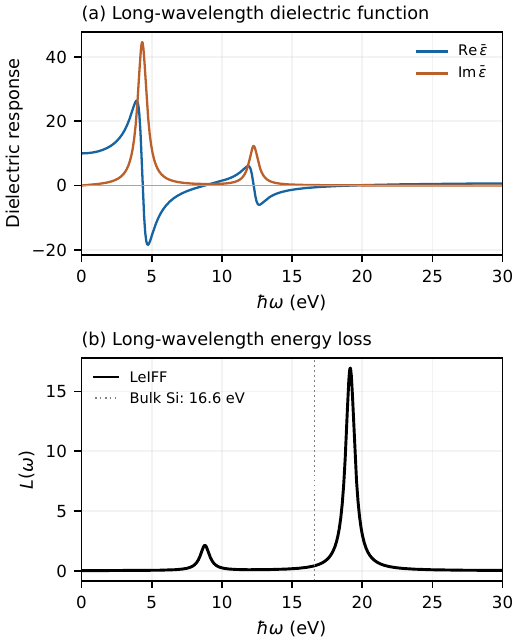}
  \caption{Collective electronic response in Si with pairwise dispersion included.
  (a) Long-wavelength dielectric function.
  (b) Long-wavelength energy-loss spectrum; the dotted line marks the experimental bulk-plasmon energy~\cite{chen1975siliconloss}.
  Numerical damping and averaging procedures are specified in the Supplementary Information.}
  \label{fig:si-summary}
\end{figure}

Silicon provides a test of transfer from molecular bonding to the structural, vibrational, and electronic properties of a crystal.
Using the LeIFF parameterization presented in this work and without re-fitting to any solid-state property,
we first calculate the diamond-Si equation of state and adiabatic phonon dispersion.
Although LeIFF predicts a bound diamond-lattice minimum and the qualitative acoustic and optical branch structure of silicon, it does not
produce good quantitave results.
The equilibrium lattice constant of $5.20$~\AA\ and bulk modulus of $238$~GPa are too compact and too stiff compared with experiment ($a_{\mathrm{exp}} = 5.43$~\AA, $B_{\mathrm{exp}} = 170$~GPa), and the optical phonon frequencies are overestimated.
Nonetheless, given the qualitative physics agreement, these tests establish a baseline for examining the electronic dynamics of the same model in a solid
and allows us to compare and test how far the current model can 
go in the description of coupled electronic and ionic response.
We first study the dielectric response, which probes a purely electronic mechanism with the ions held fixed.
We do so by analyzing the dynamic response of electrons to a weak perturbation.
In particular, a spatially uniform velocity kick of $1.25\times10^{-3}$ atomic units is applied to every electron center along \(x\), \(y\), and \(z\) in separate trajectories
and the system is allowd to evolve for 8000 steps of 0.1 atomic units.
We substract the unperturbed current response from these trajectories in order to compute the dielectric function. 
Details are provided in the SI.
Figure~\ref{fig:si-summary} shows the resulting dielectric function and longitudinal energy-loss function, $-\operatorname{Im}[1/\varepsilon(\omega)]$.
The structured dielectric spectrum demonstrates electronic screening, while the prominent loss feature shows that the localized-electron representation supports a collective excitation in crystalline silicon.

The dominant broadened loss peak lies near $19.1$~eV, above the measured bulk-Si plasmon near $16.6$~eV~\cite{chen1975siliconloss}.
Thus the plasmon-like feature emerges without solid-state training, although its energy indicates an excessively stiff response of the electrons participating in this excitation.
The electronic normal-mode analysis in the SI provides complementary evidence of collective longitudinal density motion.

\begin{figure*}[t]
  \centering
  \resizebox{0.88\textwidth}{!}{\input{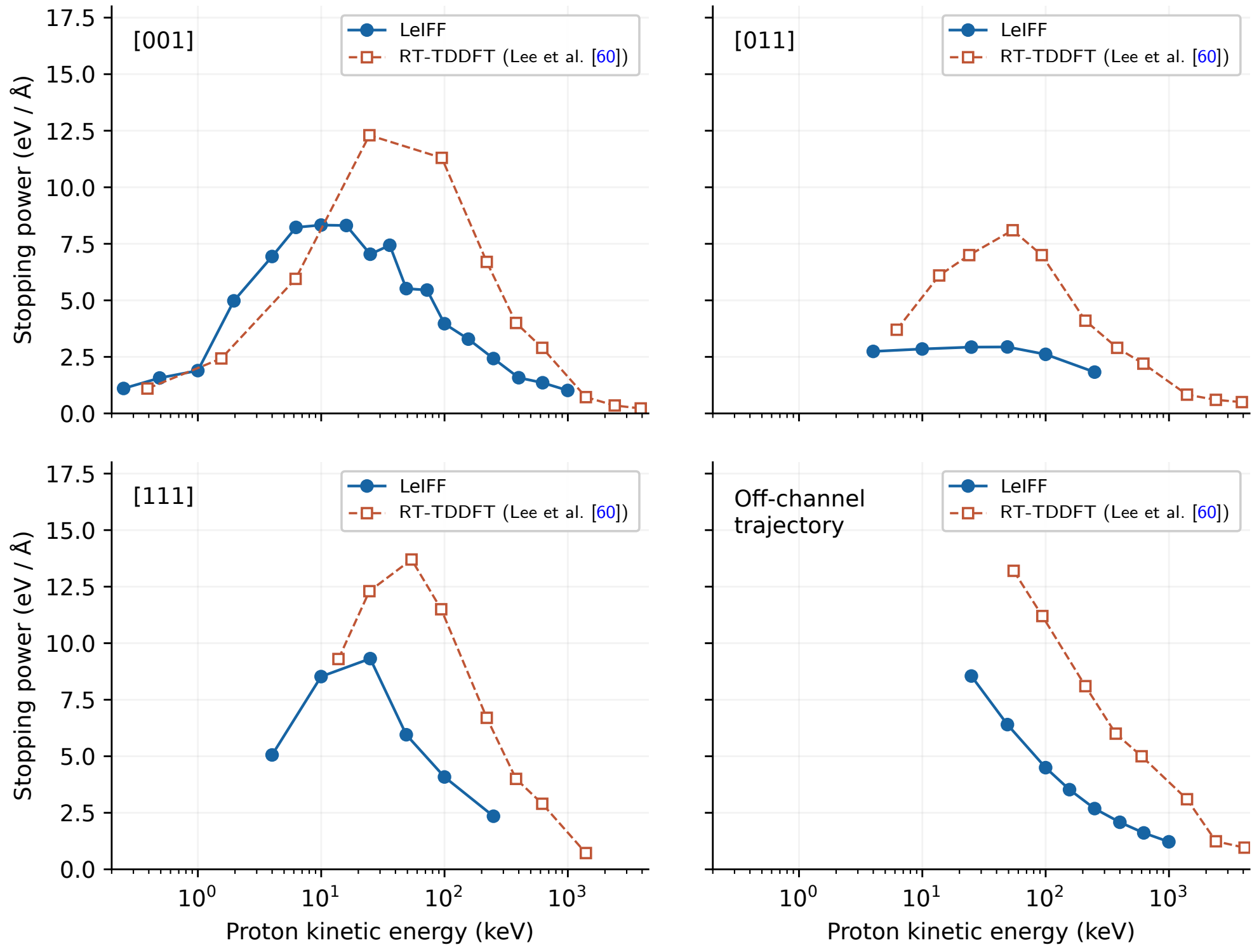}}
  \caption{Proton electronic stopping in silicon calculated with LeIFF (filled circles), compared with digitized RT-TDDFT results from Lee et al.~\cite{lee2020stopping} (open squares).
  The panels show three crystallographic channels and a sampled off-channel trajectory.
  Lines guide the eye. Numerical qualifications, including width-envelope exits in eight trajectories, are documented in the Supplementary Information.}
  \label{fig:si-stopping-main}
\end{figure*}

Electronic stopping provides a further test of this response under nonadiabatic conditions.
As an ion moves through the crystal, it transfers energy to electronic excitations, a mechanism absent from ground-state Born--Oppenheimer dynamics~\cite{lee2020stopping}.
The loss function connects to this process through the electronic excitations available to absorb energy from the moving projectile.
Silicon is particularly suitable for testing this mechanism because published RT-TDDFT calculations 
provide reference proton-stopping curves over a broad range of projectile energies and crystallographic directions~\cite{lee2020stopping}.
We therefore use proton stopping to assess whether the electronic degrees of freedom that support screening and collective oscillations also describe electronic energy deposition.
We used supercells containing 1000 Si atoms, compared with 216 in
Lee et al.~\cite{lee2020stopping}, while adopting the same proton
channeling directions and integration time step of $0.0145$ atomic units.
Figure~\ref{fig:si-stopping-main} compares the LeIFF stopping curves with the  RT-TDDFT results.
The calculations reproduce the characteristic rise and fall with projectile energy and qualitative channel dependence, including the reduced stopping along $\langle011\rangle$.
The stopping magnitude is generally underestimated, particularly near the maximum.
This trend is consistent with the elevated collective excitation energy: an electronic response that is too stiff may absorb less energy from the projectile, 
although excitation strengths and projectile coupling also affect stopping.
The comparison therefore demonstrates the energy-transfer mechanism while identifying quantitative response as a target for refinement.

Taken together, lattice binding and vibrations, electronic screening, a plasmon-like collective excitation, and 
electronic stopping provide a strong proof of concept for LeIFF beyond its molecular training domain.
Their emergence within a single framework motivates refinement of the interactions for solid-state environments, 
rather than fitting each observable through a separate phenomenological model.
Training on solid-state data is ongoing and will form part of the next version of LeIFF.
With improved quantitative interactions, its explicit electronic dynamics and computational efficiency offer a route to 
large-scale nonadiabatic simulations beyond the practical reach of current first-principles methods.
All additional silicon results, including the equation of state, phonon bands, electronic-mode analysis, 
and detailed stopping tests, are presented in the SI, together with computational procedures and numerical qualifications.

\section{Conclusions}
\label{sec:conclusions}

LeIFF demonstrates that explicit, localized electrons can be treated within a semiclassical model that describes both chemical bonding and electronic response.
We attribute LeIFF's robustness to the combined development of the interaction model, the electronic reference data, and the optimization procedure.
In the benchmarks presented here, LeIFF combines broader chemical coverage and more reliable relaxation with smaller structural and response errors than the earlier electron force-field implementations tested.
These capabilities extend beyond the systems used in fitting.
Tests on a large set of organic molecules demonstrate molecular transferability, while crystalline silicon provides a more demanding change of environment.
Without solid-state training, LeIFF produces dielectric screening, collective electronic oscillations, and electronic stopping.
The molecular normal modes also capture qualitative trends in excitation energies, although their correspondence with specific excited states requires further study.
Together, these results show that the explicit electronic variables can support both ground-state bonding and nonequilibrium energy transfer within the same model.

Although substantial quantitative errors remain, LeIFF robustness provides a foundation for building more accurate models of coupled electron--ion dynamics.
Environment-dependent many-body interactions, including neural-network representations, offer a way to extend the present spherical, pairwise description while retaining the explicit electronic variables and the physical constraints of LeIFF.

Further progress also requires first-principles reference data that probe the electronic energy surface away from its ground-state minimum.
Conventional ground-state datasets constrain equilibrium electronic configurations and their response, but do not map the energy cost of general changes in electron positions and widths.
This calls for further development of electronic-structure methods and data-generation strategies that sample controlled departures from equilibrium electron distributions.
Such data would allow the electronic landscape explored by LeIFF to be constrained and tested more directly.
LeIFF thus opens a route toward large-scale nonadiabatic simulations in which chemical bonding, electronic excitation, and energy transfer are described through the coupled motion of electrons and ions.

\section*{Acknowledgments}

We thank E.~Artacho, E.~Hernandez, B.~Levine,
and C.~Dreyer for many discussions and ideas.

This work was supported by Grant
PGC2018-096955-B-C42, funded by
MICIU/AEI/10.13039/501100011033 and by
ERDF (``A way of making Europe'');
Grant PID2022-139776NB-C64, funded by
MCIN/AEI/10.13039/501100011033 and by ERDF/EU;
and the ``Unidad de Excelencia Mar\'ia de Maeztu''
Grant CEX2023-001316-M, funded by
MICIU/AEI/10.13039/501100011033.

J.J. acknowledges financial support from Grant
PID2022-139776NB-C63, funded by
MCIN/AEI/10.13039/501100011033 and by ERDF/EU.


This work was also supported by the National Science Foundation under award DMR-2427902 and by a seed award from the Institute for Advanced Computational Science (IACS) at Stony Brook University.
The authors would like to thank Stony Brook Research Computing and Cyberinfrastructure and IACS at Stony Brook University for access to the high-performance SeaWulf computing system, funded by the National Science Foundation (awards 1531492 and 2215987) and matching funds from Empire State Development's Division of Science, Technology and Innovation (NYSTAR) program (contract C210148).

The original Matlab programs to fit LeIFF \cite{losada2022thesis,mannino2026thesis} were entirely coded by the authors. The authors used the AI assistants Codex (OpenAI) and Claude (Anthropic) to help translate the code to JAX, perform some of the test comparisons, and draft parts of this article. All AI-assisted code and text were reviewed and verified by the authors, who take full responsibility for the content of this article.




\bibliographystyle{unsrt}
\bibliography{references}

\onecolumn
\newcommand{\SIincluded}{}          
\section*{Supplementary Information}
\setcounter{section}{0}\setcounter{figure}{0}\setcounter{table}{0}\setcounter{equation}{0}
\renewcommand{\thesection}{S\arabic{section}}
\renewcommand{\thefigure}{S\arabic{figure}}
\renewcommand{\thetable}{S\arabic{table}}
\renewcommand{\theequation}{S\arabic{equation}}
\renewcommand{\theHsection}{S\arabic{section}}
\renewcommand{\theHfigure}{S\arabic{figure}}
\renewcommand{\theHtable}{S\arabic{table}}
\renewcommand{\theHequation}{S\arabic{equation}}

\maketitle

\section{Width dependence of kinetic energy}
\label{sec:si-ck}

The first term of main-text Eq.~\eqref{eq:Etot} is the kinetic self-energy of the independent electrons, and the $w^2$ denominator captures its main width dependence.

\begin{figure}[h]
  \centering
  \includegraphics[width=0.6\columnwidth]{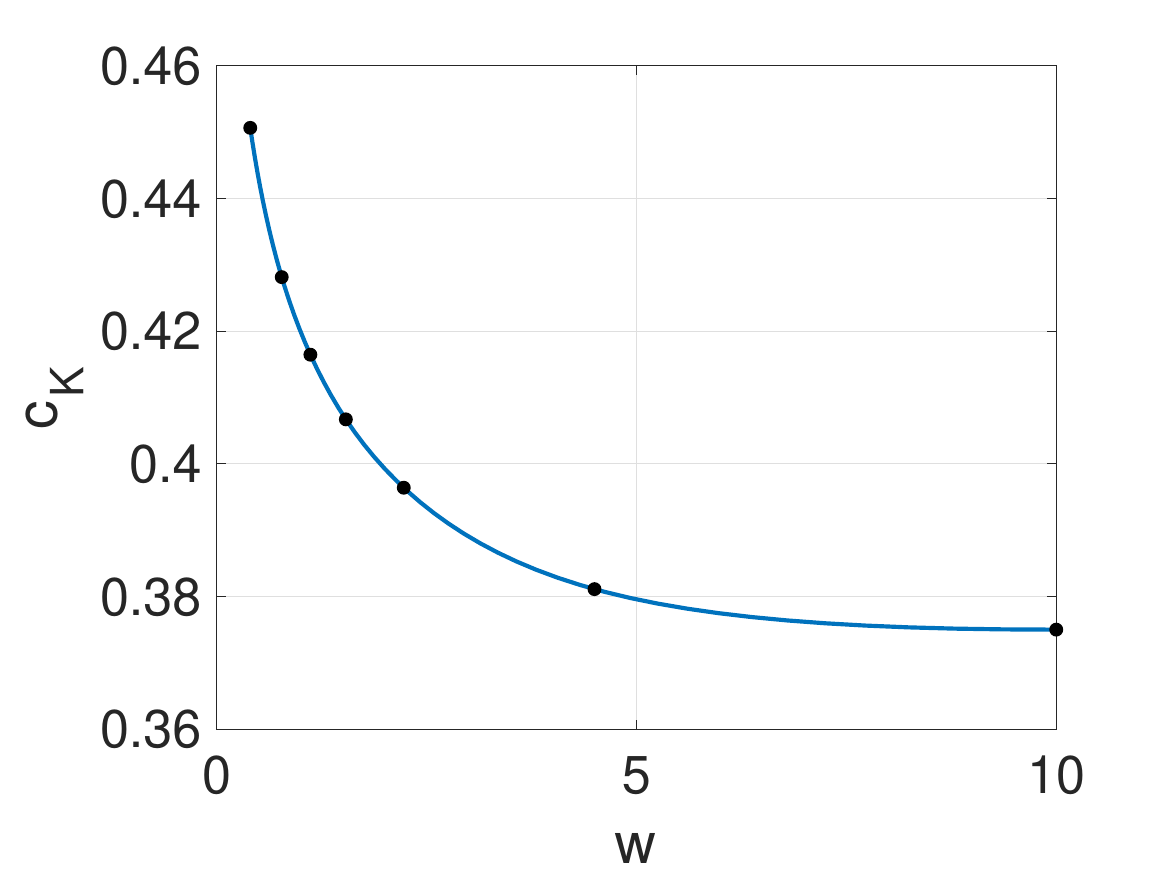}
  \caption{Width dependence of the kinetic energy coefficient $c_\mathrm{K}(w)$ of the first term of main-text Eq.~\eqref{eq:Etot}. The dots are the knots of the cubic spline interpolation. The value at $w=10$ is fixed by the lower bound of the kinetic energy of a wave packet, given by that of a Gaussian ($c_\mathrm{K}=3/8=0.375$). For a hydrogenic, exponential function, $c_\mathrm{K}=1/2$.}
  \label{fig:ck}
\end{figure}

For an exponential function with density $\rho(r)=\psi^2(r)=\exp(-2r/w)/(\pi w^3)$, one finds $c_K=1/2$ and for a spherical Gaussian with $\rho(r)=\exp(-(r/w)^2/2)/(2\pi w^2)^{3/2}$, $c_K=3/8$.
We allow an effective width dependence of the kinetic coefficient to accommodate deviations from these idealized radial shapes, for example between bonding and lone electrons.
This dependence is accounted for by the numerator $c_\mathrm{K}(w)$, parameterized with a cubic spline interpolation shown in Fig.~\ref{fig:ck}.

\FloatBarrier

\section{Detailed derivation of van der Waals energy term}
\label{sec:si-vdw}

\begin{figure}[t]
  \centering
  \includegraphics[width=0.6\columnwidth]{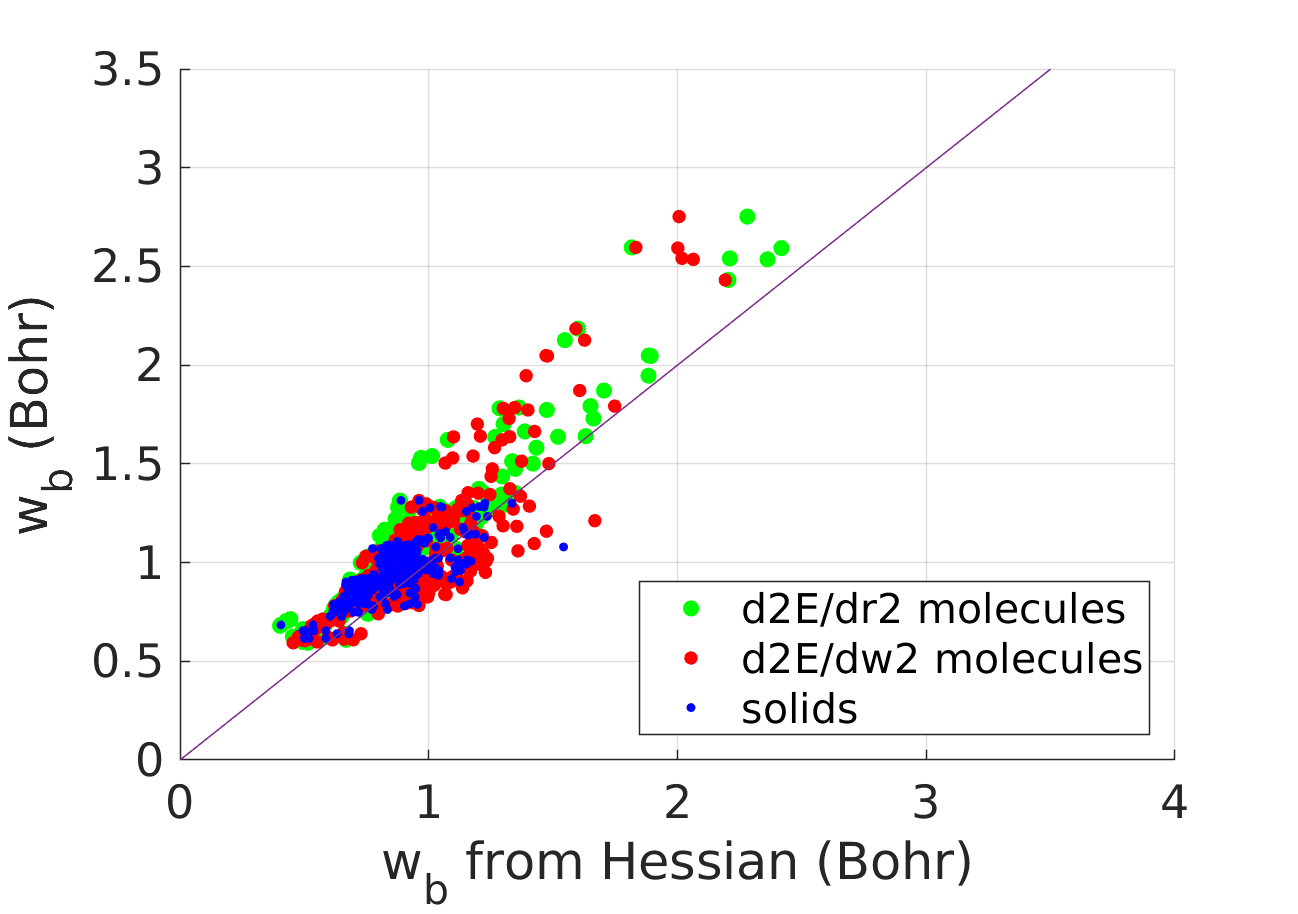}
  \caption{Comparison between relaxed e-ball widths and harmonic widths of the EFF Hessian, for a set of 186 atoms, molecules, and solids.}
  \label{fig:wb_consistency}
\end{figure}

First, we consider each e-ball as an electron in an effective harmonic potential whose frequency $\Omega$ and force constant $K$  can be obtained from the e-ball width $w$ as $\Omega=1/(2 w^2), K=\Omega^2$. 
Figure~\ref{fig:wb_consistency} assesses this approximation, where we compare the actual e-ball widths at the relaxed geometry with those obtained from the relationship $w_i=(4K_i)^{-1/4}$, where $K_i$ is the harmonic force constant of the $i$th e-ball in the EFF potential of the remaining electrons and ions. 

The frequencies of the collective electron modes are given by the eigenvalues $(\Omega_i+\Delta\Omega_i)^2$ of the dynamical matrix $D_{ij\alpha\beta} = \partial^2 E / (\partial r_{i\alpha} \partial r_{j\beta})$, where $(\alpha,\beta)$ label Cartesian coordinates.
The diagonal terms are approximated by $D_{ii\alpha\beta}=\Omega_i^2\delta_{\alpha\beta}$ and the effect of the non-diagonal terms is approximated by second-order perturbation theory.
For a pair of oscillators, the coupling Hamiltonian is
$\hat V_{ij}=\sum_{\alpha\beta}D_{ij\alpha\beta}\hat q_{i\alpha}\hat q_{j\beta}$.
Its matrix element between the uncoupled ground state and a state with one excitation in each of the coupled directions is $D_{ij\alpha\beta}/(2\sqrt{\Omega_i\Omega_j})$.
Second-order perturbation theory therefore gives a pair energy
$-D_{ij}^2/[4\Omega_i\Omega_j(\Omega_i+\Omega_j)]$.
The change in total zero-point energy is then
\begin{eqnarray}
    E_\mathrm{vdW} &=& \frac{1}{2} \sum_n \Delta\omega_n  \nonumber \\
        &\simeq& -\frac{1}{2} \sum_{i\ne j}\sum_{\alpha,\beta} 
          \frac{ |D_{ij\alpha\beta}|^2 }
               { 4\Omega_i \Omega_j (\Omega_i+\Omega_j) } \nonumber \\
        &=& -2\sum_{i<j} \frac{w_i^4 w_j^4}{w_{ij}^2} D_{ij}^2 \nonumber \\
        &=& -2\sum_{i<j} \frac{w_i^4 w_j^4}{w_{ij}^2}
             \left[ \left( \frac{d^2 V^C_{ij}}{d r_{ij}^2} \right)^2 + \right. \nonumber \\
        && \hspace{2cm} \left. 2\left( \frac{1}{r_{ij}} \frac{dV^C_{ij}}{dr_{ij}} \right)^2 \right] \nonumber \\
        &=& -2\sum_{i<j} \frac{w_i^4 w_j^4}{w_{ij}^8} f\left(\frac{r_{ij}}{w_{ij}}\right)    \label{eq:si-VdW}
\end{eqnarray}
where $n$ labels the $3N_e$ oscillator modes, $V^\mathrm{C}_{ij}=\erf(r_{ij}/(\sqrt{2}w_{ij}))/r_{ij}$ is the Coulomb interaction between Gaussian e-balls, main-text Eq.~\eqref{eq:Etot}, $D_{ij}^2 = \sum_{\alpha\beta} D_{ij\alpha\beta}^2$, and
\begin{eqnarray}
    f(x) &=& \frac{1}{x^6} \left( 6\erf^2\left(\frac{x}{\sqrt{2}}\right) \right. \nonumber \\
         &&  -4\sqrt{\frac{2}{\pi}}e^{-x^2/2}\erf\left(\frac{x}{\sqrt{2}}\right)(x^3+3x) \nonumber \\
         &&  \left. + \frac{2}{\pi}e^{-x^2}(x^6+4x^4+6x^2) \right)
\end{eqnarray}
In Eq.~\eqref{eq:si-VdW} we have assumed that $D_{ij\alpha\beta}=\partial^2 V^\mathrm{C}_{ij}/\partial r_{i\alpha} \partial r_{j\beta}$, i.e.\ that the coupling between $\rr_i$ and $\rr_j$ is purely electrostatic.
For $r_{ij}=0$, $f(0)=2/(3\pi)$ and $V^\mathrm{vdW}_{ij}(r=0)=-[4/(3\pi)]w_i^4 w_j^4/w_{ij}^8$.
If $w_i=w_j$, $V_{ij}^\mathrm{vdW}(0)=-1/(12\pi)$.
For $r_{ij}\rightarrow\infty$, $V^C_{ij} \rightarrow 1/r_{ij}$, $f(x)\rightarrow6/x^6$, and we recover the London formula~\cite{London1930} for harmonic oscillators:
\begin{equation}
E_\mathrm{vdW} \simeq -12\sum_{i<j}\frac{w_i^4 w_j^4}{w_{ij}^2}\frac{1}{r_{ij}^6}, 
        ~~\text{for} ~~r_{ij}>>w_{ij}
\label{eq:London}
\end{equation}

A problem with the above scheme arises in a periodic system when $w\rightarrow\infty$ and each e-ball overlaps with a continuum of other e-balls.
It can be verified that in this case the energy diverges as $E_{vdW}\sim -w^3$.
This pathology arises from equating the diagonal terms of the Hessian to $D_{ii}\propto w_i^{-4}\rightarrow0$ while assuming that $|D_{ij}|<<\min(|D_{ii}|,|D_{jj}|)$ in second-order perturbation theory.
To avoid it, we add a small regularization term to $D_{ii}$, taking $\Omega_i=\tfrac12\sqrt{w_i^{-4}+w_0^{-4}}$, with $w_0\simeq 3$ Bohr, in the second line of Eq.~\eqref{eq:si-VdW}.
The regularized expression is
\begin{equation}
    E_\mathrm{vdW} = -\frac{1}{4}\sum_{i<j}\frac{D_{ij}^2}{\Omega_i\Omega_j(\Omega_i+\Omega_j)}
    \label{eq:si-Evdw}
\end{equation}
with $D_{ij}^2 = f(r_{ij}/w_{ij})/w_{ij}^6$.
The corresponding long-range coefficient is
\begin{equation}
C_{6,ij}=\frac{3}{2\Omega_i\Omega_j(\Omega_i+\Omega_j)},
\label{eq:C6regularized}
\end{equation}
since $D_{ij}^2\rightarrow6/r_{ij}^6$.
In the limit $w_0\rightarrow\infty$, this reduces to
$C_{6,ij}=12w_i^4w_j^4/(w_i^2+w_j^2)$, consistent with Eq.~\eqref{eq:London}.

\section{Molecular Fitting Data and Reference Sources}
\label{sec:si-training}

The molecular fit contains 731 atoms, molecules, cations, and anions with elements through argon.
Reference nuclear geometries were taken from the NIST Computational Chemistry Comparison and Benchmark Database, using the B3LYP calculation with the largest available basis for each system~\cite{NIST_CCCBDB_2022,Becke1993,Lee1988}.

Energies and atomic forces were evaluated with SIESTA/PBE using a double-zeta plus polarization basis at these fixed nuclear coordinates~\cite{soler2002siesta,garcia2020siesta,perdew1996generalized}.
The nuclear geometries were not reoptimized with PBE.
Consequently, the reference PBE nuclear forces need not vanish.
Localized-orbital centers and spreads were obtained from Wannier90 calculations based on the SIESTA/PBE electronic states and supply the electronic-position and width targets~\cite{marzari1997mlwf,marzari2012review}.
The reference level for the nuclear structure is therefore distinct from that used for the energy, force, and localized-orbital targets.

The fitted observables include molecular structure, energies, multipole moments, static polarizabilities, vibrational frequencies, and electronic normal-mode energy scales, together with forces at the prescribed reference configurations and spline-smoothness regularization.
The widths of the (orthogonal) DFT Wannier functions were multiplied by an ad hoc factor 0.9 to account for the fact that the LeIFF localized electrons are not orthogonal.
The QUEST excitation energies used for the electronic normal-mode comparisons were included in fitting~\cite{veril2021questdb,loos2025quest}.
Agreement with those targets measures the fitted representation; the QM9 and silicon calculations provide separate assessments of transferability.

This fitting dataset is distinct from the QM9 evaluation set described below.
In particular, the reference sources and relaxation protocols for the QM9 tests must be specified separately from the fitting-data sources.

\section{Single-Target Loss Function}
\label{sec:si-loss}

\begin{figure}[t]
  \centering
  \resizebox{0.6\textwidth}{!}{\input{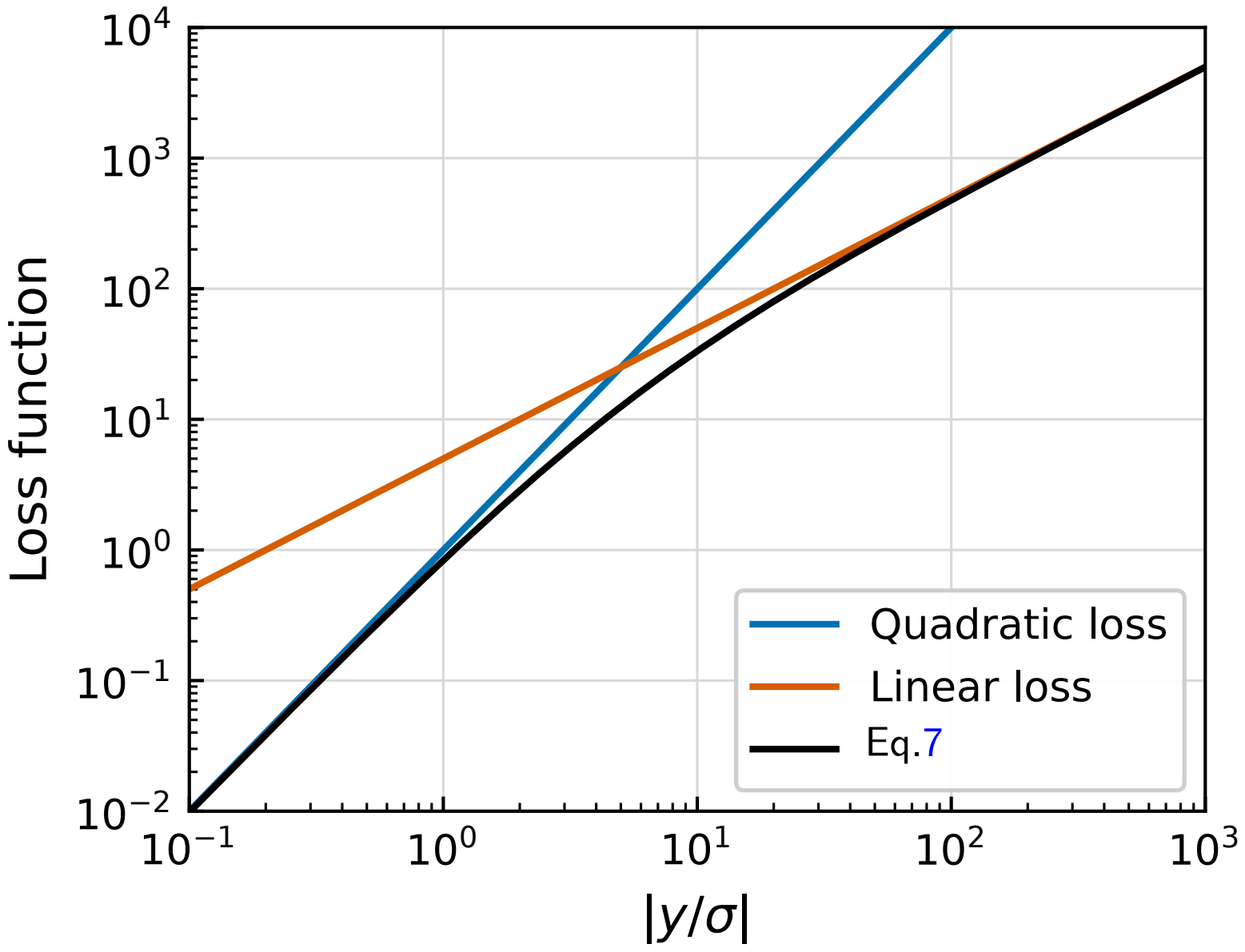}}
  \caption{Log--log plot of the loss function in main-text Eq.~\eqref{eq:loss} with $\lambda=5$.
  The function is quadratic for $|y/\sigma|\ll\lambda$ and linear for $|y/\sigma|\gg\lambda$, reducing the influence of large outliers.}
  \label{fig:cutoff}
\end{figure}

The single-target loss in Eq.~(\ref{eq:loss}) of the main text interpolates smoothly between quadratic and linear behavior, as illustrated in Fig.~\ref{fig:cutoff}.

\FloatBarrier
\section{Core--Valence Interaction Potentials}
\label{sec:si-core-interaction}

\begin{figure}[t]
  \centering
  \includegraphics[width=0.85\textwidth]{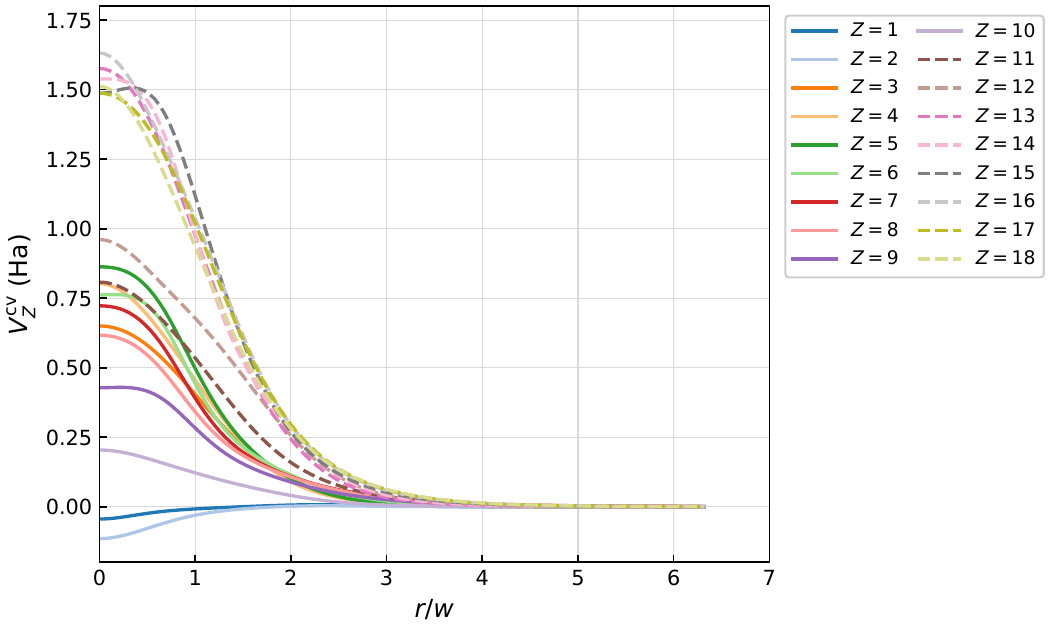}
  \caption{Short-range core--valence interaction potentials $V^{\mathrm{cv}}_Z$ for $Z=1,\ldots,18$, plotted as functions of $r/w$ at $w=0.95$ Bohr.}
  \label{fig:Vcv}
\end{figure}

Figure~\ref{fig:Vcv} shows the fitted short-range core--valence interaction potentials
$V^{\mathrm{cv}}_{Z}$ for $Z = 1,\dots,18$, plotted as functions of $r/w$ at a fixed electron width
$w = 0.95$~Bohr.

\FloatBarrier
\section{Molecular Training-Set Parity Plots}
\label{sec:si-training-parity}

Reference nuclear geometries are taken from the NIST Computational Chemistry Comparison and Benchmark Database (CCCBDB)~\cite{NIST_CCCBDB_2022}.

\begin{figure}[!htbp]
  \centering
  \begin{minipage}[t]{0.48\textwidth}\vspace{0pt}\centering
    \includegraphics[width=\linewidth]{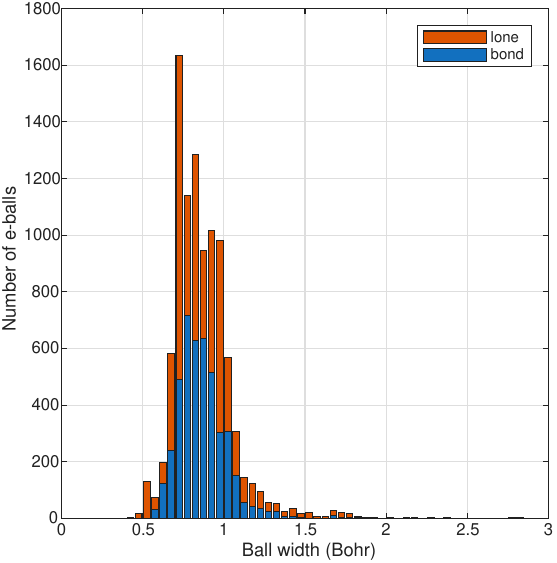}
    \caption{Distribution of LeIFF electron widths in the relaxed structures of the training set, separated into bonding (blue) and lone (red) electrons.}
    \label{fig:widthHist}
  \end{minipage}\hfill
  \begin{minipage}[t]{0.48\textwidth}\vspace{0pt}\centering
    \includegraphics[width=\linewidth]{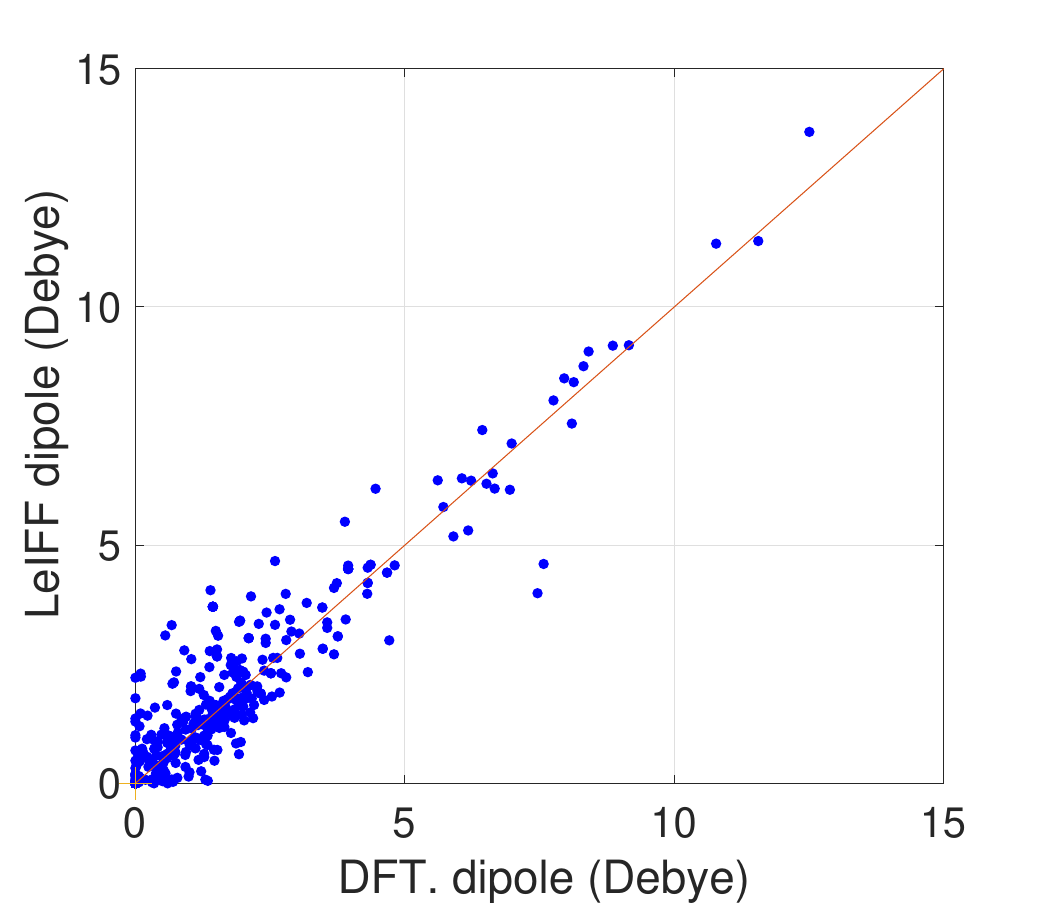}
    \caption{Comparison of LeIFF and DFT molecular dipoles.}
    \label{fig:dipole}
  \end{minipage}
\end{figure}

The following figures compare LeIFF predictions with reference data for properties included in fitting. Figure~\ref{fig:widthHist} additionally shows the electron-width distributions in the relaxed training-set structures. The RMS errors and fitting tolerances are summarized in Table~\ref{tab:spline-rms} of the main text.

\begin{figure}[!htbp]
  \centering
  \begin{minipage}[t]{0.48\textwidth}\vspace{0pt}\centering
    \includegraphics[width=\linewidth]{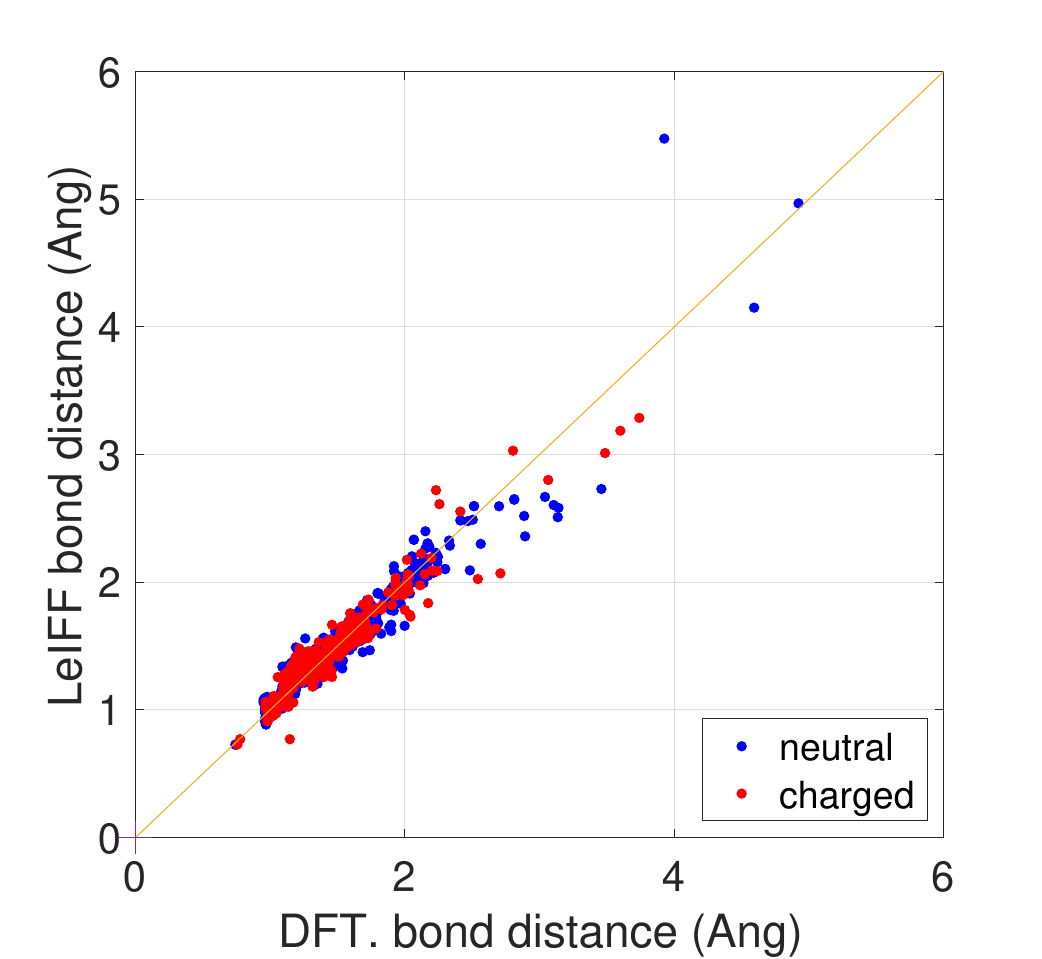}
    \caption{Comparison of NIST CCCBDB B3LYP reference~\cite{NIST_CCCBDB_2022} and relaxed LeIFF bond lengths for the training set of 731 atoms, molecules, and ions.}
    \label{fig:bondLength}
  \end{minipage}\hfill
  \begin{minipage}[t]{0.48\textwidth}\vspace{0pt}\centering
    \includegraphics[width=\linewidth]{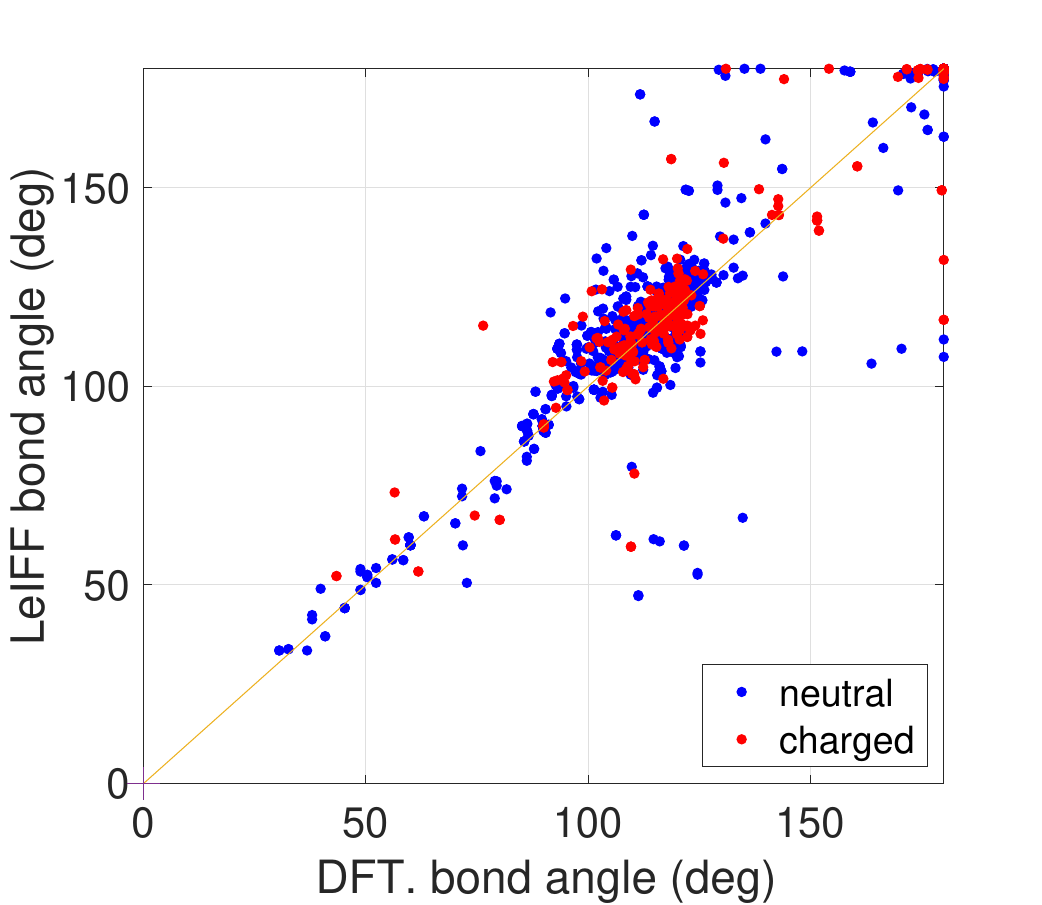}
    \caption{Comparison of NIST CCCBDB B3LYP reference~\cite{NIST_CCCBDB_2022} and relaxed LeIFF bond angles.}
    \label{fig:bondAngle}
  \end{minipage}
\end{figure}

\begin{figure}[!htbp]
  \centering
  \begin{minipage}[t]{0.48\textwidth}\vspace{0pt}\centering
    \includegraphics[width=\linewidth]{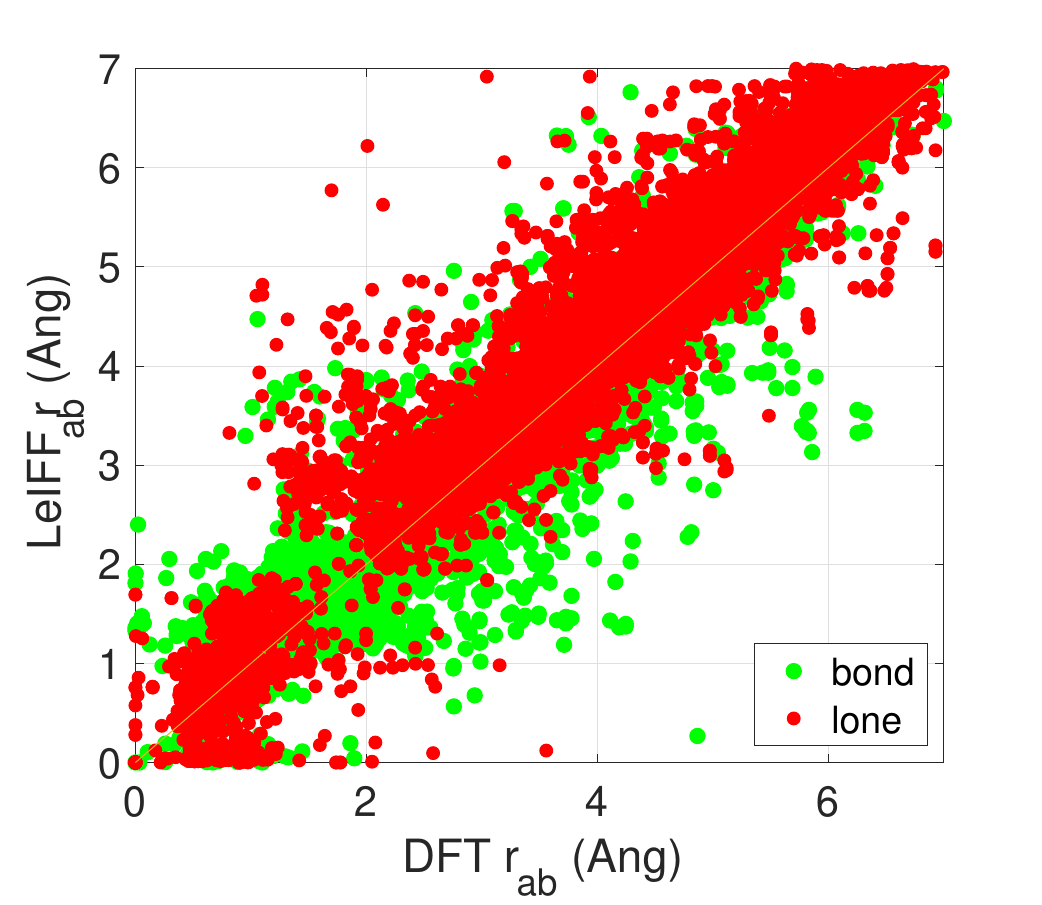}
    \caption{Comparison of reference Wannier-center distances and relaxed LeIFF electron distances to their closest atom. The Wannier data were calculated at the fixed NIST CCCBDB B3LYP geometries~\cite{NIST_CCCBDB_2022}.}
    \label{fig:rabCorr}
  \end{minipage}\hfill
  \begin{minipage}[t]{0.48\textwidth}\vspace{0pt}\centering
    \includegraphics[width=\linewidth]{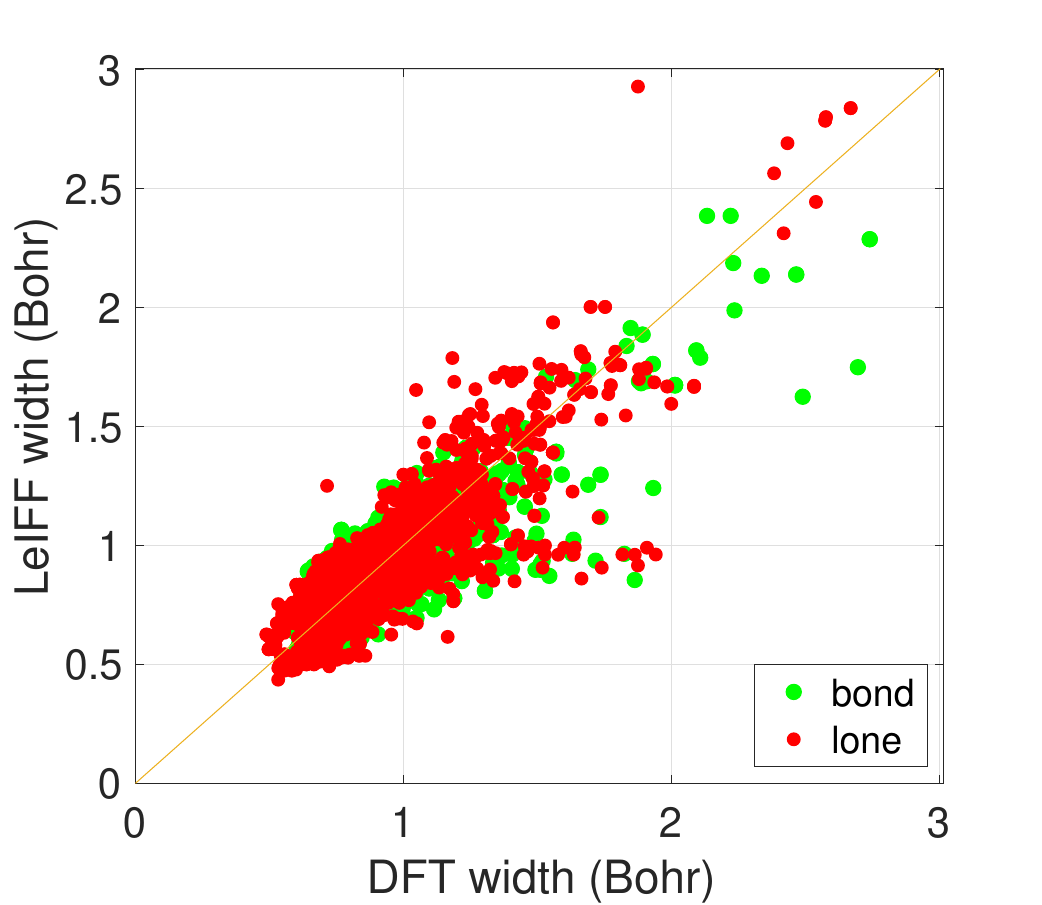}
    \caption{Comparison of relaxed LeIFF electron widths and reference Wannier-function widths calculated at the fixed NIST CCCBDB B3LYP geometries~\cite{NIST_CCCBDB_2022}.}
    \label{fig:ballWidthCorr}
  \end{minipage}
\end{figure}

\begin{figure}[!htbp]
  \centering
  \begin{minipage}[t]{0.48\textwidth}\vspace{0pt}\centering
    \includegraphics[width=\linewidth]{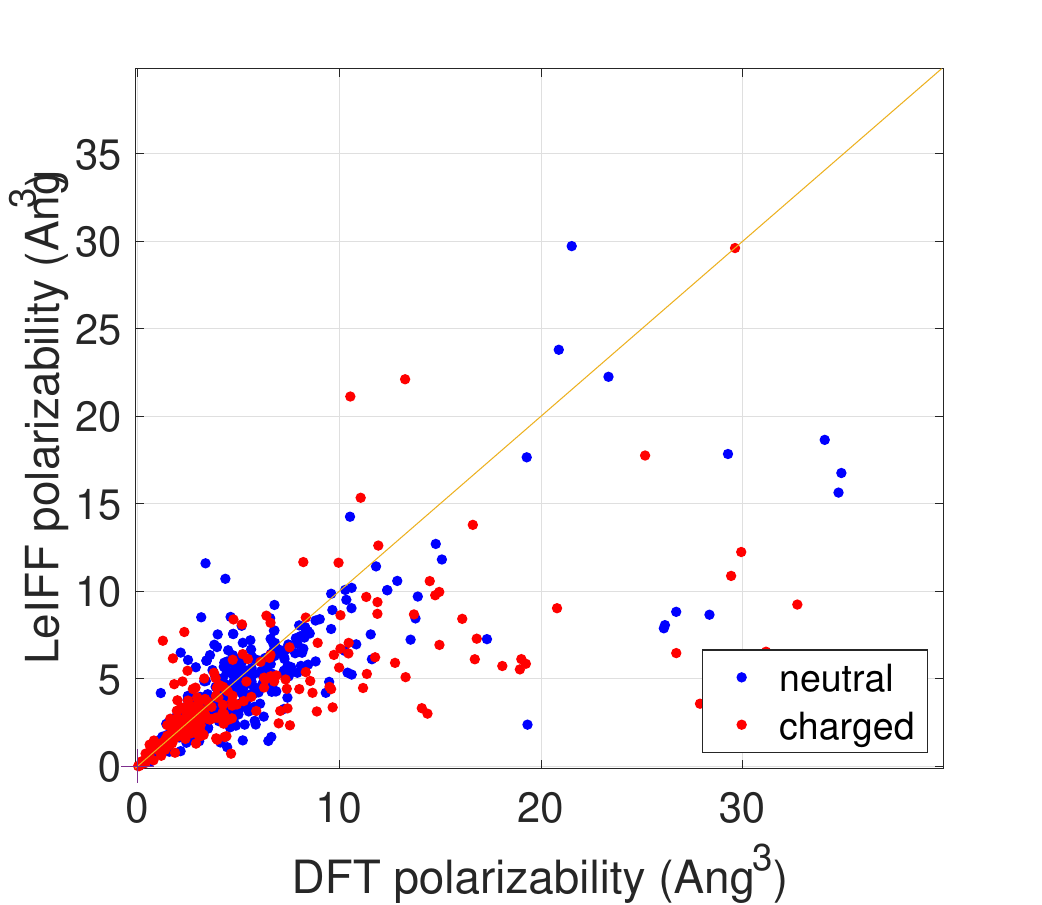}
    \caption{Comparison of LeIFF and DFT molecular polarizabilities.}
    \label{fig:polarizability}
  \end{minipage}\hfill
  \begin{minipage}[t]{0.48\textwidth}\vspace{0pt}\centering
    \includegraphics[width=\linewidth]{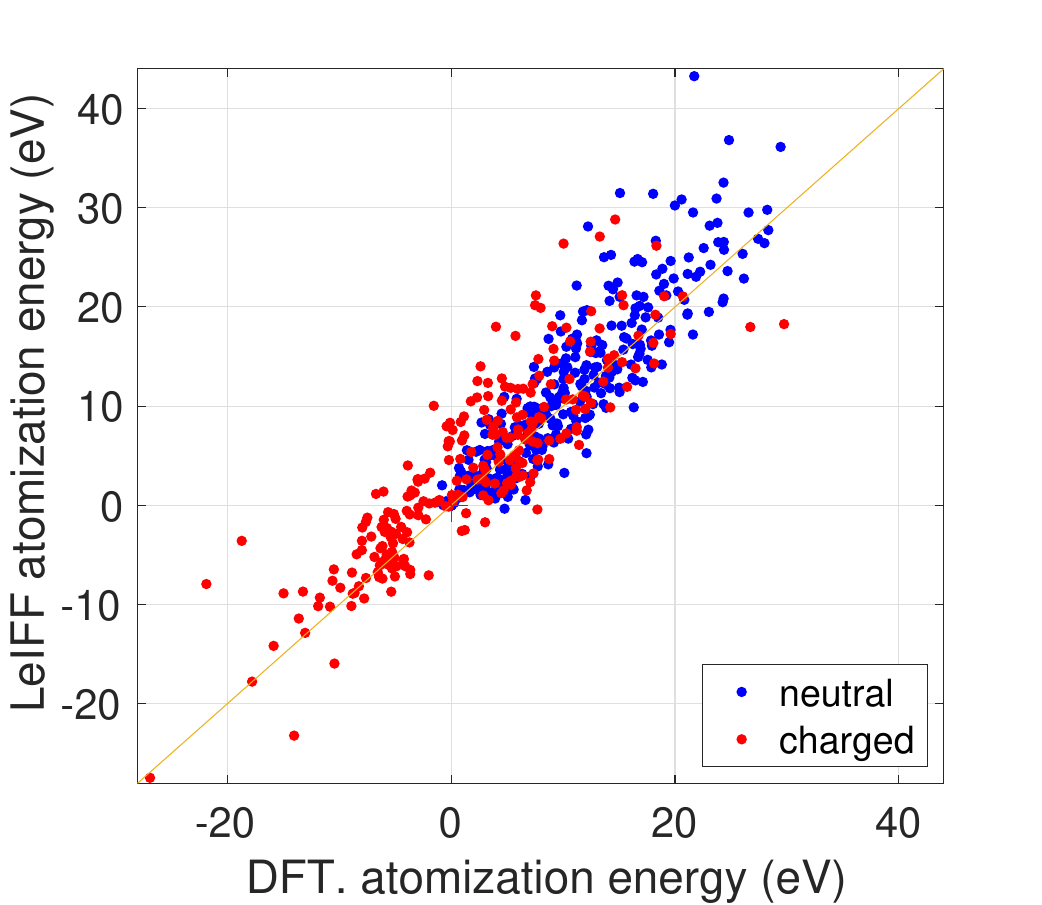}
    \caption{Comparison of LeIFF and DFT atomization energies.}
    \label{fig:atomizationEnergy}
  \end{minipage}
\end{figure}

\begin{figure}[!htbp]
  \centering
  \begin{minipage}[t]{0.48\textwidth}\vspace{0pt}\centering
    \includegraphics[width=\linewidth]{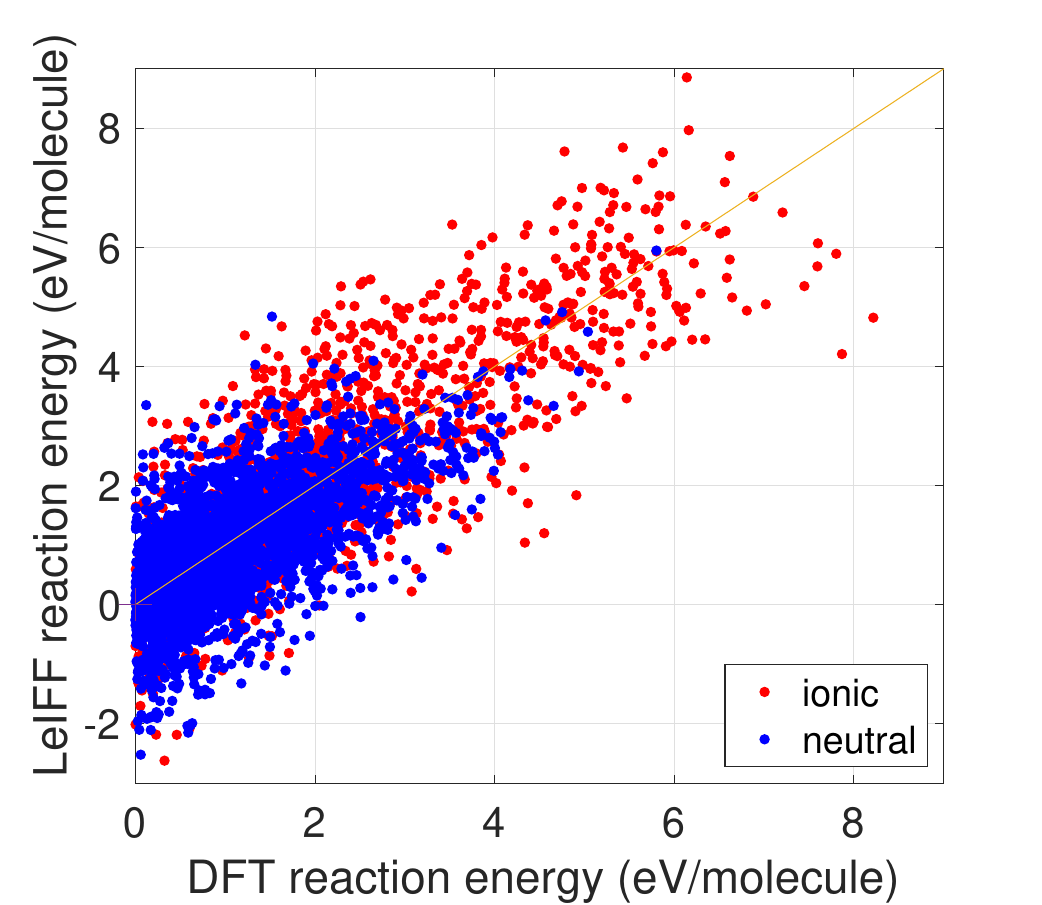}
    \caption{Comparison of LeIFF and DFT reaction energies, defined as the difference between product and reactant energies, not as energy barriers. All possible reactions of up to three molecules of the training set are included, with direction set to make the DFT reaction exothermic. Reactions with different initial and final total charge are plotted as red dots.}
    \label{fig:reactionEnergy}
  \end{minipage}\hfill
  \begin{minipage}[t]{0.48\textwidth}\vspace{0pt}\centering
    \includegraphics[width=\linewidth]{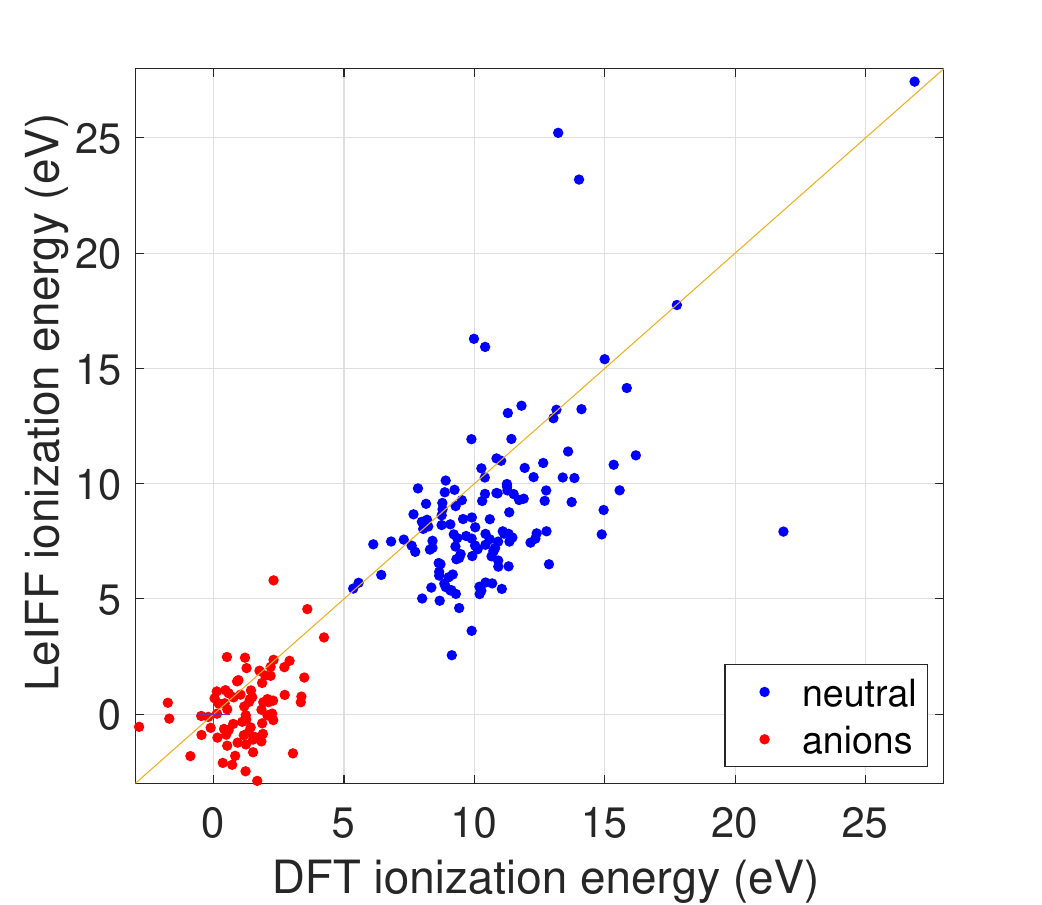}
    \caption{Comparison of LeIFF and DFT ionization energies and electron affinities, calculated as energy differences between molecules and ions. Electron affinities are plotted as ionization energies of anions.}
    \label{fig:ionizationEnergy}
  \end{minipage}
\end{figure}

\begin{figure}[!htbp]
  \centering
  \includegraphics[width=0.48\textwidth]{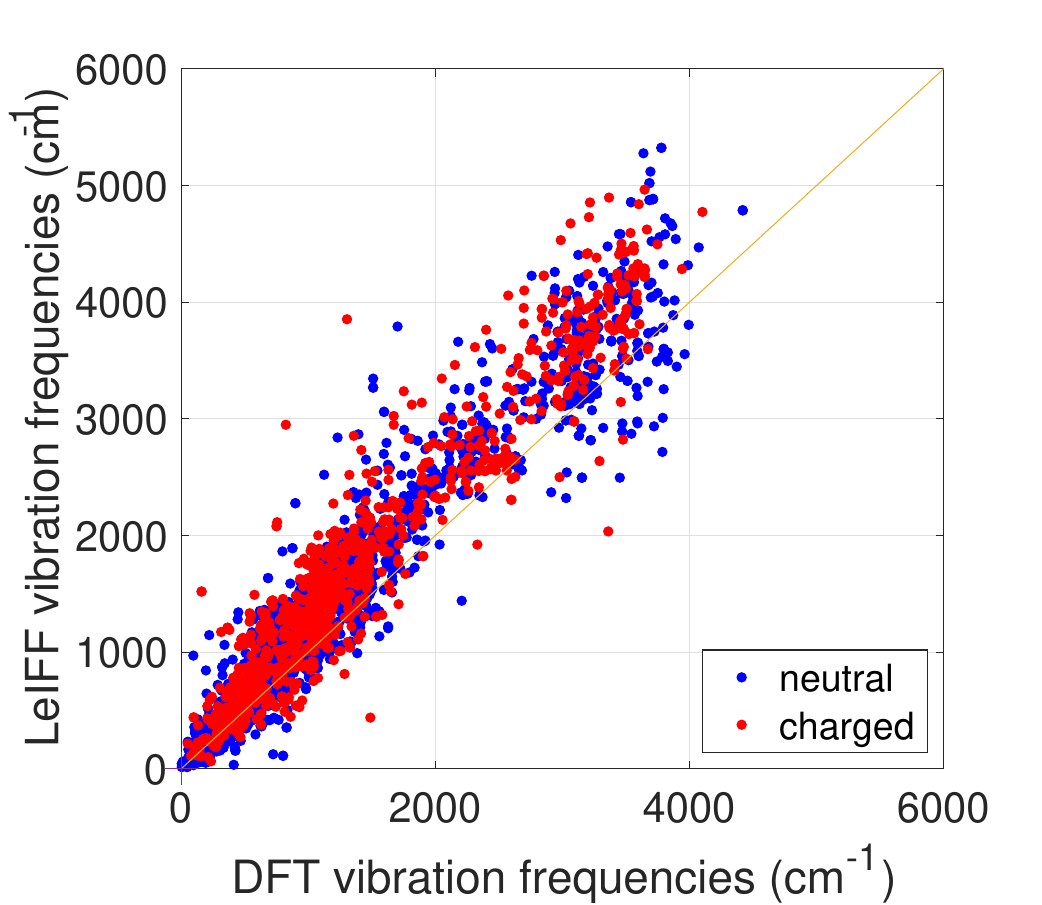}
  \caption{Comparison of LeIFF and DFT vibrational frequencies. The vibration modes are ordered according to their frequency but no comparison is done of their polarization vectors.}
  \label{fig:vibrations}
\end{figure}

\FloatBarrier

\section{Linear-Response Derivatives of Relaxed Properties}
\label{sec:si-linear-response}

\subsection{Response along a stationary branch}

Let $\mathbf{x}_*(\mathbf{f})$ collect the relaxed nuclear positions, electron centers, and electron widths for a fixed choice of local minimum, where $\mathbf{f}$ denotes the force-field parameters.
Define
\begin{equation}
g_i=\frac{\partial E}{\partial x_i},\qquad
H_{ij}=\frac{\partial^2E}{\partial x_i\partial x_j},\qquad
b_{i\alpha}=\frac{\partial^2E}{\partial x_i\partial f_\alpha}.
\end{equation}
Differentiating $g_i(\mathbf{x}_*(\mathbf{f}),\mathbf{f})=0$ gives
\begin{equation}
\sum_j H_{ij}\frac{d x_{*,j}}{d f_\alpha}+b_{i\alpha}=0.
\end{equation}
Thus the coordinate response is
\begin{equation}
\frac{d\mathbf{x}_*}{df_\alpha}
=-H_{\mathrm{reg}}^+\mathbf{b}_\alpha,
\label{eq:si-stationary-response}
\end{equation}
where $H_{\mathrm{reg}}^+$ denotes the regularized pseudoinverse used to suppress rigid-motion zero modes and control ill-conditioned directions~\cite{golub1973pseudoinverse}.
This derivative applies locally on a stable stationary branch and for an unchanged retained Hessian subspace.
A switch between competing minima, a change of particle assignment, or a change in the retained subspace is not described by this local derivative.

For a property $A(\mathbf{x}_*(\mathbf{f}),\mathbf{f})$, the total parameter derivative is
\begin{equation}
\frac{dA}{df_\alpha}
=\left.\frac{\partial A}{\partial f_\alpha}\right|_{\mathbf{x}}
-\left(\nabla_{\mathbf{x}}A\right)^{\!T}
H_{\mathrm{reg}}^+\mathbf{b}_\alpha.
\label{eq:si-property-gradient}
\end{equation}
The second term includes the response of the relaxed configuration without repeating its geometry optimization for every parameter perturbation.
For an observable evaluated at a prescribed, parameter-independent reference configuration, the coordinate-response term is absent.
This distinction applies to the force targets evaluated at the reference nuclear and Wannier configurations.

If a target uses aligned coordinates $\widetilde{\mathbf{x}}=a(\mathbf{x},\mathbf{f})$, its coordinate response includes the alignment Jacobian,
\begin{equation}
\frac{d\widetilde{\mathbf{x}}}{df_\alpha}
=\frac{\partial a}{\partial\mathbf{x}}
\frac{d\mathbf{x}_*}{df_\alpha}
+\frac{\partial a}{\partial f_\alpha}.
\end{equation}
The derivative is evaluated for a fixed equivalent-particle assignment and a smoothly varying alignment.

\subsection{Response derivatives and the fitting loss}

Properties depending on the Hessian require its total parameter derivative along the relaxed branch,
\begin{equation}
\frac{dH_{ij}}{df_\alpha}
=\frac{\partial^3E}{\partial x_i\partial x_j\partial f_\alpha}
+\sum_k
\frac{\partial^3E}{\partial x_i\partial x_j\partial x_k}
\frac{dx_{*,k}}{df_\alpha}.
\label{eq:si-hessian-gradient}
\end{equation}
This explains the third energy derivatives required for the polarizability and frequency gradients.
For an isolated nonzero normal mode satisfying
$H\mathbf{u}_\nu=\omega_\nu^2M\mathbf{u}_\nu$, with fixed masses and normalization
$\mathbf{u}_\nu^TM\mathbf{u}_\nu=1$, the frequency derivative is
\begin{equation}
\frac{d\omega_\nu}{df_\alpha}
=\frac{1}{2\omega_\nu}\,
\mathbf{u}_\nu^T\frac{dH}{df_\alpha}\mathbf{u}_\nu.
\end{equation}
This expression assumes an isolated eigenvalue; it does not assign separate derivatives to arbitrarily chosen vectors within a degenerate eigenspace.

For the robust single-target loss used in the main manuscript, write $t_I=y_I/\sigma_I$ and
\begin{equation}
l(y_I,\sigma_I)=\frac{t_I^2}{1+|t_I|/\lambda}.
\end{equation}
Its residual derivative is
\begin{equation}
\frac{\partial l}{\partial y_I}
=\frac{1}{\sigma_I}\,
\frac{t_I(2+|t_I|/\lambda)}{(1+|t_I|/\lambda)^2},
\end{equation}
with derivative zero at $t_I=0$.
For fixed tolerances, the loss gradient follows from
\begin{equation}
\frac{d\mathcal{L}}{df_\alpha}
=\sum_I\frac{\partial l}{\partial y_I}\frac{dy_I}{df_\alpha},
\end{equation}
using the appropriate fixed-configuration or relaxed-property derivative for each target.

\section{Su--Goddard Implementation Benchmarks}
\label{sec:si-sg-benchmark}

\subsection{Reference inventory, model coverage, and minimization}

The benchmark contains 731 named states: 673 molecular species and 58 isolated atomic states, with charge and spin states counted separately.
The complete state list accompanies the machine-readable benchmark data.
An earlier saved subset contains 725 states; the six states outside that subset are BF$_3^-$, H$_2$CC$^-$, HF$^-$, He$^-$, LiO$^-$, and Mg$^-$.
All 731 states enter the full benchmark denominator, including unsuccessful relaxations.
This inventory defines the implementation benchmark; the reported 731-system fitting-loss summary in the main manuscript is a separate evaluation, and its residuals are not pooled with these benchmark errors.

The LeIFF parameterization contains 606 coefficients.
The benchmark uses the unchanged molecular parameters, without a width-confinement potential or parameter refitting.
SG calculations use LAMMPS release \texttt{stable\_22Jul2025}, the EFF package, and \texttt{pair\_style eff/cut}~\cite{thompson2022lammps,jaramillo2011eff,lammpsEffManual}.
SG-ECP uses the default C, N, O, Al, and Si effective-core parameters with explicit-electron H and He, giving 262 supported states.
The all-electron comparison, SG-eFF1, is restricted to the selected $Z\leq6$ domain and contains 85 supported states.
These are the specified LAMMPS variants; neither comparison evaluates the later eFF2 parameterization.
The calculation inputs use a nonperiodic box and a $1000a_0$ interaction cutoff.
The Gaussian width conventions are related by $s=2w$, where $s$ is the LAMMPS radius and $w$ is the LeIFF Gaussian standard deviation.

The initial benchmark uses three reproducible electronic starts with reference nuclei and radii $s=1.7$, 2.1, and $2.5a_0$.
Electron centers and log radii relax first at fixed nuclei, followed by joint nuclear--electronic conjugate-gradient and L-BFGS-B refinement.
LeIFF additionally uses the native analytic-Hessian minimizer to examine initially rejected cases, with the available production starting pool capped at 20 configurations per system and separate refinements from the original starts and endpoints.
The native search allows 500 iterations, 300 line-search steps, and a maximum coordinate step of $0.1a_0$.
The final LeIFF force-convergence flags are checked against native analytic gradients.
Because the minimizers and additional searches differ, the tabulated rates are not an equal-computational-cost comparison.

Force convergence requires finite coordinates and maximum physical Cartesian/radius force components below $10^{-5}$~Ha/$a_0$.
The LeIFF width gradient is converted to the common radius convention before applying this threshold.
Force convergence is reported separately from connectivity, boundedness, and curvature diagnostics.
For example, the LeIFF inventory has 727 force-converged states but 713 endpoints retained by the additional structural/stability screen.
Changed-connectivity or screened endpoints remain in every computable geometric, dipole, and energy error; they are not removed because they disagree with the reference.
Neither force convergence nor these finite searches establish a global minimum.

\subsection{Property definitions and prediction coverage}

Reference values follow the supplied molecular property records, preserving their individual DFT or experimental provenance.
Reference total energies are evaluated at the prescribed reference geometries; model energies use the selected relaxed states.
Bond and angle RMSEs weight each reference feature equally.
Each matched-model comparison uses exactly the same species, features, and targets; the full LeIFF column uses every available LeIFF prediction for that property.
Dipoles are magnitudes for neutral species, avoiding an origin convention for charged systems.
For atomization energies, each neutral molecule is compared with separately force-converged neutral atomic states calculated with the same model, without fitted elemental offsets.
This differs from the five-offset energy calibration used for QM9.
Failed atomic calculations invalidate dependent atomization predictions and remain in the eligible denominator.

Table~\ref{tab:si-sg-coverage} separates computable predictions from their reference-eligible populations.
Structural properties are inapplicable to isolated atoms, and angles are inapplicable to diatomics; missing reference values are also excluded from eligibility.
Unsupported chemistry is recorded separately from failed predictions.
For SG-ECP, 11/88 eligible atomization energies are computable: 61 cases fail molecular force convergence, and 16 additional molecular endpoints lack converged atomic dependencies.
These are prediction failures, not missing reference energies.
The supplied per-system ledger retains these categories and all property outliers.

\begin{table}[htbp]
\centering
\caption{Computable predictions divided by reference-eligible systems within each model's supported domain. Unsupported elements and cases without an applicable reference are excluded from eligibility; failed predictions remain in the denominators. Original benchmark availability is followed by the updated atomization estimates on the 88-molecule common LeIFF/SG-ECP set.}
\label{tab:si-sg-coverage}
\begin{tabular}{lrrr}
\toprule
Property & LeIFF & SG-ECP & SG-eFF1 \\
\midrule
Bond length (\AA) & 668/670 & 60/236 & 55/65 \\
Bond angle ($^\circ$) & 460/461 & 51/190 & 33/42 \\
Atomization energy (eV) & 310/310 & 11/88 & 26/27 \\
Dipole magnitude (D) & 422/422 & 39/136 & 40/41 \\
Polarizability (\AA$^3$) & 639/642 & 58/222 & 59/68 \\
\midrule
\multicolumn{4}{l}{Updated atomization comparison (88 common molecules)} \\
Energy estimates & 88/88 & 88/88 & -- \\
Force-checked central states & 88/88 & 38/88 & -- \\
\bottomrule
\end{tabular}
\par\smallskip
\begin{minipage}{0.97\textwidth}\footnotesize
The updated estimates retain all 88 eligible molecular predictions with propagated atomic/molecular energy-sensitivity ranges.
They use the expanded SG-ECP energy audit, including its force-sign-corrected evaluation, without changing the scalar energy model or parameters.
Of these 88 SG-ECP estimates, 38 have force-converged central atomic and molecular states; the other 50 remain provisional.
No expanded estimate calculation was performed for SG-eFF1, indicated by dashes.
\end{minipage}
\end{table}

\subsection{Fixed-nuclei polarizability and soft electronic modes}

Polarizabilities follow the native V38 fitting/reporting convention for all three models.
Let $\mathbf{q}$ contain electron centers and physical widths at fixed nuclei, $H_{ee}$ be the electronic Hessian, and $Q_{ia}=\partial\mu_a/\partial q_i$ be the dipole derivative.
The reported tensor and scalar are
\begin{equation}
\boldsymbol{\alpha}=Q^T H_{ee,\tau}^{+}Q,
\qquad
\alpha_{\mathrm{iso}}=\tfrac13\operatorname{Tr}\boldsymbol{\alpha},
\label{eq:si-sg-polarizability}
\end{equation}
where the pseudoinverse uses an absolute singular-value cutoff $\tau=10^{-6}$~Ha/$a_0^2$ in the common physical center/width coordinates.
This is the convention implemented by \texttt{pinv(H,1e-6)} in V38.
It suppresses all sufficiently small-curvature directions, rather than an explicitly identified set of rotational modes.
Fixed nuclei exclude rigid rotation of the whole molecule; electron-only rotations can couple to the dipole and are not generally gauge motions.
Consequently, the finite response defined by Eq.~\eqref{eq:si-sg-polarizability} can differ from the unrestricted weak-field limit when dipole-active soft modes are removed.

All 639 reference-eligible, force-converged LeIFF states use this definition; 233 previously identified cutoff-sensitive cases received electronic-only refinement at unchanged nuclear geometries.
SG responses are calculated from stored force-derived Hessians in the same physical width coordinates and with the same cutoff.
Retained negative electronic curvature or inconsistent full-tensor responses at two finite-difference steps make a response unavailable independently of its agreement with the reference.
No additional exclusion is imposed solely because a truncated soft mode couples to the dipole.
Direct native-function and rotation-covariance checks verify the implementation of the reported convention, while the numerical cutoff remains part of the definition.
Electronic-response exclusions and numerical uncertainties are distinct from molecular-relaxation failures.

\subsection{Comparison with all-electron SG-eFF1}

In the 85-state all-electron domain, LeIFF reaches force convergence for 83 states (97.6\%) and SG-eFF1 for 75 (88.2\%).
Table~\ref{tab:si-sg-eff1} shows the full LeIFF results and identical available property subsets for the two models.
LeIFF has smaller matched errors for all five properties in this comparison.
In particular, the polarizability RMSEs are 10.6 and 144~\AA$^3$ for 59 common states.
The conditional errors are accompanied by the full eligibility counts in Table~\ref{tab:si-sg-coverage}.

\begin{table}[htbp]
\centering
\caption{Force convergence and property RMSEs for LeIFF and SG-eFF1. The full LeIFF column covers the 731-state benchmark inventory. For convergence, the matched LeIFF column uses the chemical domain supported by the comparator; for each property, the last two columns use identical available systems and features. Parentheses give system counts.}
\label{tab:si-sg-eff1}
\small
\setlength{\tabcolsep}{5pt}
\begin{tabular}{lrrr}
\toprule
Quantity & LeIFF: full & LeIFF: matched & SG-eFF1 \\
\midrule
Force convergence & 99.5\% (727/731) & 97.6\% (83/85) & 88.2\% (75/85) \\
Bond length (\AA) & 0.126 (668) & 0.104 (55) & 0.221 (55) \\
Bond angle ($^\circ$) & 12.3 (460) & 15.2 (33) & 22.4 (33) \\
Atomization energy (eV) & 4.08 (310) & 3.66 (26) & 11.2 (26) \\
Dipole magnitude (D) & 1.59 (422) & 0.667 (40) & 3.58 (40) \\
Polarizability (\AA$^3$) & 17.5 (639) & 10.6 (59) & 144 (59) \\
Polarizability availability & 639/642 & -- & 59/68 \\
\bottomrule
\end{tabular}
\par\smallskip
\begin{minipage}{0.97\textwidth}\footnotesize
Force-convergence and stability diagnostics are defined in the Supplementary Material.
Property errors retain every computable prediction in the stated subset, including outliers and changed-connectivity states.
Dipoles and atomization energies use neutral species; atomic reference energies are calculated with the same model without fitted offsets.
Polarizabilities use the common fixed-nuclei response convention described in the Supplementary Material.
Availability entries give computable/reference-eligible predictions within each model's supported domain.
\end{minipage}
\end{table}

\FloatBarrier
\subsection{LAMMPS implementation limitations and reproducibility}

The numerical comparison retains the unmodified SG implementation used for the original calculations.
A source-level audit of that release identified a Cartesian-force defect in the small-distance branch at separations at or below $10^{-6}a_0$: component magnitudes were returned without restoring the signs of negative displacement components.
At affected configurations this violates inversion covariance while leaving the scalar energy unchanged.
An additional numerical issue arises near coincident same-spin packets from cancellation and small overlap denominators in the Pauli terms.
Nonconvergence in this implementation is therefore not, by itself, evidence of physical instability of the underlying SG energy model.
The separate sign-corrected, extended-search analysis changes the implementation and search budget; its recovered endpoints are not mixed into the tabulated comparisons.
Higher-energy atomic local minima are not substituted for unresolved atomic ground-state references.

The reproducibility files contain the benchmark inventory, convergence checks, property outcomes, per-feature predictions, input hashes, and a table-generation script.
They reproduce the reported aggregates without refitting or rerunning molecular optimizations.
The source calculation directories retain the minimization inputs, logs, and native-Hessian and finite-field checks needed for a complete simulation release.

\FloatBarrier
\section{QM9 Reference Data and Evaluation Protocols}
\label{sec:si-qm9}

\subsection{Reference sources}

The evaluation contains 112,292 molecules drawn from QM9, with up to nine heavy atoms chosen from C, N, O, and F~\cite{ramakrishnan2014qm9}.
The LeIFF interaction parameters are not refitted to this evaluation set.

The reference data have two sources.
Wannier centers and spreads, energies, and dipole moments are obtained from SIESTA/PBE calculations.
Reference nuclear geometries, polarizabilities, and vibrational frequencies are taken from the published QM9~\cite{ramakrishnan2014qm9} dataset.
Table~\ref{tab:si-qm9-sources} lists the reference source for each group of observables.

\begin{table}[htbp]
\centering
\caption{Reference sources for the QM9~\cite{ramakrishnan2014qm9} evaluation.}
\label{tab:si-qm9-sources}
\small
\begin{tabular}{p{0.40\textwidth}p{0.50\textwidth}}
\toprule
Observable & Reference source \\
\midrule
Bond lengths and angles & Published QM9~\cite{ramakrishnan2014qm9} geometries \\
Wannier centers and spreads & SIESTA/PBE localized orbitals \\
Energies & SIESTA/PBE \\
Dipole moments & SIESTA/PBE \\
Polarizabilities & Published QM9~\cite{ramakrishnan2014qm9} dataset \\
Vibrational frequencies & Published QM9~\cite{ramakrishnan2014qm9} dataset \\
\bottomrule
\end{tabular}
\end{table}

\subsection{Atomic energy references}

For molecule $m$, the LeIFF atomization energy is
\begin{equation}
E_{\mathrm{at},m}^{\mathrm{LeIFF}}
=\sum_Z n_{mZ}\varepsilon_Z-E_m^{\mathrm{LeIFF}},
\label{eq:si-qm9-atomization}
\end{equation}
where $n_{mZ}$ is the number of atoms of element $Z$ and the five $\varepsilon_Z$ correspond to H, C, N, O, and F.
These atomic reference energies are determined by a least-squares fit to the reference atomization energies.
The reported atomization-energy error is therefore evaluated after this five-parameter atomic-energy calibration.
The calibration changes the elemental energy reference without modifying the LeIFF interactions or the calculated geometries and response properties.
It is distinct from refitting the force-field parameters.

\subsection{Vibrational frequencies}

Reference vibrational frequencies are taken from the published QM9~\cite{ramakrishnan2014qm9} dataset.
The supplementary parity plots and the RMS errors in the main manuscript use the reference sources listed above.

\subsection{Property comparison plots}

\begin{figure}[!htbp]
\centering
\begin{minipage}[t]{0.48\textwidth}
\captionsetup{font=small}
\centering
\includegraphics[width=\linewidth]{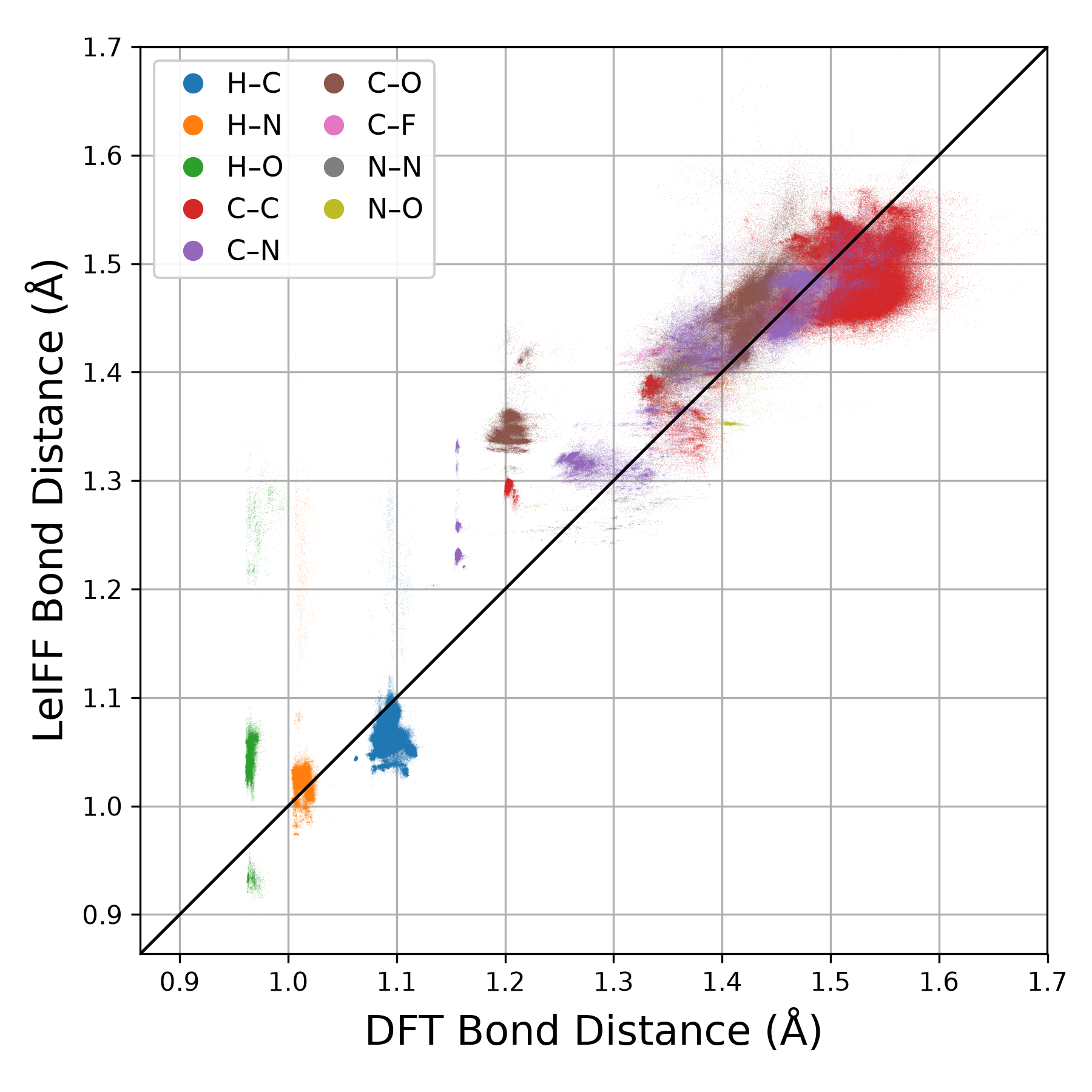}
\caption{Bond distances calculated with LeIFF compared with published QM9~\cite{ramakrishnan2014qm9} reference structures.}
\label{fig:si-bondDistanceQM9}
\end{minipage}
\hfill
\begin{minipage}[t]{0.48\textwidth}
\captionsetup{font=small}
\centering
\includegraphics[width=\linewidth]{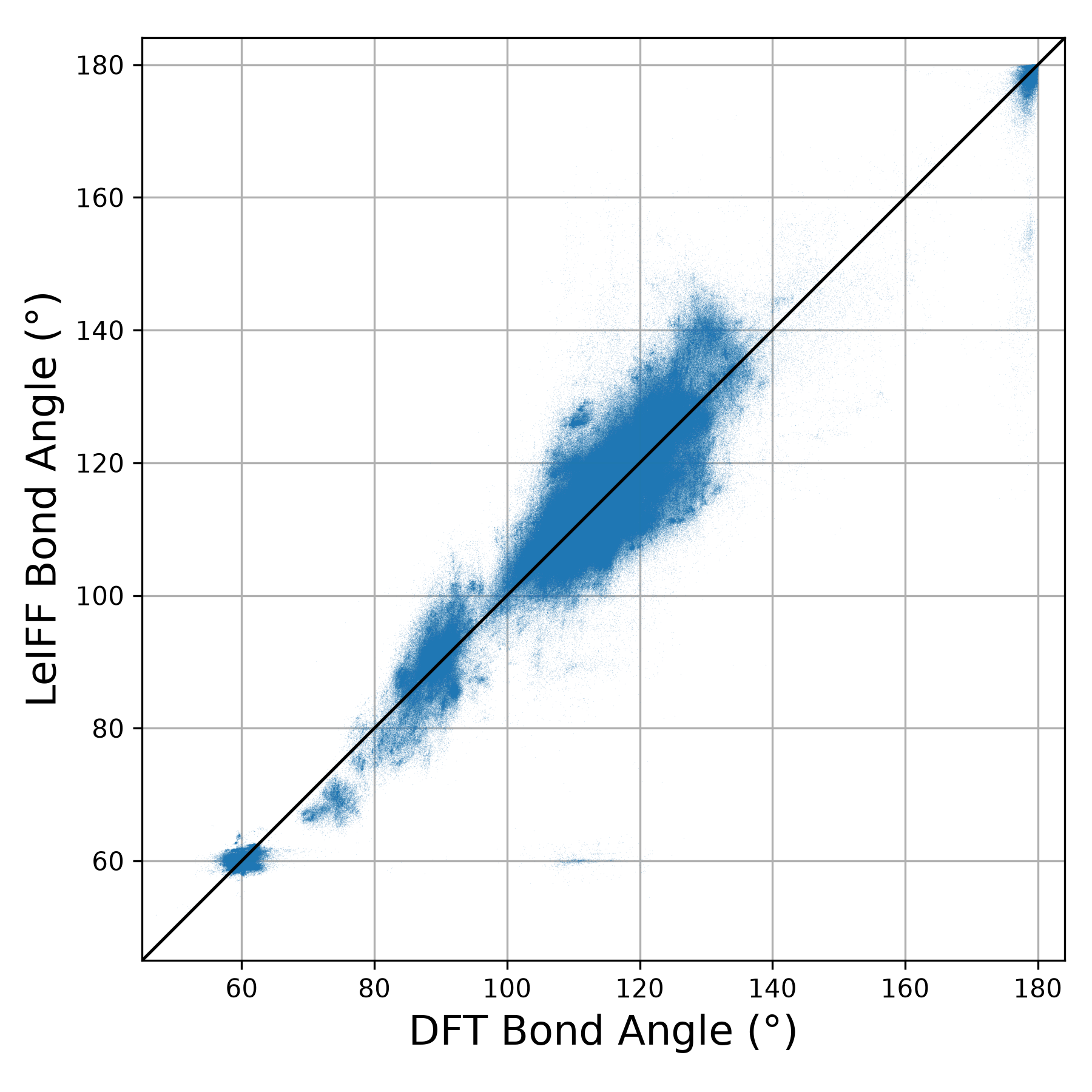}
\caption{Bond angles calculated with LeIFF compared with published QM9~\cite{ramakrishnan2014qm9} reference structures.}
\label{fig:si-bondAngleQM9}
\end{minipage}
\end{figure}

\begin{figure}[!htbp]
\centering
\begin{minipage}[t]{0.48\textwidth}
\captionsetup{font=small}
\centering
\includegraphics[width=\linewidth]{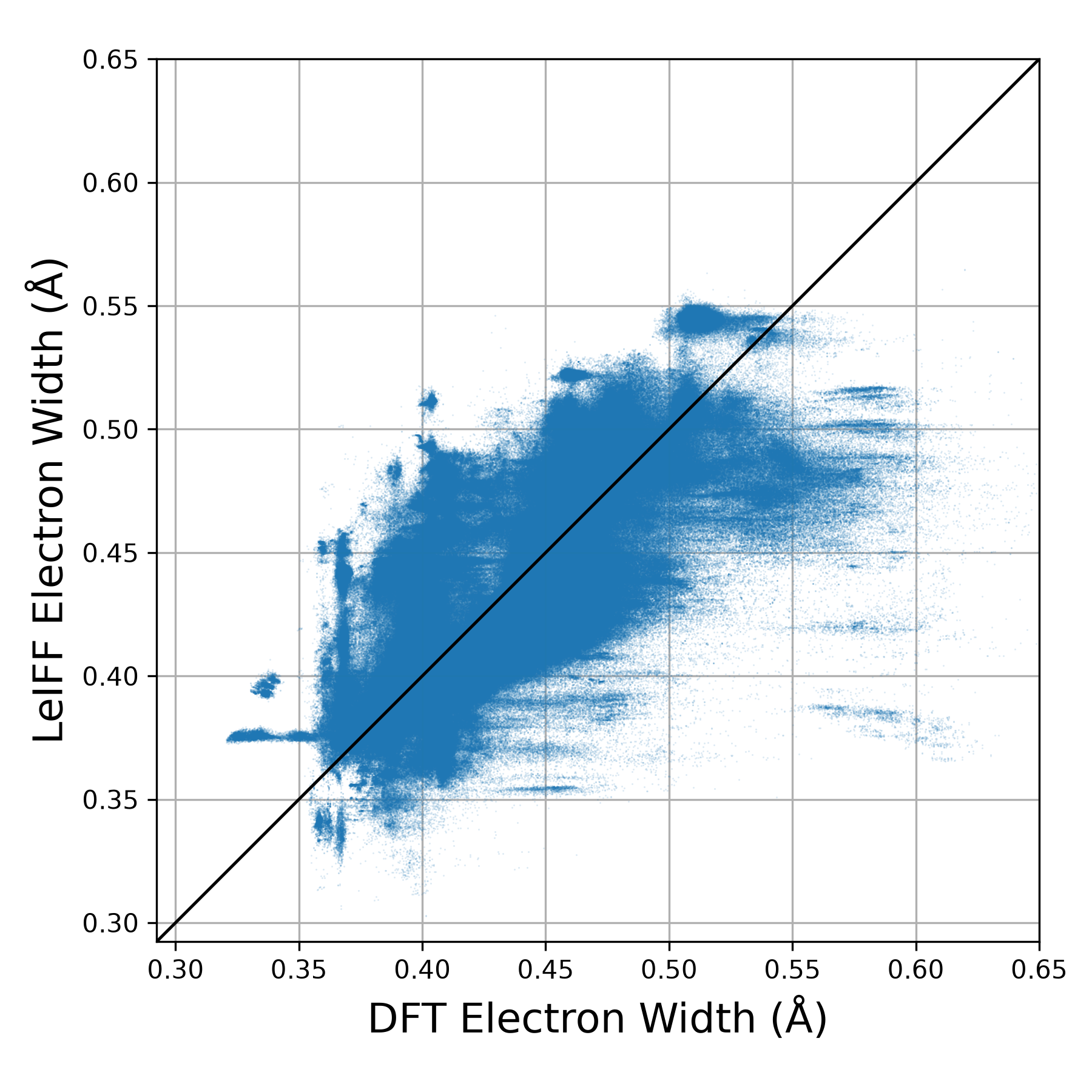}
\caption{Electron widths calculated with LeIFF compared with DFT Wannier-function radii of the published QM9~\cite{ramakrishnan2014qm9} structures.}
\label{fig:si-electronWidthQM9}
\end{minipage}
\hfill
\begin{minipage}[t]{0.48\textwidth}
\captionsetup{font=small}
\centering
\includegraphics[width=\linewidth]{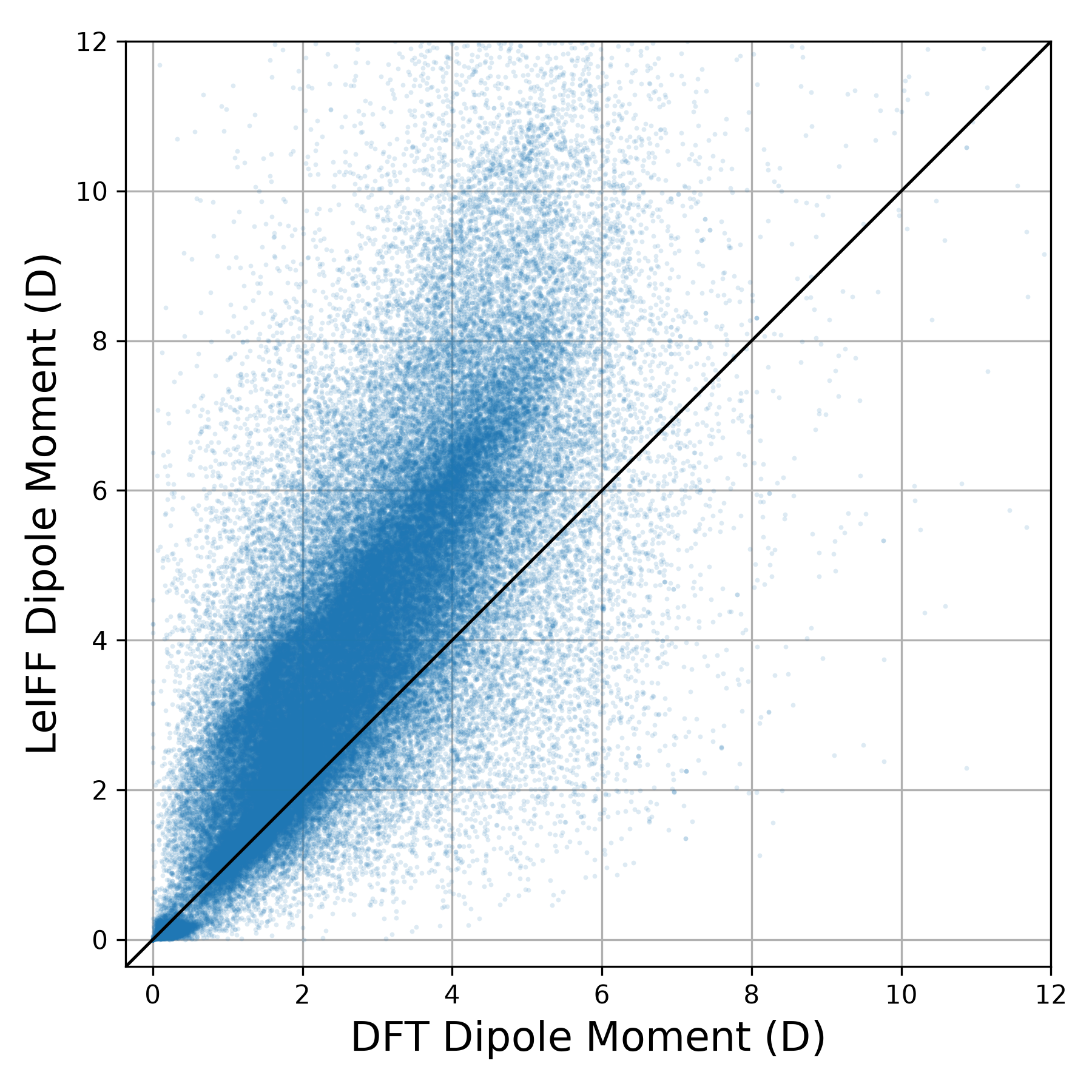}
\caption{Dipole moments calculated with LeIFF compared with dipole moments calculated from the published QM9~\cite{ramakrishnan2014qm9} reference structures.}
\label{fig:si-dipoleQM9}
\end{minipage}
\end{figure}

\begin{figure}[!htbp]
\centering
\begin{minipage}[t]{0.48\textwidth}
\captionsetup{font=small}
\centering
\includegraphics[width=\linewidth]{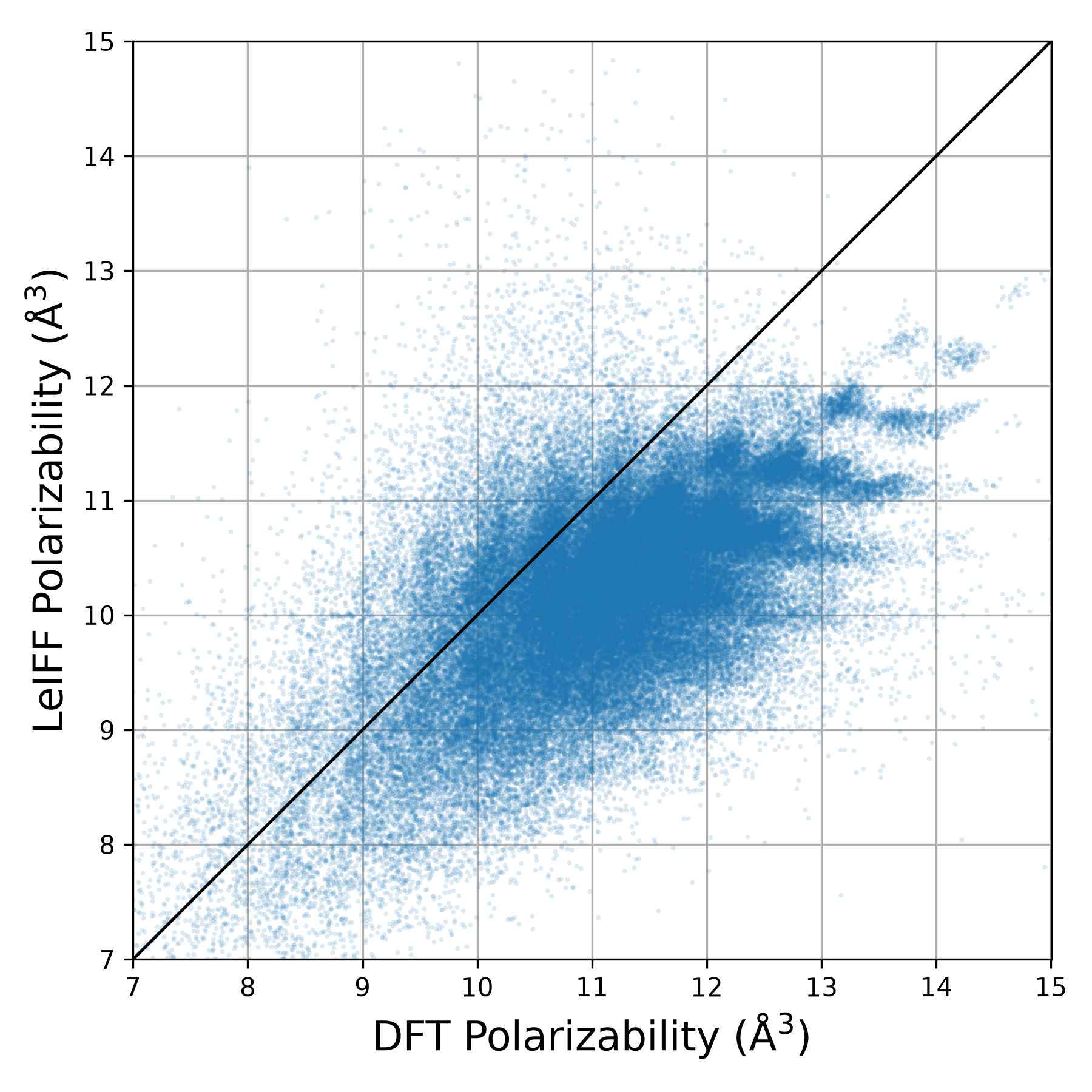}
\caption{Polarizabilities calculated with LeIFF compared with published QM9~\cite{ramakrishnan2014qm9} reference values.}
\label{fig:si-polarizabilityQM9}
\end{minipage}
\hfill
\begin{minipage}[t]{0.48\textwidth}
\captionsetup{font=small}
\centering
\includegraphics[width=\linewidth]{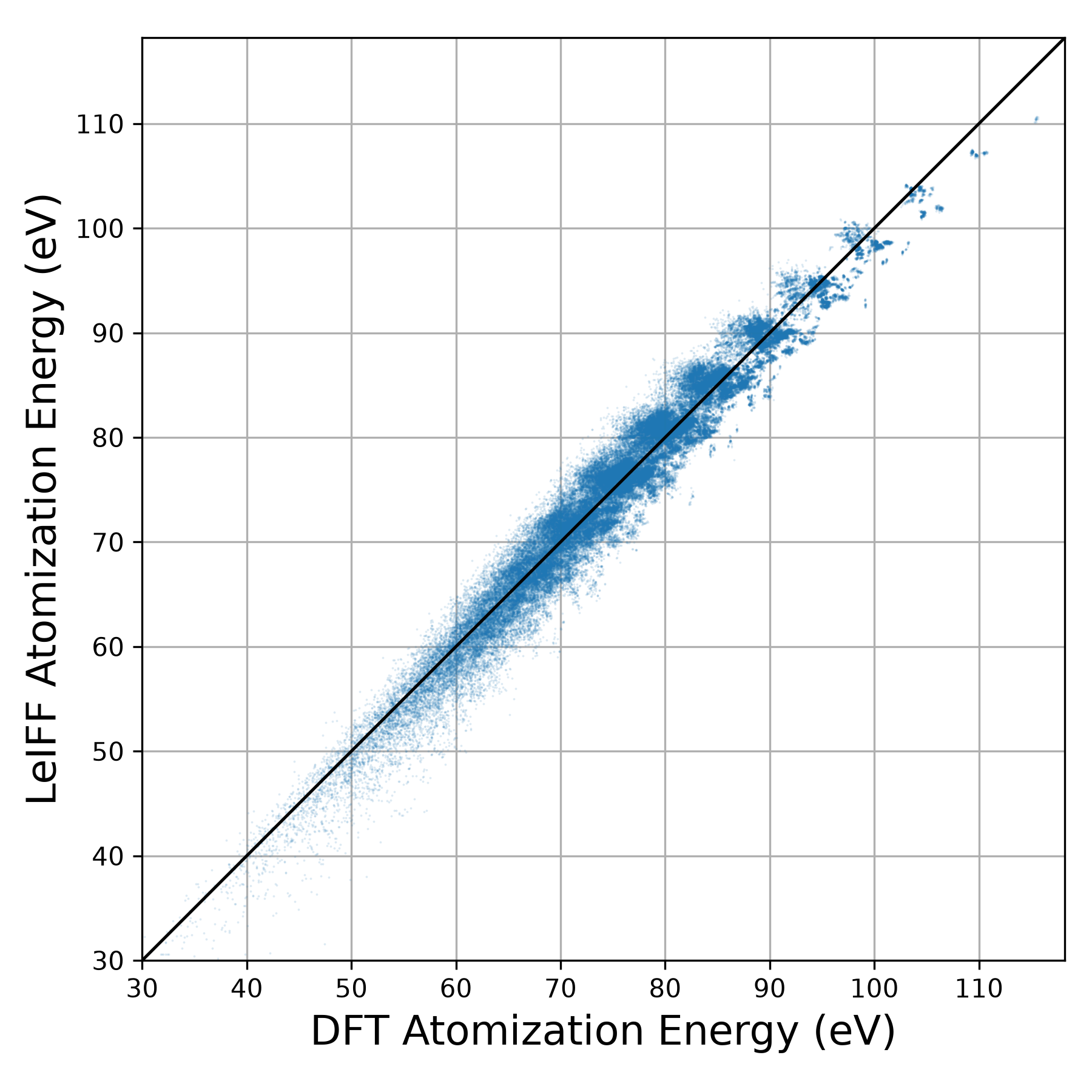}
\caption{Atomization energies calculated with LeIFF compared with SIESTA/PBE reference values for the QM9 molecules~\cite{ramakrishnan2014qm9}. LeIFF atomic reference energies for H, C, N, O, and F are fitted by least squares to the reference atomization energies.}
\label{fig:si-atomizationQM9}
\end{minipage}
\end{figure}

\begin{figure}[!htbp]
\centering
\captionsetup{font=small}
\centering
\includegraphics[width=0.6\columnwidth]{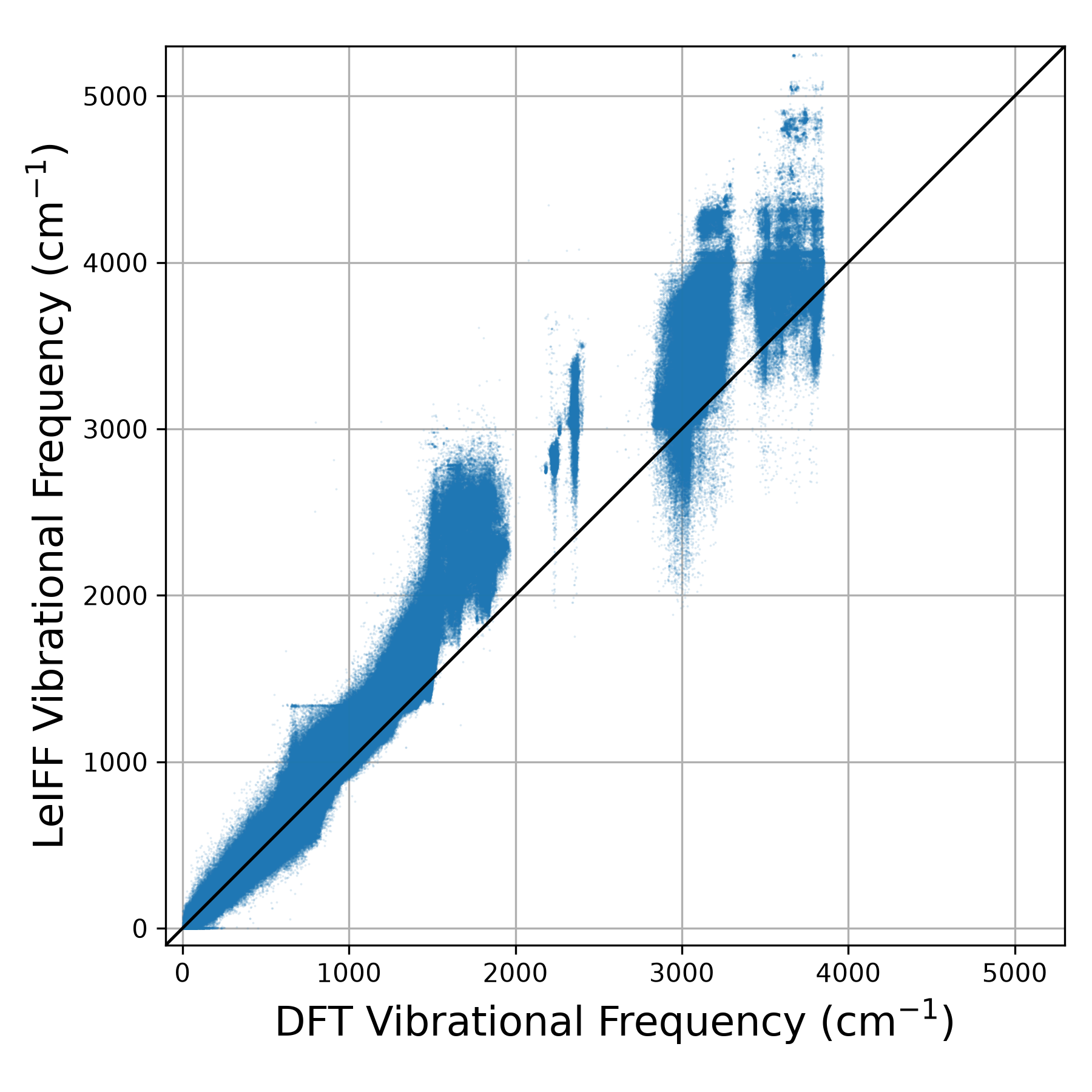}
\caption{Vibrational frequencies calculated with LeIFF compared with published QM9~\cite{ramakrishnan2014qm9} reference frequencies. For each molecule, both sets are sorted and compared position by position.}
\label{fig:si-vibrationsQM9}
\end{figure}

Figures~\ref{fig:si-bondDistanceQM9}--\ref{fig:si-vibrationsQM9} show the seven QM9~\cite{ramakrishnan2014qm9} property comparisons summarized by the RMS errors in the main manuscript.
In each plot, the color indicates the number of data points in each bin on a logarithmic scale, and the solid line marks perfect agreement.

\FloatBarrier

\section{Structural, Vibrational, and Electronic Tests in Silicon}
\label{sec:si-silicon}

Silicon tests whether electronic degrees of freedom constrained by molecular data can also describe collective response in a solid.
The small-amplitude response connects LeIFF electronic normal modes to dielectric screening and energy-loss spectroscopy.
Neither the dielectric nor energy-loss spectra were included in fitting.

The calculations use the LeIFF molecular parameterization with periodic boundary conditions and fixed diamond-Si ions, with the dielectric-response and stopping calculations performed at a conventional lattice constant of approximately $5.44$~\AA.
This prescribed lattice constant is distinct from the equilibrium value predicted by the equation-of-state calculation below.
Electron centers and widths evolve with masses $m_e$ and $3m_e$, respectively.
A smooth large-width confinement term, centered at $w=3a_0$, is retained in the periodic implementation.
Here $a_0$ denotes the Bohr radius.

\subsection{Dielectric response and the long-wavelength loss spectrum}

For the LeIFF response calculations, a $2\times2\times2$ conventional supercell contains 64 Si ions and 256 valence electrons.
All electron centers and widths are re-relaxed at fixed ionic positions using the molecular parameters, with maximum electronic force components below $3\times10^{-7}$~Ha/$a_0$.
An unperturbed trajectory remains essentially stationary for 800 atomic units ($19.35$~fs), with a total-energy variation below $4\times10^{-11}$~Ha and a maximum electron-center displacement below $5\times10^{-6}a_0$.
This reference is used for the long-wavelength spectra reported here.

\begin{figure*}[t]
  \centering
  \includegraphics[width=0.99\textwidth]{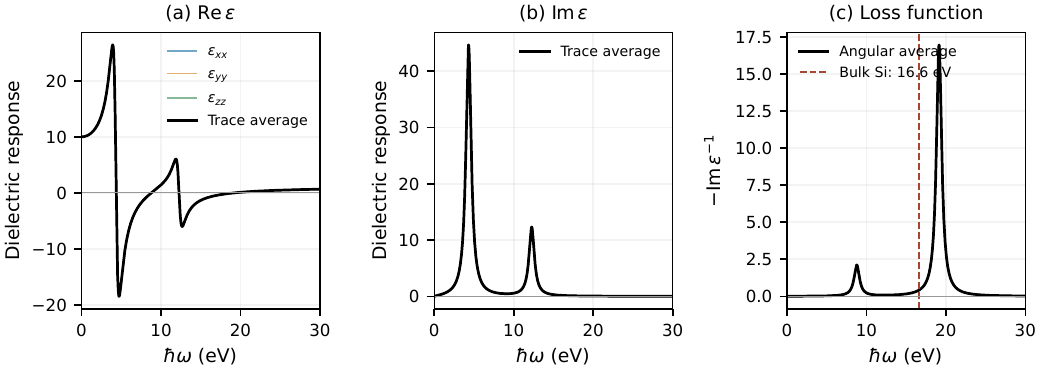}
  \caption{Long-wavelength electronic response of the relaxed 64-atom Si supercell using the LeIFF molecular parameterization.
  Panels (a) and (b) show the real and imaginary dielectric response; thin colored curves denote Cartesian diagonal components and black curves their trace average.
  Panel (c) shows the corresponding directional loss spectra and their angular average in black.
  The dashed line marks the experimental bulk-plasmon energy near $16.6$~eV~\cite{chen1975siliconloss}.
  The complex-frequency damping is $\hbar\eta=0.40$~eV.}
  \label{fig:si-supp-optical}
\end{figure*}

Small uniform velocity kicks of $0.00125$ atomic units along the three Cartesian directions generate the current response.
After subtracting the unperturbed reference current, a causal Fourier transform yields the conductivity tensor $\sigma_{ij}(z)$ and, in atomic units,
\begin{equation}
\varepsilon_{ij}(z)=\delta_{ij}
+\frac{4\pi i}{z}\sigma_{ij}(z),
\qquad z=\omega+i\eta.
\label{eq:si-supp-dielectric}
\end{equation}
We propagate for 8000 steps of $0.1$ atomic units and apply exponential damping with $\hbar\eta=0.40$~eV.
The same complex frequency is used in the Fourier transform and in the conductivity-to-dielectric conversion, providing consistent numerical broadening of the dielectric function and its inverse.
This damping resolves the finite-time response and is not interpreted as a physical linewidth.
The Cartesian diagonal responses coincide to numerical accuracy and the off-diagonal response is negligible.
Figure~\ref{fig:si-supp-optical} shows the trace-averaged dielectric function and the orientational average of the directional longitudinal loss, $-\operatorname{Im}\{[\hat{\mathbf{q}}\cdot\boldsymbol{\varepsilon}(z)\cdot\hat{\mathbf{q}}]^{-1}\}$.

The loss spectrum has a dominant collective feature near $19.15$~eV and a lower-energy feature near $8.8$~eV.
The dominant feature remains above the measured bulk-Si plasmon near $16.6$~eV~\cite{chen1975siliconloss}.
Integrating to 100~eV recovers $99.49\%$ and $99.42\%$ of the dielectric and loss sum rules, respectively, relative to the valence-density plasma energy of $16.576$~eV.
Extending the integration to 150~eV raises these fractions to $99.66\%$ and $99.60\%$.
The remaining integrated deficits depend on the finite energy cutoff and numerical broadening; full convergence is not claimed.
The Fourier and inversion procedure also reproduces the analytic complex-frequency response of a single harmonic oscillator to a relative dielectric error of $2.7\times10^{-5}$ over 1--40~eV.

Larger kicks along $z$, of $0.0025$ and $0.005$ atomic units, change the full normalized current trace by $1.47\%$ and $10.3\%$, respectively, relative to the smallest kick.
The smallest Cartesian kicks are therefore used throughout; further amplitude convergence requires a smaller perturbation.
The dominant broadened peak is stable near $19.14$~eV under damping and time-window variations that suppress endpoint artifacts, while cell-size and longer-trajectory convergence remain to be established.
The structured dielectric and loss spectra connect the model's electronic dynamics to the observables used in first-principles response theory~\cite{weissker2010silicon}.

These calculations demonstrate that molecularly calibrated electron--ion interactions support dielectric screening and collective density motion in a solid.
The response spectra provide concrete observables for assessing and improving the electronic energy landscape beyond equilibrium molecular fitting.

\subsection{Proton electronic stopping: proof-of-method comparison}
\label{sec:si-supp-stopping}

A $5\times5\times5$ conventional diamond-Si supercell contains 1000 fixed Si ions and 4000 valence electrons.
A proton follows a prescribed constant-velocity path along the $\langle001\rangle$, $\langle011\rangle$, and $\langle111\rangle$ channels or a sampled off-channel path.
The electronic centers and widths evolve using the same LeIFF molecular parameters and the smooth width-confinement term.
The time step is $0.0145$ atomic units; trajectories sample six path periods.
The plotted stopping estimate is the arithmetic mean of the projectile-drag estimate and the slope of electronic energy versus path length.
The off-channel result is a single sampled trajectory, not an orientational average, and its two estimators show appreciable disagreement.

\begin{figure*}[t]
\centering
\resizebox{0.98\textwidth}{!}{\input{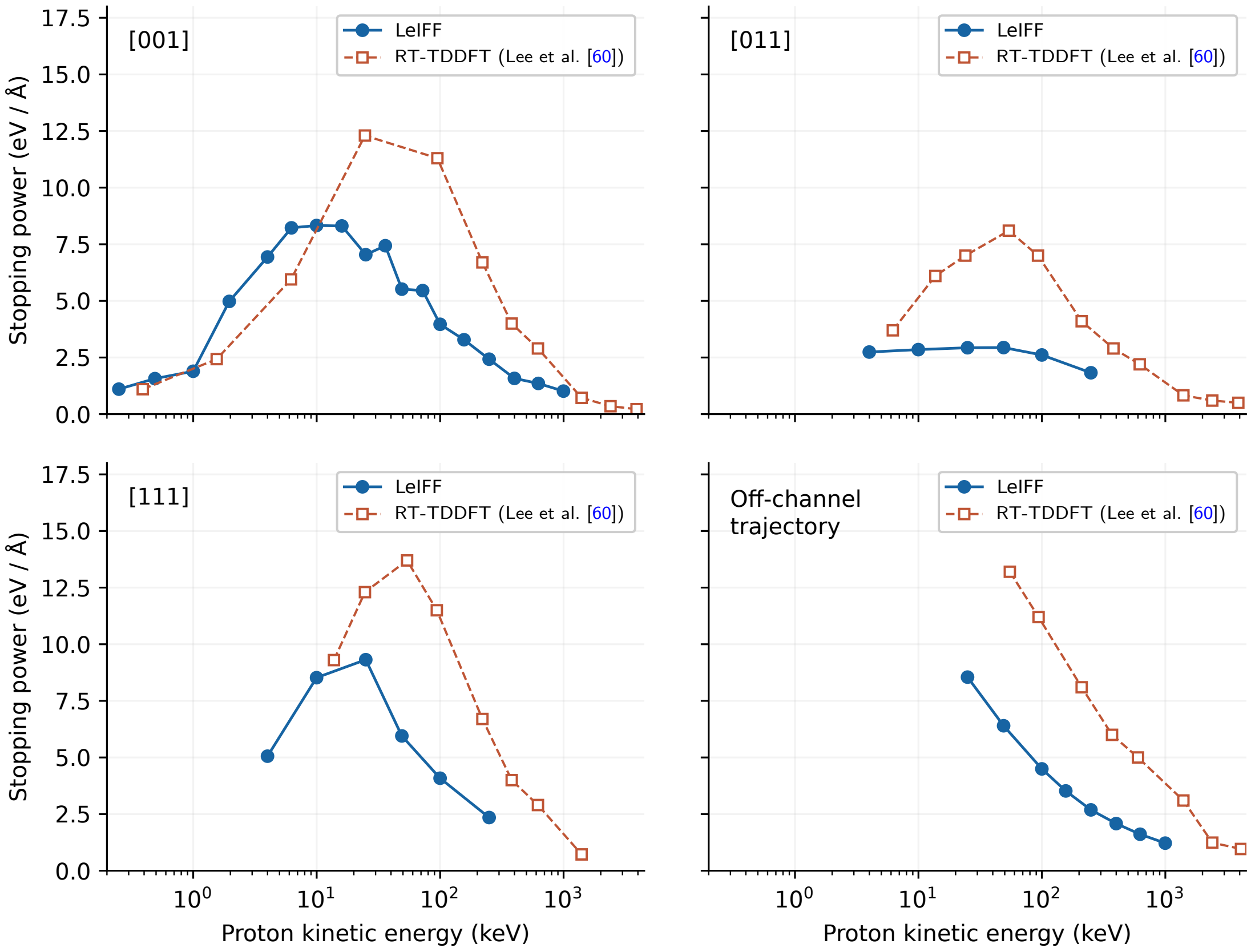}}
\caption{Proton electronic stopping in silicon with LeIFF (filled circles) and RT-TDDFT results from Lee et al.~\cite{lee2020stopping} (open squares).
Panels show three crystal channels and a sampled off-channel trajectory.
LeIFF values average the drag and electronic-energy-slope estimates; connecting lines guide the eye.
All 38 points use the same symbols, including eight numerical width-envelope exits discussed in the text.
This comparison is a preliminary demonstration of the method, not a converged Si stopping benchmark.}
\label{fig:si-supp-stopping}
\end{figure*}

Figure~\ref{fig:si-supp-stopping} compares all 38 trajectories with the RT-TDDFT results of Lee et al.~\cite{lee2020stopping}.
The comparison is intended to assess qualitative energy and channel dependence outside the molecular fitting domain.
Peak magnitudes differ substantially from the reference curves.

Eight trajectories exceed the numerical width envelope of approximately $2.15a_0$, even though the soft confinement centered near $3a_0$ is active.
That envelope reflects the finite-cell interaction-range criterion, rather than failure of the confining potential.
These points are retained with the same symbols for the proof-of-method comparison; their stopping estimates are not numerically certified.
Thus qualitative trends and the scale of electronic energy deposition are reproduced, while peak magnitudes, trajectory sampling, width-range validity, and cell-size convergence require further work before quantitative Si predictions can be claimed.

\subsection{Equation of state and adiabatic lattice vibrations}
\label{sec:si-supp-ground-state}

The equation-of-state and phonon calculations use the same LeIFF molecular parameter file as the response calculations.
The equation of state follows the diamond-lattice branch in a two-ion primitive cell, with electron centers and widths relaxed independently at eleven conventional lattice constants between $5.00$ and $5.85$~\AA.
A third-order Birch--Murnaghan fit over $5.00$--$5.43$~\AA\ gives an equilibrium lattice constant of $5.197$~\AA\ and a bulk modulus of $237.8$~GPa.
The wider-volume points are displayed but not included in this local fit.
The fit does not constitute a search over competing crystal structures or electronic configurations.

For lattice vibrations, a $2\times2\times2$ primitive supercell contains 16 Si ions and 64 electrons.
The electronic coordinates and then internal ionic coordinates are relaxed at fixed volume to maximum force components below $10^{-6}$~Ha/$a_0$.
Central finite differences of the full ionic and electronic energy gradient, with displacement $10^{-6}a_0$, yield the full Hessian.
The electrons follow ionic motion adiabatically through the Schur complement
\begin{equation}
K_{\mathrm{ion}}=H_{aa}-H_{ae}H_{ee}^{-1}H_{ea},
\end{equation}
where the electronic block includes both center and width coordinates.
The electronic Hessian is positive definite at the reference structure.
The maximum Hessian asymmetry is $8.3\times10^{-9}$~Ha/$a_0^2$; projecting out uniform translations changes the ionic force-constant matrix by $2.8\times10^{-9}$ in relative Frobenius norm.
Shortest-image averaging, including equally weighted periodic boundary ties, is used in the real-space interpolation of the force constants.

\begin{figure*}[t]
\centering
\includegraphics[width=0.99\textwidth]{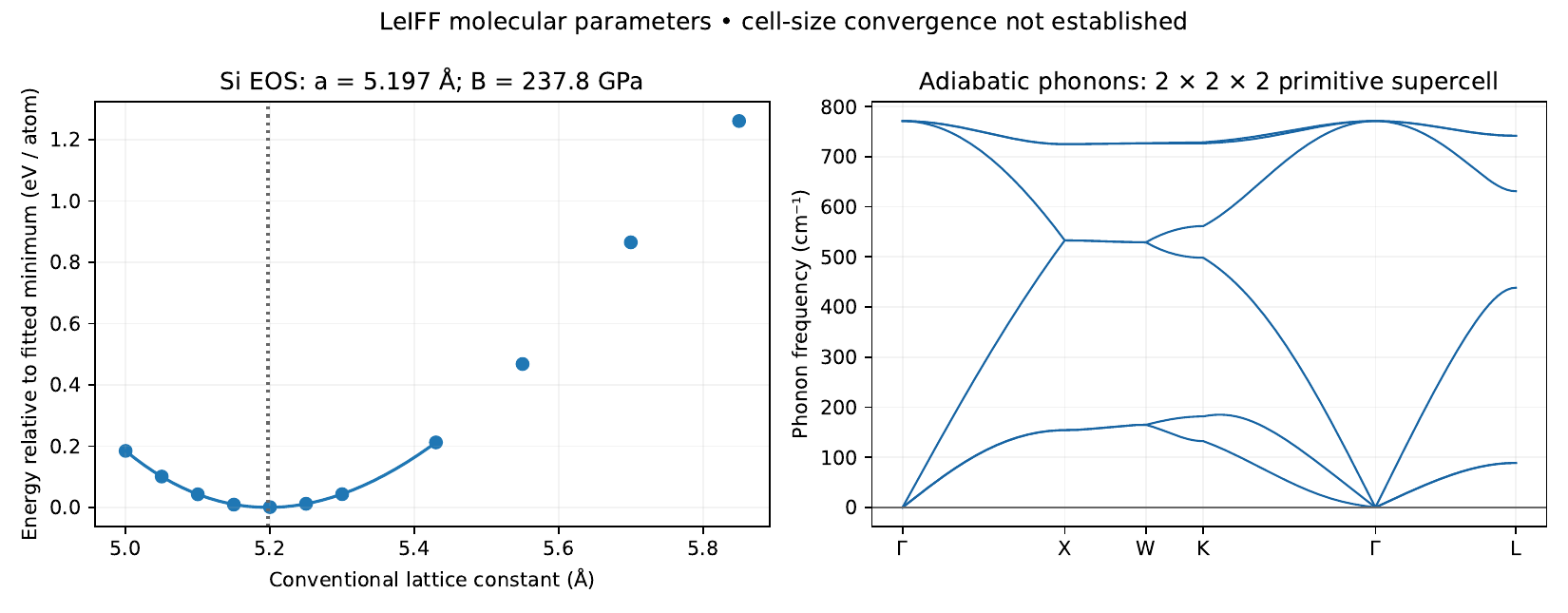}
\caption{LeIFF calculations with the molecular parameterization.
Left: diamond-Si energy versus conventional lattice constant; points are independently relaxed electronic states, and the curve is the local equation-of-state fit.
Right: adiabatic phonon bands interpolated from a $2\times2\times2$ primitive-supercell Hessian at the predicted equilibrium lattice constant.
These preliminary calculations demonstrate lattice binding and vibrational branches; quantitative agreement and cell-size convergence are not established.}
\label{fig:si-supp-ground-state}
\end{figure*}

Figure~\ref{fig:si-supp-ground-state} shows acoustic and optical branches without significant imaginary frequencies; the residual acoustic frequencies at $\Gamma$ are below $0.004$~cm$^{-1}$.
The $\Gamma$ optical triplet lies near $771$~cm$^{-1}$.
These first-pass results establish a bound diamond-lattice branch supporting lattice vibrations, but the predicted structure is too compact and stiff for quantitative Si predictions.
Finite-difference step, primitive-cell EOS, and phonon supercell-size convergence remain to be established.
In particular, absence of imaginary frequencies along the sampled path is not proof of global structural stability.

\subsection{Electronic normal modes and longitudinal density response}

With fixed ions, electronic normal modes include both Gaussian center and width coordinates, with masses $m_e$ and $3m_e$.
The full mass-normalized electronic Hessian is diagonalized for supercells containing 8, 64, and 216 Si ions at the prescribed lattice constant near $5.44$~\AA.
For each mass-normalized mode, the Coulomb stiffness is its projection onto the Coulomb Hessian; the remainder is the projection onto all other energy terms.
These signed stiffness contributions add to the total mode stiffness and are not probabilities.
Density-response residues and the distribution of induced-field energy over reciprocal vectors provide complementary measures of longitudinal character.
Near-degenerate modes are grouped to reduce arbitrary eigenvector-basis dependence in the highlighted responses.

\begin{figure*}[!]
\centering
\includegraphics[width=0.99\textwidth]{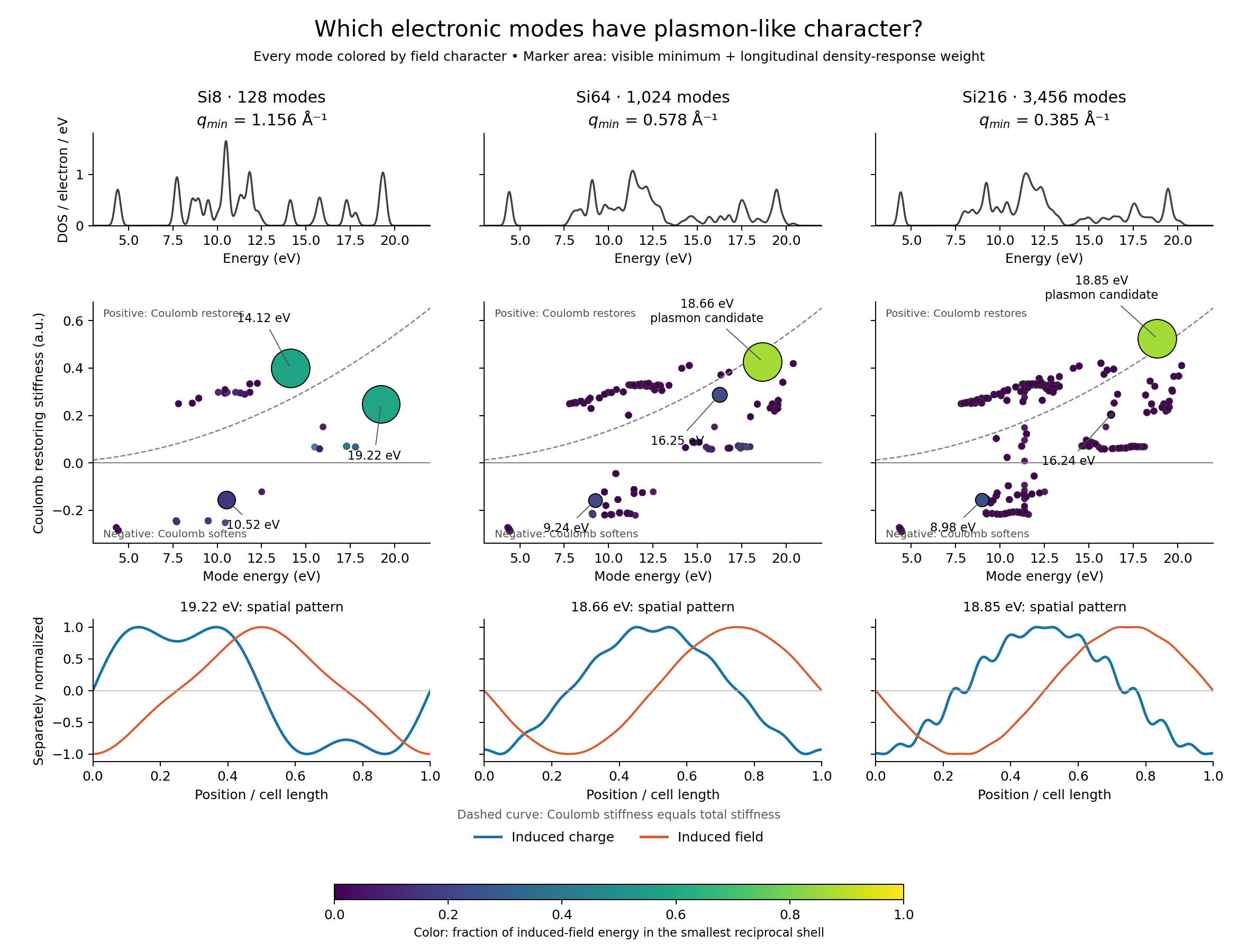}
\caption{Electronic DOS, Coulomb mode stiffness, and longitudinal density character for three Si supercells.
Every electronic mode is colored by the fraction of induced-field energy in the smallest reciprocal shell; marker area has a visible minimum plus a contribution proportional to the relative smallest-wavevector density-response weight within each cell.
The three strongest near-degenerate response groups are highlighted.
The dashed curve marks Coulomb stiffness equal to total stiffness.
Bottom panels show separately normalized plane-averaged induced charge and field patterns of selected density-bright standing-wave combinations; their spatial offset is not a temporal delay.}
\label{fig:si-supp-modes}
\end{figure*}

Figure~\ref{fig:si-supp-modes} identifies prominent finite-wavevector plasmon candidates near $18.66$~eV and $18.85$~eV for Si64 and Si216, respectively.
Their smallest-wavevector longitudinal density responses are strong, their Coulomb stiffnesses are positive, and approximately $86\%$ of their induced-field energy lies in the smallest reciprocal shell.
Uniform dipole coupling can vanish for these finite-wavevector density waves; they are therefore distinct from the directly uniform-kick-accessible modes.
Because each supercell probes a different smallest wavevector, the comparison does not establish fixed-wavevector size convergence or a bulk zero-wavevector plasmon energy.
The original implemented near-coincidence Gaussian branch is retained; its curvature affects some modes and remains a numerical-model qualification for the full electronic DOS.

\end{document}